\documentclass[%
 reprint,
 amsmath,amssymb,
 aps,
]{revtex4-2}

\usepackage{graphicx}
\usepackage{dcolumn}
\usepackage[Symbolsmallscale]{upgreek}
\usepackage{tikz}
\usetikzlibrary{decorations.pathmorphing}
\usepackage{verbatim}
\usepackage{enumitem}
\usepackage{bm}
\usepackage{dsfont}
\usepackage{mathrsfs}
\usepackage{hyperref}

\def\Xint#1{\mathchoice
   {\XXint\displaystyle\textstyle{#1}}%
   {\XXint\textstyle\scriptstyle{#1}}%
   {\XXint\scriptstyle\scriptscriptstyle{#1}}%
   {\XXint\scriptscriptstyle\scriptscriptstyle{#1}}%
   \!\int}
\def\XXint#1#2#3{{\setbox0=\hbox{$#1{#2#3}{\int}$}
     \vcenter{\hbox{$#2#3$}}\kern-.5\wd0}}

\def\dashint{\Xint-}

\newcommand{\wh}[1]{\widehat{#1}}

\newcommand{\mc}[1]{\mathcal{#1}}

\newcommand{\tl}{\tilde}

\newcommand{\ul}[1]{\underline{#1}}
\renewcommand{\i}{\text{i}}
\newcommand{\e}{\text{e}}

\newcommand{\ud}{\text{d}}

\newcommand{\cD}{\mathcal{D}}

\begin{document}


\title{A gauge invariant Hamiltonian evolution across the black hole horizon \\ in asymptotically AdS spacetimes}

\author{Anurag Kaushal}
\email{anuragkaushal314@gmail.com}
\author{Naveen S. Prabhakar}%
\email{naveen.s.prabhakar@gmail.com}
\author{Spenta R. Wadia}
\email{spenta.wadia@icts.res.in}
\affiliation{International Centre for Theoretical Sciences-\\
 Tata Institute of Fundamental Research, Shivakote, Bengaluru 560089, India.
}%




\begin{abstract}
	We study the quantum dynamics of a probe scalar field in the background of a black hole in AAdS spacetime in the Hamiltonian formulation of general relativity in the maximal slicing gauge. The black hole solution in this gauge is expressed in terms of `wormhole coordinates', a smooth coordinate system with constant time slices that cut across the horizon, and asymptote to the Killing time slices at the boundaries.
	
	The quantum scalar field is expanded in terms of normalized solutions of the Klein-Gordon equation, that are valid at all points in spacetime. The operators that appear in the expansion are in the product space of the CFTs on the two spacetime boundaries, which are by definition gauge invariant under small bulk diffeomorphisms. The entangled Hartle-Hawking (HH) state arises naturally from this construction.
	
	One of our main results is a well defined formula for the time dependent Hermitian Hamiltonian of the probe scalar in the product space of the two CFTs, which describes the time development of operators/states along the maximal slices. This Hamiltonian acting on the HH state creates a state of finite norm. Consequently there is a unitary description of horizon crossing scalar field excitations on top of the HH state. We also present a bulk reconstruction formula that evaluates an order parameter that signals horizon crossing in the boundary theory.
	
	We calculate various bulk Wightman two-point functions on the two-sided BTZ black hole. We recover Hawking's thermodynamic results in the exterior region when expressed in terms of BTZ coordinates that are related to the wormhole coordinates by a transformation which is singular at the horizon. We compute the two-point function with one insertion in the future/past interior and the other in the exterior. Both are related by a time reflection symmetry and asymptote to a non-zero constant as the coordinate time between the two points becomes large.
\end{abstract}

\maketitle

\tableofcontents

\section{Introduction and Summary}

The quantum dynamics of matter in the presence of a black hole is an
old subject beginning with Hawking’s foundational work
\cite{Hawking:1975vcx}, where he calculated the temperature of the
Schwarzschild black hole as a function of its mass,
$T_{\rm BH} =\frac{\hbar}{8\pi MG_N}$.  A simpler exposition of the
same result was given many years later by Haag and Fredenhagen
\cite{Fredenhagen:1989kr} (also see the review by Witten
\cite{Witten:2024upt}). This formula plus the first law of
thermodynamics leads to the Bekenstein-Hawking formula for the 
entropy of the black hole, $S= \frac{A_{\rm hor}}{4G_{N} \hbar}$. 
One of the key inputs in these calculations, which involves matter 
propagating in the black hole background, is the fact that the 
spacetime is restricted to be outside the horizon of the black hole 
and described by coordinates (e.g. Schwarzschild), 
that do not extend beyond the horizon.  

One of the important milestones of string theory has been the explanation of black hole thermodynamics and Hawking radiation (for a class of supersymmetric black holes) in terms of the statistical mechanics of microscopic degrees of freedom that are unitary quantum mechanical systems \cite{Strominger:1996sh, David:2002wn}. The AdS/CFT correspondence \cite{Maldacena:1997re, Witten:1998qj, Gubser:1998bc} made this explanation precise.

The Ryu-Takayanagi formula and its generalisations \cite{Ryu:2006bv, Hubeny:2007xt, Lewkowycz:2013nqa, Faulkner:2013ana, Engelhardt:2014gca} for the von Neumann entropy enabled a new diagnostic of black hole evaporation in terms of the 
Page curve for an evaporating black hole in an AdS spacetime that is coupled to an external bath (a quantum field theory) 
at its boundary \cite{Penington:2019npb, Almheiri:2019psf, Almheiri:2019hni, Almheiri:2019yqk, Penington:2019kki, Almheiri:2020cfm}. These are semi-classical bulk calculations in effective field theory without any reference to the dual degrees of freedom that reside on the boundary of AdS spacetime.

One wonders about the possibility, within the effective field theory description of the bulk, of a \emph{unitary} Hamiltonian description of the crossing of the horizon by a wave-packet built out of a finite number of diffeomorphism invariant matter field excitations on top of the Hartle-Hawking state. In this paper we answer this question in the
affirmative. The present work was partly motivated by the work of Leutheusser and Liu \cite{Leutheusser:2021qhd, Leutheusser:2021frk}, where a CFT operator which describes the time evolution of infalling observers in the black hole background was obtained using algebraic QFT methods. Ideas and questions related to infalling observers have been discussed previously in \cite{Papadodimas:2012aq, Papadodimas:2013wnh, Papadodimas:2013jku}. 

We study the quantum dynamics of a probe scalar field in the background of an eternal two-sided black hole in a spacetime that is asymptotically AdS (AAdS) paying special attention to gauge invariance. Though we present explicit formulas for the case of $2+1$ dimensions, our main results are valid for $d+1$ dimensions with $d > 2$.

We work in the framework of Hamiltonian General Relativity. An
important result in mathematical relativity is that AAdS spacetimes
can always be foliated by spatial slices that have zero or constant
mean curvature \cite{Andersson:1992yk, Andersson:1996xd,
Sakovich:2009nb, Chrusciel:2022cjz, Witten:2022xxp, Wittentalk}. In
this paper we will work the $K=0$ gauge, which is also known 
to maximize the spatial volume in the Lorentzian
signature of spacetime; hence the name maximal slicing gauge.

It is possible to transform the two-sided $d+1$ dimensional
AdS-Schwarzschild black hole solution to the maximal slicing gauge, in
which the $d$ dimensional spatial slices are cylinders
($\mathbf{R} \times \mathbf{S}^{d-1}$) or \emph{wormholes}. We
specialize to $d=2$ in this paper in which case we work with the
two-sided BTZ black hole. We obtain a smooth spacetime metric in this
gauge in what we call \emph{wormhole coordinates}
$(\bar{t}, x, \varphi)$. The time coordinate $\bar{t}$ labels the
maximal slices and has the property that it is linearly related to the
asymptotic AdS times at the two AAdS boundaries.\footnote{The time
  $\bar{t}$ can be identified with the formula for time presented in
  \cite{Kaushal:2024xob}, where it was shown that the notion of time
  derived from the Einstein-Hamilton-Jacobi equation in terms of the
  configuration space data precisely matches with the time in ADM
  decomposition of the spacetime metric.} The spatial coordinate $x$
is globally defined on the entire spatial slice.  The two AAdS
boundaries are at $x \to \pm \infty$, with $x$ behaving as an areal
radial coordinate in the asymptotic regions. The coordinate $\varphi$
is an angular coordinate with period $2\pi$. In higher dimensions, it
is replaced by the usual set of angular coordinates for a $d-1$
dimensional sphere $\mathbf{S}^{d-1}$.

These maximal slices have the following properties:
\begin{enumerate}[topsep=5pt,itemsep=0pt]
\item They smoothly cut across the horizon and the lapse function is
  non-zero, smooth and of one sign on the entire slice, including the
  bifurcate point. This is to be contrasted with the static BTZ or
  AdS-Schwarzschild solution.
\item They asymptote to the usual Schwarzschild slices near the
  boundaries of the AAdS spacetime.
\end{enumerate}
Section \ref{BTZreview} reviews the maximal slicing gauge and its
properties including various coordinate systems: Schwarzschild,
Kruskal-Szekeres, areal and wormhole coordinates in detail.

In Section \ref{scalarquant}, we quantize a probe free scalar field of
mass $m$ in this background.  Using the extrapolate dictionary near
the two boundaries, the scalar field
can be expressed in terms of generalised free fields in the two dual
CFTs associated to the two AAdS boundaries with conformal dimension
$\Delta_{+} = \frac{d}{2} + \sqrt {\frac{d^2}{4} +
  \ell^{2}m^2}$. These generalized free field operators in the CFT
are, by definition, invariant under small diffeomorphisms in the
bulk. It is important to note that this description of the bulk scalar
field in terms of gauge invariant generalized free fields on the two
boundary CFTs is possible since we are considering a two sided
description of the extended eternal black hole solution.
If we work with a one sided description with one AAdS boundary, 
then the gauge invariance of the scalar modes behind the horizon needs 
to be discussed using gravitational `Wilson lines' extending to the boundary.
  
We review Unruh's procedure \cite{Unruh:1976db} of obtaining the
scalar field solutions in the interior regions of the black hole by
analytically continuing the exterior scalar field solutions in the
lower-half Kruskal $U$ and $V$ planes. This procedure naturally gives
a mode expansion of the scalar field in terms of the Hartle-Hawking
modes, and picks out the Hartle-Hawking state as the state
annihilated by these modes.

The Hartle-Hawking mode functions $h_{\omega q}$ are labelled by a
continuous real parameter $\omega$, which can take both positive and
negative values, and a quantized angular momentum mode $q$. Though the
parameter $\omega$ is allowed to take arbitrary large values, since we
are working in the probe limit of the quantum scalar field, one should
(as we will see later) work with $\omega$ of the order of $\eta$, the surface gravity of the
black hole, and even then, only a finite number of such excitations
about the Hartle-Hawking state to avoid backreaction effects.

If one looks closely at the intermediate steps of solving the
Klein-Gordon equation, it can be seen that the $\omega$-continuum is
due to the presence of the null horizon in the black hole
solution. Thus, there is a continuous spectrum of scalar field
excitations about the Hartle-Hawking state. By the AdS/CFT
extrapolate dictionary, this implies that, in the dual CFT, there is a
continuum of energies in the neighbourhood the large energy state that
is dual to the black hole in the bulk. This continuum is a result
of working at leading order in large $N$ in the dual CFT, where it is
well-known that the spacing between energy levels in a neighbourhood
of a state with energy $N^2$ goes as $\e^{-c N^2}$, which is zero in
the strict large $N$ limit.

It turns out that the Hartle-Hawking modes $h_{\omega q}$, when
expressed in Kruskal-Szekeres coordinates, have derivatives that are
singular at the horizon. Hence, as defined, the Hartle-Hawking modes
are not smooth solutions of the Klein-Gordon equation in the
neighbourhood of the horizons. We remedy this by smearing these modes 
in the $\omega$-space with Hermite functions and
show that the new mode functions are smooth at the horizons. Thus,
the final expression for the scalar field that satisfies the
Klein-Gordon equation at all spacetime points including at the
horizons is expressed as a linear combination of the smeared
Hartle-Hawking modes. These are in turn related to the operators of
the two boundary CFTs by a Bogoliubov transformation. We also write down a bulk reconstruction formula in terms of two boundary to bulk kernels (corresponding to each boundary) in the wormhole coordinates.

In Section \ref{HamSF} we discuss the scalar field Hamiltonian in 
this background, expressed in terms of the frequency-smeared
Hartle-Hawking modes. The Hamiltonian is a time-dependent Hermitian
operator that describes the unitary time evolution of operators/states
as they traverse the horizon. We argue that it is a well defined
operator as it's action on the Hartle-Hawking state creates a state of
finite norm. 

The above facts would imply that the Wightman two-point correlators 
of the scalar field in wormhole coordinates retain information as in 
a unitary theory analogous to a quantum field theory in Minkowski 
spacetime. Indeed, we explicitly show that correlation functions in 
the interior remain finite and non-zero, even at late times.

In Section \ref{sec:OP} we propose an order parameter that signals horizon crossing from the boundary perspective. This order parameter uses a bulk reconstruction formula presented in \ref{sec:BR} and the explicit form of the time evolution operator. Such an order parameter has been discussed in \cite{Leutheusser:2021qhd} in the framework of algebraic QFT. It would be interesting to know the relation between these two constructions.


Section \ref{TPFunction} discusses the two-point function of the
scalar field in the wormhole coordinates for various insertion points
in spacetime, including the case in which the points are located on
the boundaries of the spacetime.

Hawking's result of thermal correlation functions in the bulk region
outside the horizon is recovered only after we make a coordinate
transformation from the wormhole coordinates to Schwarzschild
coordinates. The formula for the black hole temperature as a function
of its mass is recovered. It is important to note that this coordinate
transformation is singular at the horizon.

In the situation where the two point function is for one scalar field
insertion on each asymptotic boundary, we recover the result obtained
in \cite{Maldacena:2001kr} which has the stark feature of decaying to
zero beyond times of the order of the inverse temperature $\beta$ of
the black hole. As explained in \cite{Maldacena:2001kr}, this is an
artefact of working in the strict large $N$ limit, in which case the
dual CFT has a continuous spectrum of energies even though it is
defined on a compact space ($\mathbf{S}^{d-1}$). In particular, at
finite but large $N$, it has been conjectured in
\cite{Maldacena:2001kr} that additional saddle points other than the
black hole may contribute to the scalar field two point function such
that the decay to zero is modified to a small non-decaying answer 
which is consistent with unitarity of the dual CFT. However, we cannot
see these effects in the current work since we are working in the
infinite $N$ limit.


It is worth noting that at large but finite $N$, our description of the scalar field dynamics in the BTZ black hole background would be valid only for times $\bar{t}$ much smaller than the scrambling time $t_*$. This is because, at times of order $t_*$, we expect that the given black hole background may not be the dominant saddle point due to backreaction effects becoming important \cite{Shenker:2013pqa}. We can also infer this from the observation that the dual CFT four-point out-of-time-ordered correlator ceases to factorize into the product of two two-point functions at times of order $t_* \sim \frac{\beta}{2\pi} \log N$ \cite{Maldacena:2015waa}. This indicates that the large $N$ (corresponding to small $G_N$ in the bulk) saddle point analysis in the dual CFT and in the bulk is no longer valid at times of order $t_*$. Since our analysis is in the large $N$ limit, the scrambling time $t_*$ is sent to infinity and the black hole remains the dominant saddle point for all times.

In a future work, it would be nice to study the black hole evaporation in our setup upon coupling the unitary scalar field theory in the bulk to a bath. This can be achieved by connecting the AAdS space to flat space regions similar to the studies in JT gravity  \cite{Penington:2019npb, Almheiri:2019psf, Almheiri:2019hni, Almheiri:2019yqk, Penington:2019kki, Almheiri:2020cfm} and the dual SYK model \cite{Maldacena:2019ufo, Almheiri:2019jqq, Chen:2020wiq, Gaikwad:2022jar}.

\section{A review of the maximally sliced two-sided BTZ black
  hole}\label{BTZreview}

The (AdS)-Schwarzschild black hole can be put in the in the maximal
slicing gauge in any dimension $d\ge 2$. This is obtained by starting
with a spherically symmetric ansatz for the metric and then solving
the Einstein's equations together with the maximal slicing gauge
condition, along the lines of \cite{Estabrook:1973ue}. The maximal
slicing condition is a local condition on the extrinsic curvature
$K_{ij}$ of a given spatial slice:
\begin{equation}
  K \equiv  g^{ij} K_{ij} = 0\, ,
\end{equation}
where $g^{ij}$ is the inverse of the spatial metric on the given
spatial slice. The maximal slicing condition implies that the local
spatial volume is a maximum under arbitrary small deformations of the
slice.

In this paper, we mainly work with the maximally sliced $2+1$
dimensional BTZ black hole for concreteness. This solution was
obtained in the companion paper \cite{KPWI} by systematically solving
Einstein's equations in the Hamiltonian formulation. We also present
an independent, self-contained derivation of the maximally sliced BTZ
black hole solution in Appendix \ref{msbtz-app}. We review the
necessary aspects of this solution in this section and refer the
reader to Appendix \ref{msbtz-app} and the companion paper \cite{KPWI} for more details. First, we quickly present some
well-known facts about the BTZ black hole and its Kruskal extension.

\subsection{The fully extended two-sided BTZ black
  hole}\label{BTZrecall}

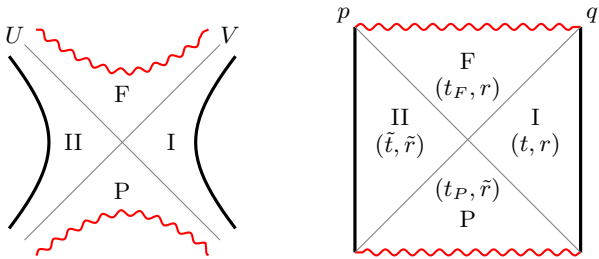
\begin{figure}[!htbp]
  \centering
  \begin{tikzpicture}[scale=0.65]
    \draw[black, very thick] plot[domain=-1.:1.] ({1.5*cosh(\x)},{1.5*sinh(\x)});
    \draw[black, very thick] plot[domain=-1.:1.] ({-1.5*cosh(\x)},{1.5*sinh(\x)});
    \draw[decorate, decoration={snake, amplitude=0.4mm, segment length=2.4mm}, thick, red] plot[domain=-1.:1.] ({1.5*sinh(\x)},{1.5*cosh(\x)});
    \draw[decorate, decoration={snake, amplitude=0.4mm, segment length=2.4mm}, thick, red] plot[domain=-1.:1.] ({1.5*sinh(\x)},{-1.5*cosh(\x)});
    \draw[gray] (-2,-2) -- (2,2);
    \draw[gray] (-2,2) -- (2,-2);
    \node[black] at (2.2,2.2) {$V$};
    \node[black] at (-2.2,2.2) {$U$};
    \node[black] at (1,0) {I};
    \node[black] at (0,1) {F};
    \node[black] at (-1,0) {II};
    \node[black] at (0,-1) {P};
  \end{tikzpicture}
  \qquad\quad \begin{tikzpicture}[scale=0.75]
    \draw[black, very thick] (-2,-2) -- (-2,2);
    \draw[black, very thick] (2,-2) -- (2,2);
    \draw[decorate, decoration={snake, amplitude=0.4mm, segment length=2.4mm}, thick, red] (-2,2) -- (2,2);
    \draw[decorate, decoration={snake, amplitude=0.4mm, segment length=2.4mm}, thick, red] (2,-2) -- (-2,-2);
    \draw[gray] (-2,-2) -- (2,2);
    \draw[gray] (-2,2) -- (2,-2);
    \node[black] at (2.2,2.2) {$q$};
    \node[black] at (-2.2,2.2) {$p$};
    \node[black] at (1.2,0.4) {I};
    \node[black] at (1.2,-0.1) {$(t,r)$};
    \node[black] at (0,1.4) {F};
    \node[black] at (0,0.9) {$(t_F,r)$};
    \node[black] at (-1.2,0.4) {II};
    \node[black] at (-1.2,-0.1) {$(\tl{t},\tl{r})$};
    \node[black] at (0,-1.4) {P};
    \node[black] at (0,-0.9) {$(t_P,\tl{r})$};
  \end{tikzpicture}
  \caption{\label{BTZKruskal} The Kruskal (left) and Penrose (right)
    diagrams for the two-sided BTZ black hole solution. The two solid
    black lines are the asymptotic AdS boundaries, the diagonal gray
    lines are the event horizons and the wiggly red lines are the
    singularities. The horizons divide the two-sided black hole into
    four regions labelled I, II, F and P, each of which have a
    static Schwarzschild-like coordinate system which is displayed in
    the Penrose diagram.}
\end{figure}

The BTZ black hole metric is \cite{Banados:1992wn, Banados:1992gq}
\begin{equation}\label{BTZmet}
	\ud s^2 = - f(r) \ud t^2 + \frac{\ud r^2}{f(r)} + r^2 \ud\varphi^2\, ,\ \ f(r) = \frac{r^2}{\ell^2} -\frac{M}{2\pi}\, .
\end{equation}
The spacetime has an asymptotic AdS boundary as $r \to \infty$. This
metric has a coordinate singularity at the horizon
\begin{equation}\label{hordef}
  r = R_h = \ell \sqrt{\frac{M}{2\pi}}\ ,
\end{equation}
at which the time translation
Killing vector becomes null.  The above metric is valid for the range
\begin{equation}
	\text{Region I}:\quad R_h < r < \infty\ ,\ - \infty < t < \infty\ ,\ 0 \leq \varphi < 2\pi\, .
\end{equation}
To go past the null horizon, one first defines the tortoise coordinate
$r_*$:
\begin{equation}
	r_* = \int_\infty^r \frac{\ud \rho}{f(\rho)} = \frac{1}{2\eta} \log \frac{r - R_h}{r + R_h}\, ,
\end{equation}
where $\eta = R_h / \ell^2$ is the surface gravity of the black
hole. The Kruskal-Szekeres coordinates are defined in terms of the
Region I coordinates as
\begin{equation}
	U = -\eta^{-1} \e^{-\eta(t - r_*)}\ ,\quad V = \eta^{-1} \e^{\eta(t + r_*)}\, .
\end{equation}
In Region I, we have $U < 0$ and $V > 0$. The past horizon
$t \to -\infty$, $r \to R_h$ is described by the surface $V = 0$
whereas the future horizon $t \to \infty$, $r \to R_h$ is described by
the surface $U = 0$. The BTZ black hole metric in terms of the
Kruskal-Szekeres coordinates is
\begin{equation}\label{BTZkruskal}
	\ud s^2 = -4 \frac{R^2_h}{\ell^2} \frac{\ud U \ud V}{(1 + \eta^2 UV)^2} + R_h^2 \left(\frac{1 - \eta^2 UV}{1 + \eta^2 UV}\right)^2 \ud \varphi^2\, .
\end{equation}
This metric has no singularities at the horizons $UV = 0$, and is
valid in the range $|\eta^2UV| < 1$, with $U,V \in \mathbf{R}$. This
is the Kruskal extension of the BTZ black hole which we refer to as
the two-sided BTZ black hole. See the left panel of Figure
\ref{BTZKruskal} for the Kruskal diagram.

There are four regions corresponding to the different signs of $U$ and
$V$ where we can find BTZ-like coordinates in which the metric takes
the form \eqref{BTZmet}:
\begin{equation}\label{signs}
\text{I}:\  (-,+)\ ,\quad \text{F}:\ (+, +)\ ,\quad \text{II}:\ (+,-)\ ,\quad \text{P}:\ (-, -)\ .
\end{equation}
These four regions are shown in Figure \ref{BTZKruskal}. It is also
useful to define the Penrose coordinates $(p, q)$:
\begin{align}
	p = \arctan(\eta V)\, , \quad  q = \arctan(\eta U)\, .
\end{align}
The Penrose diagram of the fully extended BTZ black hole is displayed in the second panel in Figure \ref{BTZKruskal}.

\subsection{The maximally sliced two-sided BTZ black hole in the
  wormhole coordinate system}

\begin{figure}
	\centering 
	\includegraphics[width=0.45\linewidth]{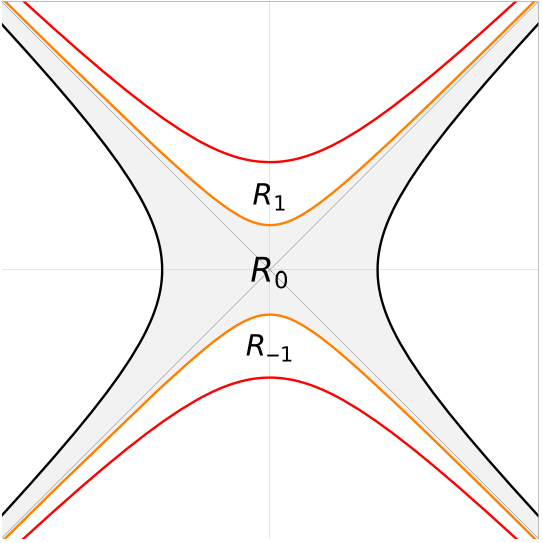}\hfill
	\includegraphics[width=0.45\linewidth]{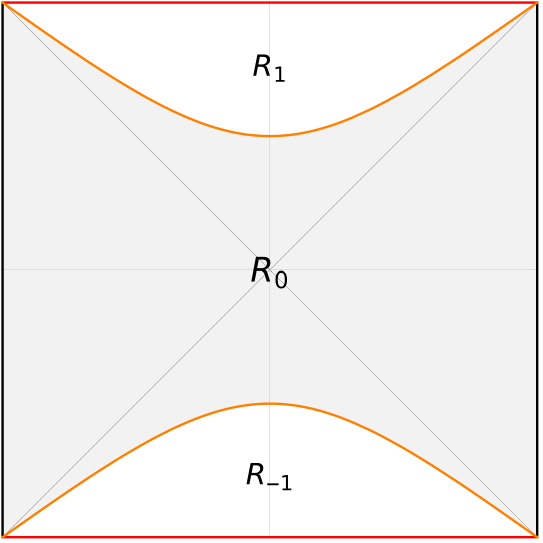}
	\caption{\label{fig:WH1Regions} The region $R_0$ (shaded gray) covered by the maximal slicing solution is demarcated by the orange lines in the BTZ Kruskal diagram (left) and Penrose diagram right) of the fully extended BTZ black hole. The three regions $R_0$, $R_1$ and $R_{-1}$ cover the full Penrose diagram. The regions $R_1$ and $R_{-1}$ (unshaded) are covered by a different family of maximal slices as is explained in the main text.}
\end{figure}


As mentioned in the introduction to this section, we derive the
maximally sliced two-sided BTZ black hole solution in Appendix
\ref{msbtz-app}. The spatial slices of this solution are
\emph{wormholes} or cylinders $\mathbf{R} \times \mathbf{S}^1$
connecting two AAdS boundaries.

This solution is presented in the \emph{wormhole} coordinate system $(\bar{t}, x, \varphi)$ with ranges
\begin{equation}\label{wormholerange}
  -\infty < \bar{t} < \infty\, ,\quad -\infty < x < \infty\, ,\quad 0 \leq \varphi \leq 2\pi\, ,
\end{equation}
where $\varphi$ is an angular coordinate with period $2\pi$. The
coordinate $\bar{t}$ is a time coordinate which labels spatial maximal
slices, and $x$ is a spatial coordinate which we call the \emph{wormhole} coordinate. The two AAdS boundaries are situated at $x = \pm \infty$ and the centre of the wormhole throat is at $x = 0$. The wormhole coordinate $x$ has the virtue that it behaves as an areal radial coordinate in the region near the boundaries $x \sim \pm \infty$.

The spacetime metric for the maximally sliced two-sided BTZ black hole
is
\begin{multline}\label{WormholeM}
  \ud s^2 = - N(x,\bar{t})^2 \ud\bar{t}^2 +  \frac{\ell^2\big(\ud x + N^x(x,\bar{t}) \ud \bar{t}\big)^2}{x^2 + R_+(\bar{t})^2 - R_-(\bar{t})^2}\\
  + \big(x^2 + R_+(\bar{t})^2\big)\ud\varphi^2\ ,
\end{multline}
with
\begin{align}\label{lapseshiftmaxl}
  	N^x(x,\bar{t}) &= \frac{1}{x} \Big(N(x,\bar{t}) T(\bar t) + R_+ \dot{R}_+\Big)\, , \\
  	N(x, \bar{t}) &= \frac{x\sqrt{x^2 + R_+^2 - R_-^2}}{\ell \sqrt{x^2 + R_+^2}} \times \nonumber\\
                  &\times\left(1 + \dot{T} \ell^3 \int_x^\infty \frac{\ud y}{y^2} \frac{\sqrt{y^2 + R_+^2}}{(y^2 + R_+^2 - R_-^2)^{3/2}}\right)\, , \\
	R_\pm(T(\bar t)) &= \ell\sqrt{\frac{M}{4\pi}} \left[1 \pm \sqrt{1-\left(\frac{4\pi T(\bar t)}{M\ell}\right)^2} \right]^{1/2}\, ,
\end{align}
and the function $T(\bar t)$ is determined by the equation
\begin{equation}\label{Tdefsym0}
	\bar t = - T(\bar t) \ell^3 \dashint^\infty_0 \frac{\ud y}{(y^2 - R_-^2)\sqrt{(y^2 + R_+^2)(y^2 + R_+^2 - R_-^2)}}\, .
\end{equation}
The above metric is time dependent as is expected in absence of a global timelike Killing vector. Further it is completely smooth in the coordinate range \eqref{wormholerange}.

\subsection{The diffeomorphism to the Kruskal extension of the BTZ
  black hole}

The smooth spacetime metric \eqref{WormholeM} is diffeomorphic to a
region $R_0$ the two-sided BTZ black hole metric \eqref{BTZkruskal}
written in Kruskal-Szekeres coordinates. The region $R_0$ is shown in
Figure \ref{fig:WH1Regions}. It is bounded by the two branches of the
hyperbola
\begin{equation}\label{limitslices}
  UV = \frac{1}{\eta^2} \frac{\sqrt{2} - 1}{\sqrt{2} + 1}\ ,
\end{equation}
one branch each in Region F and Region P. The diffeomorphism between
wormhole coordinates and Kruskal-Szekeres in the region $R_0$ is given
by
\begin{widetext}
\begin{align}\label{UVwormholediff}
  U(\bar{t},x) &= \pm \eta^{-1} \exp\left(-\eta \bar{t} - R_h \dashint_x^\infty \frac{\ud y}{(y^2 - R_-^2)\sqrt{y^2 + R_+^2}} \left(\frac{\ell T}{\sqrt{y^2 + R_+(\bar{t})^2 - R_-(\bar{t})^2}} + y\right)\right) ,\nonumber\\
  V(\bar{t},x) &= \pm \eta^{-1} \exp\left(\eta \bar{t} + R_h \dashint_x^\infty \frac{\ud y}{(y^2 - R_-^2)\sqrt{y^2 + R_+^2}} \left(\frac{\ell T}{\sqrt{y^2 + R_+(\bar{t})^2 - R_-(\bar{t})^2}} - y\right)\right) .
\end{align}
\end{widetext}

The signs in front of the expressions on the right hand side
correspond to the four regions in the two-sided black hole, see Figure
\ref{BTZKruskal} and equation \eqref{signs} for the choice of signs
for each region. The $\dashint$ symbol indicates that one has to use
the Cauchy principal value to evaluate the integral when the
integration range includes the pole at $y = \pm R_-(\bar{t})$. This
happens when $x < R_-$ or $x < -R_-$. The diffeomorphism expressions
are derived in Appendix \ref{msbtz-app}.

The coordinate transformation between the wormhole
$(\bar t, x, \varphi)$ coordinates and the Kruskal-Szekeres
$(U,V, \varphi)$ coordinates is smooth and well-defined within the
limit slices $\bar t=\pm\infty$ which are given by the equation
\eqref{limitslices} in Kruskal-Szekeres coordinates. It can be checked
that the matrix of first partials, i.e., the Jacobian matrix is
completely non-singular in the domain $R_0$ of the Kruskal diagram:
  \begin{equation}
   \det \begin{pmatrix} \frac{\partial U}{\partial \bar{t}} & \frac{\partial U}{\partial x} & 0 \\[6pt] \frac{\partial V}{\partial \bar{t}} & \frac{\partial V}{\partial x} & 0 \\[6pt] 0 & 0 & 1 \end{pmatrix} \neq 0\, .
  \end{equation}
We compute this Jacobian near the horizon where it is most suspect to vanish or diverge. In particular, near the right future horizon ($U=0, V>0$), we can expand \eqref{UVwormholediff} around $x=R_-$ (location of the $U=0$ horizon in the wormhole coordinates) to obtain
\begin{align}
	U 
	\approx - \e^{-\eta\bar t} (x-R_-)\, , \quad
	V 
	\approx \eta^{-1} \e^{\left(\eta\bar t + \frac{R_h^2 x}{2R_+^2 R_-} \right)}\, .
\end{align}
From these we can calculate the Jacobian to be
\begin{align}
	J =& \left( \frac{\partial U}{\partial \bar t} \frac{\partial V}{\partial x} - \frac{\partial U}{\partial x} \frac{\partial V}{\partial \bar t} \right) \nonumber \\
	\approx & \e^{ \frac{R_h^2 x}{2R_+^2 R_-}} \Bigg[ \left((x-R_-) + \eta^{-1}\dot{R}_- \right) \frac{R_h^2}{2R_+^2 R_-} \nonumber \\
	&\qquad\quad +\left( 1-\frac{R_h^2 x (2\dot{R}_+R_- + R_+\dot{R}_-)}{2\eta R_+^3 R_-^2} \right) \Bigg] .
\end{align}
	This clearly remains finite and non-zero even on the horizon at $x=R_-$ at all times,
\begin{equation}
	J(x=R_-) 
	= \e^{ \frac{R_h^2}{2R_+^2}} \left( 1-\frac{R_h^2 \dot{R}_+}{\eta R_+^3} \right),
\end{equation}
including at the bifurcate point at $\bar t=0$,
\begin{equation}
	J(x=0,\bar t=0) = \sqrt{\e}\, .
\end{equation}
We have used the fact that $R_+(0)=R_h$ and $\dot{R}_+(0)=0$.

\subsection{Important features of the maximally sliced two-sided BTZ
  black hole solution}\label{maxprop}

\begin{figure}[]
	\centering 
	\includegraphics[width=0.45\linewidth]{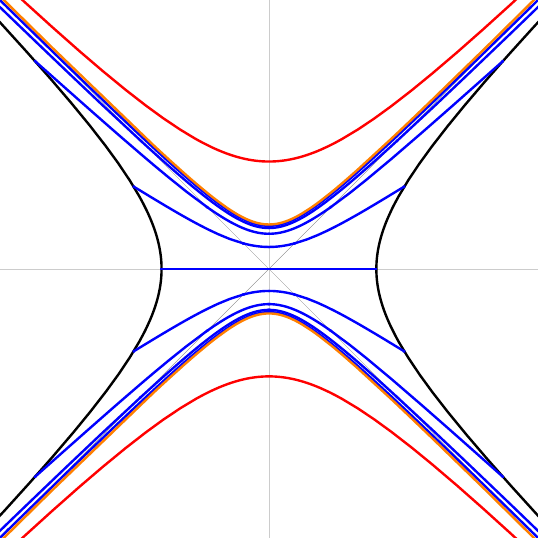}\hfill
	\includegraphics[width=0.45\linewidth]{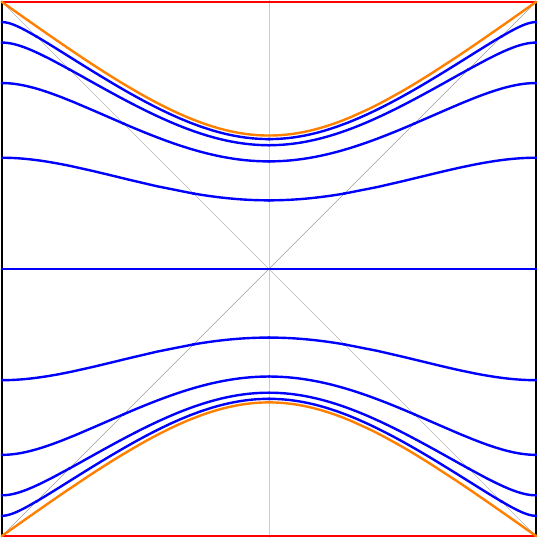}
	\caption{\label{fig:WH1F} Plots of constant $\bar t = 0, \pm 0.5, \pm 1,\pm 1.5, \pm 2$ slices, i.e., maximal slices (solid blue lines) in the BTZ Kruskal diagram (left) and Penrose diagram (right) of the fully extended BTZ black hole. As $\bar t\to\infty$, the final slice approaches $R_\infty =  \ell\sqrt{M/4\pi}$ (orange curve) from below.}
\end{figure}

Let us note some important properties of the maximally sliced
two-sided BTZ spacetime metric in wormhole coordinates.

\begin{enumerate}
\item The constant $\bar{t}$ slices are maximal slices in the
  two-sided BTZ black hole geometry and go from the left asymptotic
  boundary to the right asymptotic boundary while smoothly cutting
  across the two horizons located at $x=\pm R_-(\bar{t})$, see Figure
  \ref{fig:WH1F}. The maximal slicing time $\bar t$ matches the
  boundary time $t$ ($-\tl{t}$) on the right (left) AAdS boundary.

\item \emph{Cosmic censorship.}  The regions $R_1$ and $R_{-1}$ in Figure \ref{fig:WH1Regions} are not covered by the spacetime solution
  \eqref{WormholeM}. Thus these maximal slices do not cover a portion
  of the black hole spacetime near the singularities. A type of
  `\emph{cosmic censorship}' is automatic in the maximal slicing
  gauge. This implies that the dynamics in the dual conformal field
  theories that is dictated by the maximal slicing solution in Region
  $R_0$ cannot access the singularity. In numerical relativity, the
  exclusion of the region near the singularity is often a desired
  feature (see, for instance, \cite{Gourgoulhon:2007ue,
    Baumgarte:2010ndz}).

  We also note that a second family of maximal surfaces cover the
  regions $R_1$ and $R_{-1}$ in Figure \ref{fig:WH1Regions}. These
  maximal slices lie completely inside the horizons and do not meet
  the AAdS boundaries.  See Appendix \ref{sec:2ndFam} for a
  description of this second family.

\item \emph{Collapse of the lapse.}  An important feature of maximal
  slicing gauge is that the while the lapse $N$ is finite and positive
  (including at $x=0$ at any finite $\bar t$), it vanishes as
  $\bar t\to\infty$. This has been noted before in the literature
  \cite{Estabrook:1973ue} and is referred to as `collapse of the
  lapse'. At late times, the lapse decays exponentially,
\begin{equation}
	N(\bar t\gg \eta^{-1},x) \approx N_0(x) \e^{-\sqrt{2}\eta\bar t} 
\end{equation}
where the function $N_0(x)$ decreases for increasing values of $|x|$
and can be calculated numerically.

\end{enumerate}

\subsection{Large diffeomorphisms}\label{largediffsec}

\begin{figure}[tbp]
	\centering 
	\includegraphics[width=0.45\linewidth]{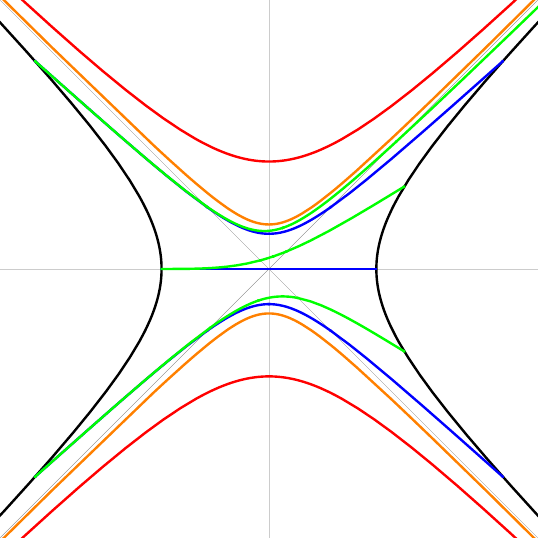}\hfill
	\includegraphics[width=0.45\linewidth]{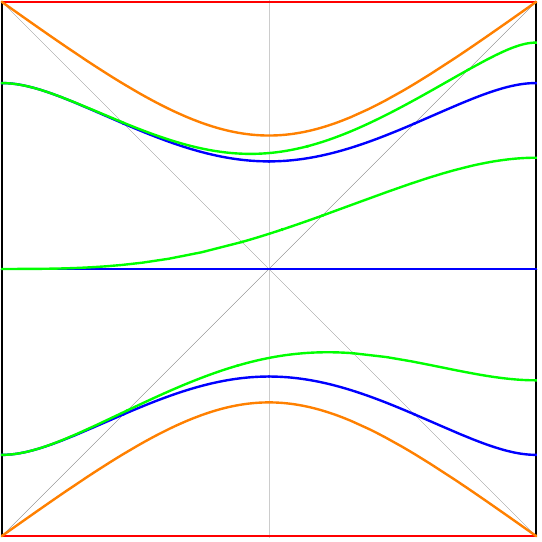}
	\caption{\label{RTfigure} The action of the large
          diffeomorphism corresponding $c = 1$, $\tl{c} = 0$, which
          translations points on the right boundary but fixes points
          on the left boundary. The blue curves are the slices of the
          symmetric foliation and the green curves attached to the
          blue curves on the left boundary are their images under the
          large diffeomorphism.}
\end{figure}

Recall the maximal slicing gauge $K = g_{ij} K^{ij} = 0$ where
$g_{ij}$ is the metric on the spatial slice and $K_{ij}$ is the
extrinsic curvature. Demanding that the maximal slicing gauge is
preserved under time evolution, i.e., $\partial_{\bar{t}} K = 0$,
gives the following differential equation for the lapse $N$:
\begin{equation}
  (D^2 - K_{ij} K^{ij} + 2\Lambda) N = 0\ .
\end{equation}
The lapse $N(\bar{t},x)$ given in \eqref{lapseshiftmaxl} is the unique
solution to the above differential equation that satisfies the AAdS
boundary conditions
\begin{equation}\label{NBCS}
  N(\bar{t},x) \sim \frac{|x|}{\ell}\ ,\quad\text{as}\quad x \to \pm \infty\ .
\end{equation}
The above boundary conditions are not the most general ones consistent
with spherical symmetry on the spatial slice, and can be generalized
to include two independent constants $c$ and $\tl{c}$:
\begin{equation}\label{cctlbc}
  N(\bar{t},x) \sim
  \begin{cases} \displaystyle c\frac{x}{\ell}, &  x \to + \infty\ ,\vspace{10pt} \\
\displaystyle    -\tl{c}\frac{x}{\ell}, & x \to -\infty\ .
  \end{cases}
\end{equation}
We reserve the notation $N(\bar{t},x)$ for the lapse corresponding to
$c = \tl{c} = 1$ which gives the symmetric maximal slicing foliation of
the two-sided BTZ black hole discussed previously. The solution with
general $c$ and $\tl{c}$ then correspond to large diffeomorphisms
which preserve the maximal slicing gauge, which translate the time
$\tl{t}$ on the left boundary by the amount $\tl{c}$ and the time $t$
on the right boundary by an amount $c$. We denote this solution by
$\zeta_\perp$. There is a corresponding shift vector
$(\zeta^x, \zeta^\varphi)$ that can be obtained by solving the
equations for the shift as in Appendix \ref{msbtz-app}. The large
diffeomorphism $\big(\zeta_\perp(\bar{t},x), \zeta^i(\bar{t},x)\big)$ on a given time slice labelled by $\bar{t}$ is given by
\begin{widetext}
\begin{align}
  	\zeta_\perp(\bar{t},x) &= 
  	\begin{cases} \displaystyle \frac{x \sqrt{x^2 + R_+^2 - R_-^2}}{\ell\sqrt{x^2 + R_+^2}} \left[c + \ell^3  \frac{c+\tl{c}}{2}\dot{T} \int_x^\infty \frac{\ud y}{y^2} \frac{\sqrt{y^2 + R_+^2}}{\left(y^2 +R_+^2 -R_-^2\right)^{3/2}} \right] & x>0 \vspace{10pt} \\
	\displaystyle                       -\frac{x \sqrt{x^2 + R_+^2 - R_-^2}}{\ell\sqrt{x^2 + R_+^2}} \left[\tl{c} +  \ell^3    \frac{c+\tl{c}}{2} \dot{T}\int_{-x}^\infty \frac{\ud y}{y^2} \frac{\sqrt{y^2 + R_+^2}}{\left(y^2 +R_+^2 -R_-^2\right)^{3/2}} \right] & x<0
    \end{cases}\ ,\nonumber \\ 
  	\zeta^x(\bar{t},x) &= \frac{1}{x}\left(\zeta_\perp(\bar{t},x) T(\bar{t}) + \tfrac{c+\tl{c}}{2} R_+ \partial_{\bar{t}} R_+ \right)\, , \nonumber \\
  	\zeta^\varphi(\bar{t},x) &= 0\, .\label{zetalarge}
\end{align}
\end{widetext}
The diffeomorphism in \eqref{zetalarge} is given in terms of normal
($\zeta_\perp$) and tangential deformations ($\zeta^i$) of the spatial
surface $\Sigma$, which is well-suited for the Hamiltonian
formulation. The spacetime vector field $\ul\zeta^\mu$,
$\mu = \bar{t}, x, \varphi$ which implements the large diffeomorphism
$(\zeta_\perp, \zeta^i)$ in the maximally sliced spacetime
\eqref{WormholeM} is given by
\begin{align}\label{spacetimediff}
   \ul{\zeta}^{\bar{t}}(\bar{t},x) &=   N(\bar{t},x)^{-1} \zeta_\perp(\bar{t}, x)\ ,\nonumber\\
  \ul{\zeta}^i(\bar{t}, x) &=\zeta^i(\bar{t}, x) -  N(\bar{t},x)^{-1} N^i(\bar{t},x) \zeta_\perp(\bar{t}, x)  \ ,
\end{align}
where $N$ and $N^i$ are the lapse and shift of the maximally sliced
spacetime \eqref{WormholeM}. The vector field above can be used to see
the action of the large diffeomorphism corresponding to given values
of $c$, $\tl{c}$ on the maximal slices of the spacetime
\eqref{WormholeM}. We display the action of the large diffeomorphism
corresponding to $c = 1$, $\tl{c} = 0$ in Figure \ref{RTfigure}.
One can generalize the boundary conditions \eqref{cctlbc} to include
angular dependence through a factor of $\cos n\varphi$ or
$\sin n\varphi$ and obtain an infinite family of large diffeomorphisms
with the above angular dependence. These are expected to generate the
Virasoro algebra which corresponds to the algebra of large
diffeomorphisms of AAdS$_3$ spacetimes.

\section{A probe scalar field in the maximally sliced BTZ black hole}\label{scalarquant}


A probe scalar field in a fixed pure gravity background is \emph{gauge
  invariant} under small diffeomorphisms at quadratic order in
perturbation theory in $\kappa^2 =16\pi G_N$.\footnote{This is
  consistent with the observations in
  \cite{Bahiru:2022oas,Antonini:2025sur} regarding the existence of
  local gauge-invariant observables in perturbation theory about
  backgrounds which do not have isometries.}  This is because (1) the
metric and scalar degrees of freedom are decoupled at quadratic order
in perturbation theory, and (2) after the background diffeomorphisms
are completely fixed, the diffeomorphisms at $\mc{O}(\kappa)$ act
purely in the metric fluctuation sector without involving the scalar
field. See Appendix \ref{pertgaugefix} for a more elaborate
discussion.

The Hamiltonian of a massive scalar field in the background of an AdS-Schwarzschild black hole is
\begin{align}\label{maxham}
  H &= \int_\Sigma \ud^{d}x\, \bigg({N} \Big(\tfrac{1}{2\sqrt{{g}}}\pi_\phi^2 + \tfrac{1}{2} \sqrt{{g}}\,  {g}^{ij} (\partial_i\phi \partial_j \phi + m^2 \phi^2)\Big)\nonumber\\
  &\qquad\qquad\quad + {N}^i \pi_\phi \partial_i \phi\bigg)\, ,
\end{align}
with the appropriate lapse and shift corresponding to maximal slicing of the black hole. The Hamiltonian equations of motion are
\begin{align}
  \partial_{\bar{t}}\phi &= \frac{{N}}{\sqrt{{g}}} \pi_\phi + {N}^i \partial_i\phi\, ,\nonumber\\
  \partial_{\bar{t}}\pi_\phi &= \partial_i({N} \sqrt{{g}} {g}^{ij} \partial_j\phi) - {N} \sqrt{{g}} m^2 \phi + \partial_i ({N}^i\pi_\phi)\, .
\end{align}

\textbf{Note:} As is usual in field theory, we work in the Heisenberg picture. Thus, the operators $\phi(x,\bar{t}), \pi_\phi(x,\bar{t})$ as well as the Hamiltonian $H$ are all in the Heisenberg picture and the states are defined for all times. After promoting the fields $\phi, \pi_\varphi$ to operators, the equations above coincide with the Heisenberg equations of motion.

Plugging in the expression for $\pi_\phi$ in terms of $\dot\phi$, we
get the Klein-Gordon equation
\begin{multline}\label{KGwave}
  \partial_t \big(-{N}^{-1} \sqrt{{g}}(\partial_t \phi -  {N}^i \partial_i \phi)\big) + \partial_i \big({N}^{-1}\sqrt{{g}}  {N}^i \partial_t\phi\big) \\ +\partial_i \big({N}\sqrt{{g}}({g}^{ij} - {N}^{-2} {N}^i {N}^j) \partial_j\phi\big) - {N} \sqrt{{g}} m^2 \phi = 0\, .
\end{multline}
It is expected that the initial value problem is well-posed for the
above linear hyperbolic differential equation when suitable
normalizable boundary conditions are imposed on the scalar field
\cite{Avis:1977yn, Breitenlohner:1982bm, Breitenlohner:1982jf,
  Hawking:1983mx, Henneaux:1984xu, Mezincescu:1984ev, Henneaux:1985tv,
  Lifschytz:1993eb, Ichinose:1994rg, Balasubramanian:1998sn,
  Keski-Vakkuri:1998gmz, Ishibashi:2004wx}. We describe these boundary
conditions briefly. In the asymptotic region, say, near the right boundary
($r \to \infty$), the spacetime metric approaches that of AdS$_d$. The
Klein-Gordon equation \eqref{KGwave} has two linearly independent
solutions $\phi_\pm$ in this region which behave in the $r \to \infty$
limit as
\begin{equation}
  \phi_\pm \sim r^{-\Delta_\pm}\, ,\quad\text{with}\quad \Delta_\pm = \frac{d}{2} \pm \sqrt{\frac{d^2}{4} + \ell^2 m^2}\, .
\end{equation}
When $m^2 > 0$, it can be shown that the solution $\phi_+$ is
normalizable with respect to the Klein-Gordon norm whereas the
solution $\phi_-$ is not (see below and Appendix \ref{BTZKGsol}). The
behaviour near the left boundary ($\tl{r} \to \infty$) is analogous to
the above.

Solving the Klein-Gordon equation directly in wormhole coordinates, analytically, turns out to be hard due to the time dependence of the coefficients in the differential equation. However we note that given Cauchy initial data for the scalar field, there is no problem in solving the differential equation numerically since everything is smooth and well-defined. To solve the Klein-Gordon equation analytically, we adopt a different strategy as described below.

As is well-known, there is typically no preferred complete set of solutions for the Klein-Gordon equation in a general curved spacetime due a lack of time translation symmetry. Depending on the context, we can choose to expand the scalar field in an appropriate complete set of mode functions. Here, we are interested in the AdS/CFT interpretation of scalar field propagation in the maximally sliced BTZ black hole. Hence, it is useful to have an expansion of the scalar field in terms of modes which are relevant for AdS/CFT.

These mode functions are obtained by quantizing the scalar field
separately in the two exterior regions of the AdS-Schwarzschild black hole in the static coordinate system which possess a timelike Killing
symmetry. We then analytically continue them to the interior regions
\cite{Unruh:1976db} (see also \cite{Papadodimas:2012aq} for a more
recent discussion) to obtain the Hartle-Hawking mode functions which
are defined not only in the exterior regions but also in the interior
regions F and P (refer to figure \ref{BTZKruskal}). These mode functions are most naturally expressed in terms of the Kruskal-Szekeres coordinate system of the fully extended
black hole. Then, to get a solution of the Klein-Gordon equation in
maximal slicing coordinates, we do a diffeomorphism from the
Kruskal-Szekeres coordinates to the maximal slicing coordinates.

One important point is that the mode functions obtained in the Killing
time slicing oscillate infinitely rapidly with constant magnitude as
one approaches the horizon. This implies that their derivatives with
respect to the Kruskal-Szekeres coordinates are singular at the
horizon. Thus we see that the expansion of the scalar field in terms of the Hartle-Hawking modes does not actually solve the Klein-Gordon equation on the horizons. 
This will present a problem when we consider derivatives of
the scalar fields on a given maximal slice -- in computing the
Hamiltonian for instance -- since the slice cuts across the
horizon. However, we show that smearing the Hartle-Hawking modes with Hermite
functions $\psi_n(\omega/\eta)$ in the frequency domain smoothens out
these singularities. The smeared modes are labelled by the discrete
index $n$ of the Hermite functions. This discreteness also is natural
from the point of view of quantization of scalar field in AAdS
spacetimes since the scalar field is effectively in a gravitational
box. 

We then express the quadratic Hamiltonian in terms of the discrete scalar field expansion that is smooth everywhere and obtain a well-defined Hamiltonian operator which evolves the scalar field along the maximal slices. This operator describes the evolution of wavepackets which smoothly cross the horizon and hence describes infalling observers.

Finally we also calculate the two-point Wightman functions in the Hartle-Hawking state in the wormhole coordinates. For concreteness, from now on we specialize to $d=2$, i.e., the BTZ black hole, where we can do explicit computations.

Before solving the Klein-Gordon equation, we note that for the BTZ black hole, there is another degree of freedom from the pure gravity sector at the same order in perturbation theory as the scalar field. This corresponds to the fluctuations in reduced phase space variables $(m,T)$ or equivalently $(Q=4\pi^2\frac{T}{m^2}, P=m)$, see the companion paper \cite{KPWI}.\footnote{The coordinates $(Q,P)$ are related to by a canonical transformation to Kuchar's mass and time shift variables \cite{Kuchar:1994zk}, for the BTZ black hole.} We can expand
\begin{align}
	Q=Q_0 + \kappa\, \delta Q, \quad P = P_0 + \kappa\, \delta P
\end{align}
where $\kappa=\sqrt{16\pi G_N}$. Here $(Q_0, P_0)$ form the on-shell black hole solution while $\delta Q$ and $\delta P$ are fluctuations on the black hole background. Since we have already solved the constraints and obtained the gauge invariant reduced phase space, there are no further constraints on these fluctuations. Quantization of $(\delta Q, \delta P)$ is expected to lead to $L^2(\mathbf{R})$, the space of square-integrable functions on the real line. However note that the scalar field does not mix with this degree of freedom at quadratic order in the semi-classical approximation.

\subsection{Klein-Gordon Mode expansions in Regions I and II}

To obtain the most general solution of the Klein-Gordon equation
\eqref{KGwave}, we start with the solutions in Regions I and II solved
in terms of the original BTZ coordinates in these regions:
\begin{multline}\label{phiBTZInearhor}
  \phi({t},{r},\varphi) =  \sum_{q\in \mathbb{Z}} \int_0^\infty \frac{\ud\omega}{\sqrt{4\pi\omega}} \Big( a_{\omega q}  F_{\omega q}(t,r,\varphi)  + \text{c.c.}\Big)\, ,
\end{multline}
where $F_{\omega q}(t, r, \varphi)$ are solutions of the Klein-Gordon
equation in Region I which satisfy the normalizability condition
$F_{\omega q}(r) \sim r^{-\Delta_+}$ in the limit $r \to \infty$. (See equations \eqref{capFdef} to \eqref{solnIinf} in
Appendix \ref{BTZKGsol} for the detailed expressions.) The $F_{\omega q}$
have the following Klein-Gordon inner products:
\begin{equation}
  (F_{\omega q}, F_{\omega'q'})_{\rm KG} = 4\pi \omega \delta(\omega - \omega')\, ,\quad (F^*_{\omega q}, F_{\omega'q'})_{\rm KG} = 0\, .
\end{equation}
where the Klein-Gordon inner product is defined on the constant $t$
surfaces with $R_h < r < \infty$ in Region I. There is a completely analogous story in Region II with the scalar field mode expansion
\begin{multline}\label{phiBTZIInearhor}
  \phi(\tl{t},\tl{r},\varphi) =  \sum_{q\in \mathbb{Z}} \int_0^\infty \frac{\ud\omega}{\sqrt{4\pi\omega}} \Big( \tl{a}_{\omega q} F^*_{\omega q}(\tl{t}, \tl{r}, \varphi)  + \text{c.c.}\Big)\, ,
\end{multline}
where we have an $F^*_{\omega q}$ in contrast to Region I since the
time $\tl{t}$ in Region II runs in the opposite sense to $t$ in Region
I.

The ladder operators $a_{\omega q}$, $a^\dag_{\omega q}$ satisfy the
commutation relations
\begin{align}
  &[a_{\omega q}, a^\dag_{\omega'q'}] =  [\tl{a}_{\omega q}, \tl{a}^\dag_{\omega'q'}] = \delta(\omega-\omega')\delta_{qq'}\, ,
\end{align}
with all other commutators being zero. Throughout this paper 
we have put $\hbar =1$. The boundary operator $\mc{O}$
in the dual CFT associated to the right boundary is obtained by the
extrapolate limit:
\begin{align}\label{bdryopI}
  &\mc{O}(t,\varphi) = \lim_{r \to \infty} r^{\Delta_+} \phi(t, r, \varphi)\, ,\nonumber\\
  & = \frac{R_h^{\Delta_+}}{\sqrt{2\pi R_h}}   \sum_{q\in \mathbb{Z}} \int_0^\infty \frac{\ud\omega}{\sqrt{4\pi\omega}}\frac{1}{N_{\omega q}} \big( a_{\omega q} \e^{-\i \omega t-\i q\varphi}  + \text{c.c.}\big)\, .
\end{align}
Similarly, the operator in the CFT associated to the left boundary
$\tl{r} \to \infty$ is
\begin{align}\label{bdryopII}
  &\tl{\mc{O}}(\tl{t},\varphi) = \lim_{\tl{r} \to \infty} \tl{r}^{\Delta_+} \phi(\tl{t}, \tl{r}, \varphi)\, ,\nonumber\\
  & = \frac{R_h^{\Delta_+}}{\sqrt{2\pi R_h}} \sum_{q\in \mathbb{Z}} \int_0^\infty \frac{\ud\omega}{\sqrt{4\pi\omega}} \frac{1}{N_{\omega q}} \big( \tl{a}_{\omega q} \e^{\i \omega \tl{t} + \i q \varphi} + \text{c.c.}\big)\, .
\end{align}

\subsection{The Hartle-Hawking mode functions}

The solutions in Region I and II above can be analytically continued
into the Regions F and P using analyticity in the lower-half $U$ and
$V$ planes following Unruh \cite{Unruh:1976db}. This process is
described in detail in the Appendix \ref{ACtoFP}. The resulting mode
functions are the Hartle-Hawking mode functions
$h_{\omega q}(U, V, \varphi)$, in terms of which the scalar field mode
expansion is
\begin{equation}\label{scalarHH}
  \phi(U,V,\varphi) = \frac{1}{\sqrt{2\pi R_h}}\sum_{q \in \mathbb{Z}} \int_{-\infty}^{\infty} \ud\omega\, \Big(c_{\omega q} h_{\omega q}(U,V,\varphi) + \text{c.c.}\Big)\, ,
\end{equation}
see equations \eqref{BTZHHmodemain}, \eqref{F12def}, \eqref{G12def} in Appendix \ref{ACtoFP} for detailed expressions for the mode function
$h_{\omega q}$. In contrast to the $F_{\omega q}$, the Hartle-Hawking
mode functions are supported everywhere in the fully extended BTZ
black hole. The Klein-Gordon inner products are
\begin{equation}
  (h_{\omega q}, h_{\omega'q'})_{\rm KG} = 2\pi R_h \delta(\omega - \omega')\, ,\quad (h^*_{\omega q}, h_{\omega'q'})_{\rm KG} = 0\, .
\end{equation}
Here, the spatial slice on which the Klein-Gordon inner product is
defined is the union of the $t = 0$ and $\tl{t} = 0$ slices in Regions
I and II respectively. This coincides with the $U + V = 0$ slice in
the Kruskal coordinates and the $\bar{t} = 0$ in maximal slicing
coordinates.

The operators $c_{\omega q}$ for $\omega < 0$ are
sometimes written as $\tl{c}_{\omega q}$:
\begin{equation}
  \tl{c}_{\omega q} = c_{-\omega, -q}\, ,\quad\text{for}\quad \omega > 0\, .
\end{equation}
The $c_{\omega q}$ and $\tl{c}_{\omega q}$ are related to the Region I
and II operators by the Bogoliubov transformations
\begin{align}\label{cctldefmain}
  c_{\omega q} &= \frac{a_{\omega q} \e^{\pi\omega/2\eta} - \tl{a}^\dag_{\omega q} \e^{-\pi\omega/2\eta}}{\sqrt{2\sinh (\pi\omega/\eta)}}\, ,\nonumber\\
  \tl{c}_{\omega q} &= \frac{\tl{a}_{\omega q} \e^{\pi\omega/2\eta} - a^\dag_{\omega q} \e^{-\pi\omega/2\eta}}{\sqrt{2\sinh (\pi\omega/\eta)}}\, ,
\end{align}
and their conjugates. The state annihilated by the $c_{\omega q}$,
$\tl{c}_{\omega q}$, for $\omega > 0$, is the Hartle-Hawking vacuum:
\begin{equation}
  c_{\omega q}|\text{HH}\rangle = \tl{c}_{\omega q}|\text{HH}\rangle = 0\ ,\quad\text{for}\quad \omega > 0,\ q \in \mathbb{Z}\, .
\end{equation}
As noted earlier, the Hartle-Hawking state exists for all times as we are working in the Heisenberg picture. The ladder operators $c_{\omega q}$, $\tl{c}_{\omega q}$ satisfy
\begin{align}
  &[{c}_{\omega q}, {c}^\dag_{\omega'q'}] = [\tl{c}_{\omega q}, \tl{c}^\dag_{\omega'q'}] = \delta(\omega-\omega')\delta_{qq'}\, ,
\end{align}
with all other commutators being zero.

\subsection{Smooth mode functions}\label{smoothmodes}

The Hartle-Hawking mode expansion \eqref{scalarHH} is defined
everywhere in the fully extended BTZ black hole geometry. However, the
mode functions $h_{\omega q}(U, V, \varphi)$ are not differentiable at
the horizons (see \cite{Unruh:1976db} for instance) and, strictly
speaking, do not satisfy the Klein-Gordon equation in a neighbourhood
of the horizons. The singularities at the horizons can be seen as
follows. In a neighbourhood of the future horizon $U = 0$, $V > 0$,
$h_{\omega q}(U,V,\varphi)$ takes the form
\begin{align}
  	&h_{\omega q} \sim  \frac{\e^{-\i q\varphi}}{\sqrt{4\pi\omega} \sqrt{2\sinh (\pi\omega/\eta)}} \times \nonumber\\
    &\times \left\{\def\arraystretch{2} \begin{array}{cl} \e^{\i\delta_{\omega q}}\e^{\frac{\pi\omega}{2\eta}} (2\eta V)^{-\frac{\i\omega}{\eta}} + \e^{-\i\delta_{\omega q}}\e^{\frac{\pi\omega}{2\eta}} (-2\eta U)^{\frac{\i\omega}{\eta}} &\quad  U < 0 \\  
    \e^{\i\delta_{\omega q}}\e^{\frac{\pi\omega}{2\eta}} (2\eta V)^{-\frac{\i\omega}{\eta}} + \e^{-\i\delta_{\omega q}} \e^{-\frac{\pi\omega}{2\eta}} (+2\eta U)^{\frac{\i\omega}{\eta}} & \quad U > 0 \, .\end{array}\right.
\end{align}
Here, $\delta_{\omega q}$ is a phase factor and depends on $\omega$,
the surface gravity $\eta$, the scalar field mass $m^2$, and the
angular momentum quantum number $q$. In the large mass limit
$m^2 \gg \ell^{-2}$, this phase simply becomes the phase of
$\Gamma(\i\omega/\eta)$, the Euler
$\Gamma$-function. It is independent of the angular momentum $q$ and
depends only on the surface gravity of the black hole. In fact, in the
large mass limit, one recovers the expression for the Hartle-Hawking
mode functions for a scalar field in Rindler space with acceleration
parameter $\eta$.

Clearly, taking derivatives of $h_{\omega q}$ with respect to $U$
brings down powers of $U^{-1}$ which blow up at the horizon. One way
to resolve this issue is to work with the Fourier transformed mode
functions \cite{Unruh:1976db}:
\begin{equation}
  \wh{h}_{\vartheta q}(U,V,\varphi) = \int_{-\infty}^\infty \frac{\ud \omega}{2\pi\eta} \e^{\i\omega \vartheta / \eta} h_{\omega q}(U,V,\varphi)\, .
\end{equation}
In the neighbourhood of the horizon and in the large mass limit of the scalar field, $\wh{h}_{\vartheta q}$ takes the form
\begin{equation}
  \wh{h}_{\vartheta q}(U,V,\varphi) \sim \frac{1}{2\pi \sqrt{2\eta}}\e^{-\i q\varphi} \left( \e^{2\i\eta V \e^{-\vartheta}} + \e^{-2\i\eta U \e^\vartheta}\right)\, .
\end{equation}
It is easy to see that the above expressions are differentiable at
$U = 0$. We expect the property of smoothness to hold for all masses
even though we do not have an explicit expression like in the above
large mass limit.

Since the scalar field is in an asymptotic AdS background which
naturally acts as a gravitational box, it may be useful to work with
modes which are labelled by a discrete index. To this end, we can
smear the Hartle-Hawking mode functions with Hermite
functions \footnote{There may be other choices for the smearing   functions which achieve similar results. For instance, we can take   Hermite functions $\psi_n(\omega / \alpha)$ for any positive real   parameter $\alpha$. The Hermite functions $\psi_n(z)$ are defined in   terms of the Hermite polynomials $H_n(z)$ as
\begin{align*}
  	\psi_n(z) &= \frac{1}{(\sqrt{\pi} 2^n n!)^{1/2}} \e^{-z^2/2} H_n(z)\, ,\\
    H_n(z) &= (-1)^n \e^{z^2} \frac{\ud^n}{\ud z^n} \e^{-z^2}\, .
\end{align*} 
They satisfy the orthogonality and completeness relations
\begin{align*}
   	&\int_{-\infty}^\infty \ud z\, \psi_n(z)\psi_m(z) = \delta_{nm}\, , \\ &\sum_{n=0}^\infty \psi_n(z) \psi_n(z') = \delta(z-z')\, .
\end{align*}}
$\psi_n(\omega /\eta)$ in
$\omega$-space and define the mode functions $g_{nq}$:
\begin{align}\label{gnBTZ}
  & g_{nq}(U,V,\varphi) := \int_{-\infty}^\infty \frac{\ud\omega}{\eta}\, h_{\omega q}(U,V,\varphi) \psi_n(\omega/\eta)\, \nonumber\\
  &= (-\i)^n \int_{-\infty}^\infty \ud\vartheta\, \wh{h}_{\vartheta q}(U,V,\varphi) \psi_n(\vartheta) \nonumber \\
  &\sim \frac{(-\i)^n \e^{-\i q\varphi}}{2\pi\sqrt{2\eta}} \int_{-\infty}^\infty \ud \vartheta\,\big(\e^{2\i\eta V \e^{-\vartheta}} + \e^{-2\i\eta U \e^{\vartheta}}\big) \psi_n(\vartheta)\, .
\end{align}
where, in going to the second line, we have used the fact that the
Fourier transform of a Hermite function $\psi_n(\omega/\eta)$ is the
same Hermite function $(-\i)^n \psi_n(\vartheta)$. The $g_{nq}$ mode
functions also have the important property that they are
differentiable at the horizons $UV=0$.

We can now expand the scalar field in terms of the $g_{nq}$ mode functions:
\begin{equation}\label{scalargn}
  \phi(U,V,\varphi) = \frac{1}{\sqrt{2\pi} \ell}\sum_{q \in \mathbb{Z}}\sum_{n=0}^\infty  \big(e_{nq}\, g_{n q}(U,V,\varphi) + \text{c.c.}\Big)\, ,
\end{equation}
where the operators $e_{nq}$ are defined in terms of $c_{\omega q}$ as
\begin{equation}\label{enqTocct}
e_{nq} =  \frac{1}{\sqrt{\eta}}\int_{-\infty}^{\infty} \ud\omega\, c_{\omega q} \psi_n(\omega/\eta)\, ,
\end{equation}
and satisfy the canonical commutation relations
\begin{equation}
  [e_{m q}, e^\dag_{n q'}] = \delta_{mn} \delta_{qq'}\ ,\quad  [e_{m q}, e_{n q'}] = [e^\dag_{m q}, e^\dag_{n q'}] = 0\, .
\end{equation}
Since $e_{nq}$ only involve the $c_{\omega q}$ and not the
$c^\dag_{\omega q}$, they also annihilate the Hartle-Hawking state
$|\text{HH}\rangle$:
\begin{equation}
  e_{nq} |\text{HH}\rangle = 0\, ,\quad n =0,1,\ldots\, .
\end{equation}
The mode expansion \eqref{scalargn} is smooth across the horizons and globally defined in the fully extended BTZ spacetime.

Finally, to obtain the general solution of the Klein-Gordon equation
in the maximally sliced BTZ spacetime \eqref{KGwave} which is smooth
everywhere, we use the diffeomorphism between maximal slicing
coordinates and Kruskal-Szekeres coordinates given in Appendix 
\ref{UVDiff} in the mode expansion \eqref{scalargn}. Define
\begin{equation}
  \hat{g}_{nq}(\bar{t}, x, \varphi) = g_{nq}(U(\bar{t}, x), V(\bar{t}, x), \varphi)\, .
\end{equation}
We then finally have the general solution of the Klein-Gordon equation
\begin{equation}\label{scalargnx}
  \phi(\bar{t},x,\varphi) = \frac{1}{\sqrt{2\pi} \ell}\sum_{q \in \mathbb{Z}}\sum_{n=0}^\infty  \big(e_{nq}\, \hat{g}_{n q}(\bar{t},x,\varphi) + \text{c.c.}\big)\, ,
\end{equation}
which satisfy the normalizable boundary conditions:
\begin{equation}\label{phiextrapol}
  \phi(\bar{t}, x, \varphi) \to \left\{\def\arraystretch{1.5}\begin{array}{cc} x^{-\Delta_+} \mc{O}({t}, \varphi) & x \to +\infty\, , \\ (-x)^{-\Delta_+} \tl{\mc{O}}(\tl{t},\varphi) & x \to -\infty\, , \end{array}\right.
\end{equation} 
where, recall that, $\mc{O}(t,\varphi)$ and
$\tl{\mc{O}}(\tl{t},\varphi)$ are generalized free field scalar
operators dual to the bulk scalar field in the right and left CFTs.

By following the chain of definitions, it is easy to compute the
Klein-Gordon inner products for the $\hat{g}_{nq}$ modes:
\begin{align}
  &(\hat{g}_{n q}, \hat{g}_{n' q'})_{\rm KG} = 2\pi \ell^2 \delta_{nn'}\delta_{qq'}\, ,\quad (\hat{g}^*_{n q}, \hat{g}_{n' q'})_{\rm KG} = 0\, ,\nonumber\\
  &(\hat{g}^*_{n q}, \hat{g}^*_{n' q'})_{\rm KG} = -2\pi \ell^2 \delta_{nn'}\delta_{qq'}\, .
\end{align}

\subsection{A bulk reconstruction formula in wormhole coordinates \label{sec:BR}}

We can view the local bulk operator $\phi(\bar t,x,\varphi)$ as extended operator in the product space of the two boundary CFTs, with $x$ as a label. More precisely, we can write down a bulk reconstruction formula for the local bulk operator $\phi(\bar t,x,\varphi)$ in the entire region $R_0$ (see figure \ref{fig:WH1Regions}), covered by the range of the wormhole coordinates $-\infty < \bar{t} < \infty,\ -\infty < x < \infty,\ 0 \leq \varphi \leq 2\pi$, as
\begin{align}\label{BRWH}
	\phi(\bar t,x,\varphi) &= \int_0^{2\pi} \ud\varphi' \int_{-\infty}^\infty  \ud t' \mathcal{K}(\bar t,x,\varphi;t',\varphi') \mathcal{O}(t',\varphi') \nonumber\\
	&+ \int_0^{2\pi} \ud\tilde{\varphi}' \int_{-\infty}^\infty \ud \tilde{t}' \tilde{\mathcal{K}}(\bar t,x,\varphi; \tilde{t}', \tilde{\varphi}') \tilde{\mathcal{O}}( \tilde{t}', \tilde{\varphi}')\, ,
\end{align}
where $\mathcal{O}(t',\varphi')$ and
$\tilde{\mathcal{O}}( \tilde{t}', \tilde{\varphi}')$ are dual
generalized free field operators in the right and left CFTs
respectively. The exact form of $\mathcal{K}$ and 
$\tilde{\mathcal{K}}$ is provided in Appendix \ref{BRinWC}, 
equation \eqref{kxktx}. This formula is derived from the discrete mode
expansion of the scalar field \eqref{scalargnx} in terms of the
$e_{n q}$ oscillators \eqref{enqTocct}, and it naturally involves
support from both the boundary CFTs, in contrast to the HKLL formula
\cite{Hamilton:2006az}. 

The boundary-to-bulk kernels $\mathcal{K}$ and $\tilde{\mathcal{K}}$ have the following important property. When the bulk spacetime point is located entirely in Region I, the kernel $\tilde{\mathcal{K}}$ vanishes. Similarly for a spacetime point in Region II, the kernel $\mathcal{K}$ vanishes. However in the interior regions F and P, both $\mathcal{K}$ and $\tilde{\mathcal{K}}$ are necessarily non-zero. Later in Section \ref{sec:OP}, this property will be crucial for defining an order parameter that detects crossing of the black hole horizon.

\section{The scalar field Hamiltonian for evolution along
  maximal slices}\label{HamSF}

\subsection{The expression for the Hamiltonian}

We now evaluate the scalar field Hamiltonian \eqref{maxham} on the
maximally sliced BTZ black hole solution \eqref{WormholeM} after
plugging in the field expansion \eqref{scalargnx} in terms of the
smooth mode functions $\hat{g}_{n q}$. The Hamiltonian is written in
condensed form as
\begin{align}\label{Hamfinal}
	H &= \sum_{\substack{n,n' \\ q,q'}}  \begin{pmatrix} e^\dag_{n'q'} & e_{n'q'}\end{pmatrix} \begin{pmatrix} A_{nn'qq'}(\bar{t})\ & B^*_{nn'qq'}(\bar{t}) \\ B_{nn'qq'}(\bar{t})\ & A^*_{nn'qq'}(\bar{t}) \end{pmatrix} \begin{pmatrix} e_{n q} \\ e^\dag_{n q}\end{pmatrix}\, ,
\end{align}
where the kernels $A$ and $B$ are defined as
\begin{align}
	&A_{nn'qq'} (\bar t)\nonumber\\
	=& \frac{1}{4\pi\ell^2} \int \ud x\ud\varphi \sqrt{g}  \Bigg[ N\Big(\cD_{\bar{t}}{\hat{g}}^*_{n'q'} \cD_{\bar{t}} \hat{g}_{n q} + g^{xx} \hat{g}'^*_{n'q'} \hat{g}'_{n q} \nonumber \\
	&\ + \left(m^2 +qq' g^{\varphi\varphi}\right) \hat{g}^*_{n'q'} \hat{g}_{n q}\Big) 
	+ 2 N^x \cD_{\bar{t}}{\hat{g}}^*_{n'q'} \hat{g}'_{n q} \Bigg]\, ,
\end{align}
and 
\begin{align}
	&B_{nn'qq'} (\bar t) \nonumber\\
	=& \frac{1}{4\pi\ell^2} \int \ud x\ud\varphi \sqrt{g} \Bigg[  N\Big( \cD_{\bar{t}}{\hat{g}}_{n'q'} \cD_{\bar{t}} \hat{g}_{n q} + g^{xx} \hat{g}'_{n'q'} \hat{g}'_{n q} \nonumber \\
	&\  + \left(m^2 +qq' g^{\varphi\varphi}\right) \hat{g}_{n'q'} \hat{g}_{n q}\Big) 
	+ 2 N^x \cD_{\bar{t}}{\hat{g}}_{n'q'} \hat{g}'_{n q} \Bigg]\, ,
\end{align}
where
$\cD_{\bar{t}} = n^\mu \partial_\mu = N^{-1}(\partial_{\bar{t}} - N^x
\partial_x)$ is the normal derivative and $' = \partial_x$. The kernel $A_{nn'qq'}(\bar{t})$ satisfies
\begin{equation}
  A^*_{nn'qq'} = A_{n'nq'q}\, .
\end{equation}
The integrals in the above kernels $A$ and $B$ are finite because the
mode functions $\hat{g}_{nq}$ are smooth everywhere in the bulk, and
fall-off at the boundary as $|x|^{-\Delta_+}$, with
$\Delta_+>1$.\footnote{We can split the $x$ integral as a bulk term
  and two terms near the boundary at $x=\pm\infty$,
  \begin{equation*}
    \int_{-\infty}^\infty \ud x = \int_{-\infty}^{-X} \ud x + \int_{-X}^{X} \ud x + \int_{X}^\infty \ud x 
  \end{equation*}
  for some large and positive value $X$. The bulk term is finite since all the quantities are smooth and finite. Near the boundary, the mass term in $A_{nn'qq'}$ goes as,
  \begin{equation*}
    m^2\int_{X}^\infty \ud x \sqrt{g}N \hat{g}^*_{n'q'} \hat{g}_{nq} \propto \int_{X}^\infty \ud x\, x^{1-2\Delta_+} \sim x^{2(1-\Delta_+)} \Big|_X^\infty ,
  \end{equation*}
  which is finite for $\Delta_+ >1$. Similarly contribution from all
  the other terms is also finite.} The Hamiltonian \eqref{Hamfinal}
is Hermitian by construction and is written entirely in terms of
the smooth scalar field modes $e_{n q}$.

The expectation value of $H$ \eqref{Hamfinal} in the Hartle-Hawking
state $|\text{HH}\rangle$ is
\begin{equation}
  \langle \text{HH} | H(\bar{t}) | \text{HH}\rangle = \sum_{n, q} A_{nnqq}(\bar{t})\ .
\end{equation}
We can now define a normal-ordered Hamiltonian operator
\begin{equation}
  	\hat{H}(\bar t) = H(\bar{t}) - \langle \text{HH} | H(\bar{t}) | \text{HH}\rangle\, ,
\end{equation}
which generates time translations along maximal slices that probe the black hole interior.

\subsection{Hamiltonian in terms of the CFT modes $a_{\omega q}$, $\tl{a}_{\omega q}$}

Recall that the $e_{nq}$ operators are related to the Hartle-Hawking
operators $c_{\omega q}$, $\tl{c}_{\omega q}$, $\omega > 0$, by the
formula
\begin{align}\label{enqToaat}
  e_{nq} =& \frac{1}{\sqrt{\eta}}\int_{0}^{\infty} \ud\omega\, c_{\omega q} \psi_n(\omega/\eta) + \nonumber \\
          & \frac{1}{\sqrt{\eta}}\int_{0}^{\infty} \ud\omega\, \tl{c}_{\omega q} \psi_n(-\omega/\eta)\, .
\end{align}
The Hartle-Hawking operators are in turn related to the operators
$a_{\omega q}$, $\tl{a}_{\omega q}$, $a^\dag_{\omega q}$,
$\tl{a}^\dag_{\omega q}$ in the two CFTs by the Bogoliubov
transformations \eqref{cctldefmain}:
\begin{align}\label{cctldefmain1}
  c_{\omega q} &= \frac{a_{\omega q} \e^{\pi\omega/2\eta} - \tl{a}^\dag_{\omega q} \e^{-\pi\omega/2\eta}}{\sqrt{2\sinh (\pi\omega/\eta)}}\, ,\nonumber\\
  \tl{c}_{\omega q} &= \frac{\tl{a}_{\omega q} \e^{\pi\omega/2\eta} - a^\dag_{\omega q} \e^{-\pi\omega/2\eta}}{\sqrt{2\sinh (\pi\omega/\eta)}}\, .
\end{align}
Hence the Hamiltonian \eqref{Hamfinal} serves as the time evolution
operator in the dual product CFT of the two boundaries which entangles
the degrees of freedom in the two CFTs. The time-dependent operators
\eqref{enqtime} demonstrate this entanglement explicitly since it is
no longer possible to separate out the left and right CFT operators
$a_{\omega q}$ and $\tl{a}_{\omega q}$ under $\bar{t}$ evolution.

One can also obtain an ill-defined expression (see below) for the
Hamiltonian \eqref{Hamfinal} in terms of the Hartle-Hawking operators
$c_{\omega q}$, $\omega \in \mathbf{R}$, by plugging in the
expressions \eqref{enqToaat} for $e_{nq}$ into \eqref{Hamfinal}. We
get
\begin{align}\label{Hamcomegaq}
  H &= \sum_{q,q'} \int_{-\infty}^\infty \ud \omega\int_{-\infty}^\infty \ud\omega' \nonumber\\
  &\qquad\begin{pmatrix} c^\dag_{\omega'q'} & c_{\omega'q'}\end{pmatrix} \begin{pmatrix} \mc{A}_{\omega\omega'qq'}(\bar{t})\ & \mc{B}^*_{\omega\omega'qq'}(\bar{t}) \\ \mc{B}_{\omega\omega'qq'}(\bar{t})\ & \mc{A}^*_{\omega\omega'qq'}(\bar{t}) \end{pmatrix} \begin{pmatrix} c_{\omega q} \\ c^\dag_{\omega q}\end{pmatrix}\, ,
\end{align}
with
\begin{align}\label{HHAkernel}
	&\mc{A}_{\omega\omega'qq'} (\bar t)\nonumber\\
	=& \frac{1}{4\pi\ell^2} \int \ud x\ud\varphi \sqrt{g}  \Bigg[ N\Big(\cD_{\bar{t}}h^*_{\omega'q'} \cD_{\bar{t}} h_{\omega q} + g^{xx} h'^*_{\omega'q'} h'_{\omega q} \nonumber \\
	&\ + \left(m^2 +qq' g^{\varphi\varphi}\right) h^*_{n'q'} h_{n q}\Big) 
	+ 2 N^x \cD_{\bar{t}}h^*_{\omega'q'} h'_{\omega q} \Bigg]\, ,
\end{align}
and
\begin{align}\label{HHBkernel}
	&\mc{B}_{\omega\omega'qq'} (\bar t)\nonumber\\
	=& \frac{1}{4\pi\ell^2} \int \ud x\ud\varphi \sqrt{g}  \Bigg[ N\Big(\cD_{\bar{t}}h_{\omega'q'} \cD_{\bar{t}} h_{\omega q} + g^{xx} h'_{\omega'q'} h'_{\omega q} \nonumber \\
	&\ + \left(m^2 +qq' g^{\varphi\varphi}\right) h_{n'q'} h_{n q}\Big) 
	+ 2 N^x \cD_{\bar{t}}h_{\omega'q'} h'_{\omega q} \Bigg]\, .
\end{align}
The kernels $\mc{A}_{\omega\omega'qq'}$ and
$\mc{B}_{\omega\omega'qq'}$ are ill-defined since the integration over
$x$ includes the horizons and the derivatives of the Hartle-Hawking
mode functions that appear in the kernels are singular at the horizon
as explained in Section \ref{smoothmodes}.

Now, the Hamiltonian \eqref{Hamcomegaq} can be further written in
terms of the $a_{\omega q}$, $\tl{a}_{\omega q}$, $a^\dag_{\omega q}$,
$\tl{a}^\dag_{\omega q}$ by using the Bogoliubov transformations
\eqref{cctldefmain1} in \eqref{Hamcomegaq}. The kernel
$\mc{A}_{\omega\omega'qq'}$ (similarly for
$\mc{B}_{\omega \omega'qq'}$) has to be separated into four different
parts corresponding to positive and negative $\omega$, $\omega'$, and
the Hamiltonian will be of the form
\begin{align}\label{Hamcomegaq}
  H &= \sum_{q,q'} \int_{0}^\infty \ud \omega\int_{0}^\infty \ud\omega' \nonumber\\
  &\qquad\begin{pmatrix} a^\dag_{\omega' q'} & a_{\omega' q'} & \tl{a}^\dag_{\omega' q'} & \tl{a}_{\omega' q'}\end{pmatrix} \bigg( \text{$4 \times 4$}\bigg)_{\omega \omega' qq'} \begin{pmatrix} a_{\omega q} \\ a^\dag_{\omega q} \\ \tl{a}_{\omega q} \\ \tl{a}^\dag_{\omega q}\end{pmatrix}\, ,
\end{align}
which has a $4 \times 4$ matrix of kernels whose entries are all
ill-defined due to the same divergences present in the kernels
$\mc{A}_{\omega\omega'qq'}$, $\mc{B}_{\omega\omega'qq'}$
\eqref{HHAkernel} and \eqref{HHBkernel}.

Thus, though one can write the Hamiltonian for maximal slicing
evolution in terms of the CFT operators, the expression has
divergences arising from the blue-shift singularities of the mode
functions at the horizons and hence is an ill-defined operator when
acting individually on the modes of the left or the right CFT.

\subsection{The time evolution operator}
      
Since this Hamiltonian is time-dependent, the time-evolution operator has to constructed as the anti time-ordered exponential \footnote{The time evolution operator satisfies the differential equation
	\begin{equation*}
		\partial_{\bar t} U(\bar t,0) 
		= -i U(\bar t,0) H(\bar t) ,
	\end{equation*}
	where $H(\bar t)$ is the Hamiltonian in the Heisenberg picture. Integrating this equation gives
	\begin{equation*}
		U(\bar t,0) = 1 - \i \int_{0}^{\bar t} \ud t' U(\bar t',0) H(\bar t') . 
	\end{equation*}
	We can iterate this further to get the Dyson series
	\begin{align*}
		U(\bar t,0) =& 1 - \i \int_{0}^{\bar t} \ud \bar t' H(\bar t') + (-\i)^2 \int_{0}^{\bar t} \ud \bar t' \int_{0}^{\bar t'} \ud \bar t'' H(\bar t'') H(\bar t') \\
		&+ \cdots  + (-\i)^n \int_{0}^{\bar t} \ud \bar t' \int_{0}^{\bar t'} \ud \bar t'' \cdots \int_{0}^{\bar t^{(n-1)}} \ud \bar t^{(n)} \times \\
		& \times H(\bar t^{(n)}) \cdots H(\bar t'') H(\bar t') + \cdots ,
	\end{align*}
	which is written compactly as anti time-ordered exponential in \eqref{UOperator}.}
	
\begin{equation}\label{UOperator}
	U(\bar t) = \bar{\mathcal{T}} \exp \left(-\i\int_{0}^{\bar t} \ud \bar u \hat{H}(\bar u) \right)\, .
\end{equation}
This is a \emph{unitary} operator which describes the evolution of the
quantum scalar field along maximal slices. In particular, one can
define the time-dependent creation operators which satisfy the
Heisenberg equations of motion: 
\begin{equation}\label{enqtime}
	e^\dag_{nq}(\bar t) = U(\bar t)^\dagger e^\dag_{nq}(0) U(\bar t)\, .
\end{equation}
We must note that the scalar field Hamiltonian \eqref{Hamfinal} goes
to zero for $\bar{t} \gg \eta^{-1}$ due to the phenomenon of the
\emph{collapse of the lapse} discussed in Section \ref{maxprop}. This
reflects the freezing of proper time in the maximally sliced BTZ
geometry as one approaches the limit slices $\bar{t} \to
\pm\infty$. To avoid spurious effects that may arise from the
lapse becoming zero, we must restrict the range of $\bar{t}$ to be of
the order of $\eta^{-1}$.

An immediate consequence of the unitarity of the time evolution
operator in the maximal slicing gauge in the Schroedinger picture is
that the time evolved Hartle-Hawking state,
\begin{align}
	|\text{HH};\bar t\rangle = U(\bar t)|\text{HH}\rangle\, ,
\end{align}
remains pure for all times. This means that 
\begin{equation}
	\rho(\bar{t})^2 = \rho(\bar{t}), \qquad \rho(\bar{t}) = |\text{HH};\bar t\rangle \langle \text{HH};\bar t|\, ,
\end{equation}
so its von Neumann entropy vanishes at all times. The Hartle-Hawking
state is supported on smooth spatial slices which cut across the
horizon and hence the crossing of the horizon by a wave-packet built out of a finite number of $e^\dag_n$ excitations on top of the Hartle-Hawking state can be smoothly described by this Hamiltonian.


\subsection{The Hamiltonian $\hat{H}(\bar t)$ is a well-defined operator}

We observe that our formalism is also capable describing the Killing time evolution in the exterior regions I and II. We simply impose a different boundary condition on the lapse, namely
\begin{equation}\label{NbcKilling}
	N \sim
	\begin{cases} \displaystyle \frac{r}{\ell}, & r \to \infty \quad (\text{Region I}) ,\vspace{10pt} \\
		\displaystyle  -\frac{\tilde{r}}{\ell}, & \tilde{r} \to \infty \quad (\text{Region II}) .
	\end{cases}
\end{equation}
With these boundary conditions, one recovers the static BTZ metric in the two exterior regions, which can be combined together by introducing a wormhole-like coordinate. The slices of this foliations are also smooth and satisfy the maximal slicing gauge but all of them pass through the bifurcate horizon at $r=\tilde{r}=R_h$ where the lapse vanishes. Following the steps of Section \ref{scalarquant}, we land up with the Hamiltonian $H_{\text{Killing}}$, which turns out to be the same as `$h_r - h_l$' \footnote{$h_r$ and $h_l$ are the bulk Hamiltonians that evolve in the Schwarzschild time $t$ and $\tilde{t}$ in Region I and Region II respectively, see equation 2.7 of \cite{Witten:2021unn} for the precise definitions.} and generates time evolution along the Killing slices. Smoothness and the fact that the Cauchy slice is not split into parts,\footnote{It is important that the Cauchy slice is not split, since there can be divergent fluctuations near the splitting point.} ensures that $H_{\text{Killing}}$ is a well-defined Hamiltonian. In fact, it annihilates the Hartle-Hawking vacuum, $H_{\text{Killing}}|\text{HH}\rangle = 0$.

The Hamiltonian $\hat{H}(\bar t)$ in \eqref{Hamfinal} corresponds to the lapse with positive boundary condition on both the asymptotic boundaries \eqref{NBCS} and generates time evolution along maximal slices that go into the black hole interior.\footnote{This is analogous to evolution with the Minkowski Hamiltonian as opposed to the Rindler time evolution. Minkowski time slices enter the Milne region while the Rindler slices do not.} The Hamiltonian $\hat{H}(\bar t)$ has finite expectation value and matrix elements in the Hilbert space of small excitations obtained by acting with a finite number of creation operators $e_{nq}^\dagger$ above the Hartle-Hawking state, which we denote by $\mathcal{H}_{\text{HH}}$. Since these maximal slices are smooth and not split into regions, we expect $||\hat{H}(\bar t)|\text{HH}\rangle||^2 = \langle \text{HH}| \hat{H}(\bar t)^2 |\text{HH}\rangle < \infty$, i.e., the state $\hat{H}(\bar t)|\text{HH}\rangle$ is normalizable. It is easy to see that
\begin{equation}
	\langle \text{HH}| \hat{H}(\bar t)^2 |\text{HH}\rangle =\sum_{\substack{n,n' \\ q,q'}} 2B^*_{nn'qq'}(\bar t) B_{n'nq'q}(\bar t) =2 \text{Tr}(B^* B) .
\end{equation}
Below we outline the reason for $\text{Tr}(BB^*)$ to be finite. 

The range of $n$ and $q$ is bounded by the following considerations. Recall that Hermite functions are eigenfunctions of the quantum harmonic oscillator in one dimension. Given this it is easy to see that the spatial spread of the $n$'th eigenstate, quantified by $\langle n|\hat{x}^2|n\rangle$, is proportional to $n$. Hence a function with support on an interval of say size $L$, can be well approximated by summing $n$ from zero to $n_{max}$, which is of the order of $L^2$. In the present problem, there is already a the natural scale in the frequency space, namely the surface gravity $\eta$ and the above estimate then gives $n_{max} \sim (\eta\ell)^2 \sim M$. \footnote{Another justification for the bound on the range of $n$ is by numerically studying the mode functions $g_{nq}$.  For example, using the near horizon expressions, one can check that their magnitude decreases with increasing $n$.}

The bound on the range of $q$ comes from the perspective of scattering off a black hole where the horizon radius, acts as an impact parameter. The larger the angular momentum, the smaller is the amplitude of a particle to fall into the black hole. In-falling modes with an `impact parameter' larger than the horizon radius are naturally cutoff. The relevant length scale here is the black hole horizon radius which in units of the AdS length $\ell$, is proportional to the square-root of the mass parameter $M$, thus effectively restricting $|q|\lesssim q_{max} \sim \sqrt{M}$. Once the range of $n$ and $q$ is bounded, $\text{Tr}(B^* B)$ is finite and the state $\hat{H}(\bar t)|\text{HH}\rangle$ has a finite norm and is in the Hilbert space $\mathcal{H}_{\text{HH}}$. 

Another reason to expect $\text{Tr}(B^* B)$ to be finite is from the lens of algebraic quantum field theory. The operator algebra $\mathcal{A}_\phi$ of the quantum scalar field associated to a complete Cauchy slice (which is the case for maximal slices considered here) is Type I because the state on the full Cauchy slice, namely the Hartle-Hawking state, is pure. Since any automorphism of a Type I factor is always an inner automorphism \cite{Witten:2021unn}, the evolution operator $U(\bar t)$ defined in equation \eqref{UOperator} belongs to the algebra $\mathcal{A}_\phi$. For a very short time interval $\epsilon$, since $U(\bar t=\epsilon) \approx \mathds{1} - \i\epsilon \hat{H}(0)$, $\hat{H}(0)$ should also belong to $\mathcal{A}_\phi$. Conjugation by the time evolution operator $U(\bar t)$ would then imply that $\hat{H}(\bar t) \in \mathcal{A}_\phi$ for all times.

\subsection{Diagonalizing the Hamiltonian}

Since the Hamiltonian $\hat{H}(\bar{t})$ is time dependent, the number operator $\mc{N}_{nq}(\bar{t}) = e^\dag_{nq}(\bar{t}) e_{nq}(\bar{t})$ is also time dependent in the Heisenberg picture. Equivalently the expectation value of the number operator $\mc{N}_{nq}(\bar{t}=0)$ in the state $|\text{HH}; \bar{t}\rangle$ changes with $\bar{t}$. We will now remedy this by introducing a new basis of operators $(b_{nq}, b^{\dagger}_{nq})$ related to the regularised Hartle-Hawking modes $(e_{nq}, e^{\dagger}_{nq})$ by a Bogoliubov transformation.

The Hamiltonian $H(\bar t)$ is Hermitian and hence can diagonalized. Rewriting \eqref{Hamfinal} more compactly as
\begin{equation}\label{HComp}
	H(\bar t) = \begin{pmatrix} e^{\dagger T} & e^T \end{pmatrix} \begin{pmatrix} A(\bar{t})\ & B^*(\bar{t}) \\ B(\bar{t})\ & A^*(\bar{t}) \end{pmatrix} \begin{pmatrix} e \\ e^\dagger \end{pmatrix} \, ,
\end{equation}
where we have arranged all the operators in a column, 
\begin{equation}
	e = \begin{pmatrix}
		e_{00} \\ e_{10} \\ \vdots \end{pmatrix} \, , \quad e^\dagger = \begin{pmatrix}
		e^\dagger_{00} \\ e^\dagger_{10} \\ \vdots \,
	\end{pmatrix}\, .
\end{equation}
After diagonalization, the Hamiltonian will take the form
\begin{equation}
	H(\bar t) = \begin{pmatrix} b^{\dagger T} & b^T \end{pmatrix} \begin{pmatrix} \mu(\bar{t})\ & 0 \\ 0\ & \mu(\bar{t}) \end{pmatrix} \begin{pmatrix} b \\ b^\dagger \end{pmatrix} \, ,
\end{equation}
where $\mu$ is a diagonal matrix. The new operators $b$ are related to the old ones by a Bogoliubov transformation which can be written in matrix form as
\begin{equation}
	\begin{pmatrix} b \\ b^\dagger \end{pmatrix} = M\begin{pmatrix} e \\ e^\dagger \end{pmatrix} = \begin{pmatrix} \alpha\ & \beta \\ \beta^*\ & \alpha^* \end{pmatrix} \begin{pmatrix} e \\ e^\dagger \end{pmatrix}\, .
\end{equation}
Pushing this relation in the Hamiltonian \eqref{HComp}, we find
\begin{equation}
	\begin{pmatrix} A(\bar{t})\ & B^*(\bar{t}) \\ B(\bar{t})\ & A^*(\bar{t}) \end{pmatrix} = M^\dagger \begin{pmatrix} \mu(\bar{t})\ & 0 \\ 0\ & \mu(\bar{t}) \end{pmatrix} M
\end{equation}
The Bogoliubov transformation matrix $M$ is determined from the eigenvectors of the matrix $\begin{pmatrix} A(\bar{t})\ & B^*(\bar{t}) \\ B(\bar{t})\ & A^*(\bar{t}) \end{pmatrix}$. Since $A(\bar{t})$, $B(\bar{t})$ are time dependent, the Bogoliubov coefficients $\alpha, \beta$ are also time dependent.

The $b$ operators define a new state $|\Omega; \bar{t}\rangle$ such that
\begin{equation}
	b_I |\Omega; \bar{t}\rangle = 0\, \quad \forall \, I,
\end{equation}
where $I$ is a label for the new modes. Using the Bogoliubov transformation
\begin{equation}
	b_I = \sum_{n,q} \big(\alpha_{I,nq} \, e_{nq} + \beta_{I,nq} \, e^\dagger_{nq}\big)\ , 
\end{equation}
we find that $|\Omega; \bar{t}\rangle$ is a squeezed state built on top of the Hartle-Hawking vacuum,
\begin{equation}
	|\Omega; \bar{t}\rangle = \mathcal{N} \exp\left(-\frac{1}{2} \sum_{\substack{n,n' \\ q,q'}} e^\dagger_{n q} (\alpha^{-1}\beta)_{nn' qq'} e^\dagger_{n' q'} \right) |\text{HH}\rangle \, ,
\end{equation}
where $\mathcal{N}$ is the overall normalization. Note that the state $|\Omega; \bar{t}\rangle$ is also time dependent because the Bogoliubov coefficients are time dependent.

The state $|\Omega; \bar{t}\rangle$ can be thought as a co-moving state in which the occupation number $b^\dagger_I b_I$ vanishes at all times,
\begin{equation}
	\langle\Omega; \bar{t}| b^\dagger_I b_I |\Omega; \bar{t} \rangle = 0\, .
\end{equation}
The occupation number of the operator $e^\dagger_{n q} e_{n q}$ in the state $|\Omega; \bar{t}\rangle $  is given by
\begin{equation}
	\langle\Omega; \bar{t}|e^\dagger_{n q} e_{n q}|\Omega; \bar{t} \rangle = \left(\beta^*\beta\right)_{nnqq} \, .
\end{equation}
It would be interesting understand this state further and its implications for black hole physics.

\subsection{Action of large diffeomorphisms on the scalar field}

As reviewed in Section \ref{largediffsec}, there is a family of large
diffeomorphisms $(\zeta_\perp, \zeta^i)$ \eqref{zetalarge} which
preserve the maximal slicing gauge and act on the maximally sliced
two-sided BTZ black hole. These are parametrized by the constants $c$
and $\tl{c}$ which correspond to the amount by which the 
large diffeomorphism translates the left and right boundary times $t$,
$\tl{t}$ respectively. The action of the large diffeomorphism
$(\zeta_\perp, \zeta^i)$ on the scalar field $\phi(\bar{t},x,\varphi)$
is obtained by substituting $\zeta_\perp$ and $\zeta^i$ in place of
the lapse $N$ and shift $N^i$ in the scalar field Hamiltonian
\begin{align}\label{maxlargediff}
  &H[\zeta_\perp, \zeta^i]\nonumber\\
  &= \int_\Sigma \ud^{d}x\, \bigg(\zeta_\perp \Big(\tfrac{1}{2\sqrt{{g}}}\pi_\phi^2 + \tfrac{1}{2} \sqrt{{g}}\,  {g}^{ij} (\partial_i\phi \partial_j \phi + m^2 \phi^2)\Big)\nonumber\\
                          &\qquad\qquad\quad + \zeta^i \pi_\phi \partial_i \phi\bigg)\, .
\end{align}
The action of the large diffeomorphism on $\phi$ and $\pi_\phi$ is
obtained by taking commutators with the Hamiltonian generator
\eqref{maxlargediff}. In particular, the original Hamiltonian
\eqref{maxham} corresponds to the large diffeomorphism labelled by
$c = \tl{c} = 1$. The above formula generalizes the action to
arbitrary values of $c$ and $\tl{c}$. One can also obtain expressions
similar to \eqref{Hamfinal} in terms of the smeared modes $e_{nq}$,
$e^\dag_{nq}$ by plugging in the mode expansion \eqref{scalargnx} for
the scalar field in the above formula \eqref{maxlargediff}.

\section{An order parameter that signals horizon crossing \label{sec:OP}}
Recall that $\mc{O}$ and $\tilde{\mc{O}}$ are the operators dual to
the scalar field $\phi$, by the extrapolate dictionary, in the right
and left CFTs, respectively. Consider an operator $A_L$ in the left
CFT such that $[A_L,\tilde{\mc{O}}]\neq 0$.  By definition
$[A_L, \mc{O}]=0$. Now consider the localized operator at $x_0>0$,
given by
\begin{equation}
	\phi_f(\bar t, x_0) = \int_0^{2\pi} \ud \varphi\int_{-\infty}^{\infty} \ud x\, \phi(\bar t,x, \varphi) f(x-x_0)\,  , 
\end{equation}
where $f(x-x_0)$ is a smooth function with compact support centred around $x=x_0$.\footnote{An example of a smooth function with compact support is given by
	\begin{equation}
		f(x) = \begin{cases}
			\exp\left(-\frac{1}{x^2-a^2}\right) & |x|<a \\
			0 & |x|>a ,
		\end{cases}
	\end{equation}
	where $a$ is positive number that controls the support of the function $f(x)$.}
Using the bulk reconstruction formula \eqref{BRWH} and recalling that $\tilde{\mathcal{K}}$ vanishes in the Region I, we see that $[ A_L, \phi_f(\bar t,x_0) ]=0$ in Region I. However it becomes non-zero, \begin{align}
	[ A_L, \phi_f(\bar t,x_0) ] = [ A_L, U(\bar t)^\dagger \phi_f(0,x_0) U(\bar t) ] \neq 0\, ,
\end{align}
as soon as the localised operator has support in the interior region F, where $\tilde{\mathcal{K}}$ is non-zero. Here $U(\bar t) $ is the unitary time evolution operator we constructed before in \eqref{UOperator}. If we define an `order parameter',
\begin{equation}\label{OrderParameter}
	\mathscr{H}(\bar t) = \langle \text{HH}| [ A_L, \phi_f(\bar t,x_0) ] |\text{HH}\rangle\, ,
\end{equation}
then it has the property
\begin{equation}
	\mathscr{H}(\bar t) = \begin{cases} =0 & (\bar t, x_0) \in \text{Region I} \\
		\neq 0 & (\bar t, x_0) \in \text{Region F} 
	\end{cases} \, .
\end{equation}
This means that for a localised operator evolving in $\bar t$ along a constant $x_0$ trajectory (that is sufficiently close to the horizon) the order parameter $\mathscr{H}(\bar t)$ becomes non-zero when the operator crosses the horizon at some $\bar t_0$. It is important to note that using the bulk reconstruction formula $\eqref{BRWH}$, the order parameter $\eqref{OrderParameter}$ can be expressed entirely in terms of CFT operators on the two boundaries and hence the transition from zero to non-zero value signals horizon crossing in the boundary theory.

\section{Scalar field correlation functions in the maximally sliced
  two-sided BTZ black hole \label{TPFunction}}

Recall the observation that under $\bar{t}$ time evolution, the state
of the scalar field in the two-sided BTZ black hole remains pure at
all times:
\begin{equation}
  \rho(\bar{t})^2 = \rho(\bar{t}), \qquad \rho(\bar{t}) = |\text{HH};\bar t\rangle \langle \text{HH};\bar t|\, .
\end{equation}
This implies that the correlation functions of the probe scalar field
in the maximally sliced background in the Hartle-Hawking state must be
that of a unitary theory, similar to that of quantum field theory in
Minkowski spacetime, and in particular exhibit no thermality.

In this section, we demonstrate this unitary nature of our time
evolution by explicit calculation of two-point Wightman functions of
the probe scalar field.
  

\subsection{An expression for the scalar field two-point function}

Recall the mode expansion of the scalar field \eqref{scalargn} in
terms of smooth mode functions ${g}_{nq}(U, V, \varphi)$:
\begin{equation}\label{scalargnx1}
  \phi(U,V,\varphi) = \frac{1}{\sqrt{2\pi} \ell}\sum_{q \in \mathbb{Z}}\sum_{n=0}^\infty  \big(e_{nq}\, {g}_{n q}(U,V,\varphi) + \text{c.c.}\Big)\, ,
\end{equation}
where the operators $e_{nq}$ annihilate the Hartle-Hawking state,
$e_{nq} |\text{HH}\rangle = 0$. From the above expression for the
two-point Wightman function of the scalar field in the maximal slicing
coordinate system can be written as
\begin{align}\label{scalar2pt}
  &G(\bar{t}, x, \varphi; \bar{t}', x', \varphi') \equiv \langle \text{HH}| \phi(\bar{t}, x, \varphi) \phi(\bar{t}', x', \varphi') |\text{HH}\rangle\nonumber\\
  &\quad= \frac{1}{2\pi\ell^2}\sum_{n,q} {g}_{nq}(U, V, \varphi) {g}^*_{nq}(U', V', \varphi')\, ,\nonumber\\
  &\quad= \frac{1}{2\pi\ell^2}\sum_{n,q} \hat{g}_{nq}(\bar{t}, x, \varphi) \hat{g}^*_{nq}(\bar{t}', x', \varphi')\, ,
\end{align}
where $U = U(\bar{t}, x)$, $V = V(\bar{t}, x)$,
$U' = U(\bar{t}', x')$, $V' = V(\bar{t}', x')$ and the second equality
is obtained by plugging in the the diffeomorphism 
from $U, V$ to $\bar{t}, x$.

The above expression for the two point function in terms of the
Kruskal-Szekeres coordinates, and equivalently, maximal slicing
coordinates, is well-defined everywhere in the maximal slicing
coordinate system since the mode functions ${g}_{nq}(U, V, \varphi)$
are differentiable solutions of the Klein-Gordon equation, and
consequently, so is the infinite sum in \eqref{scalar2pt}.\footnote{This is because the scalar field $\phi(\bar t,x,\varphi)$ in \eqref{scalargnx} itself satisfies the Klein-Gordon equation \eqref{KGwave}.} Recall that
there are explicit expressions for the mode functions
$\hat{g}_{nq}(\bar{t}, x, \varphi)$ which are obtained by starting
with the hypergeometric functions arising in Regions I and II and
performing the appropriate chain of manipulations described in Section
\ref{scalarquant}.

A very similar formula can be written down for the two-point function
of the scalar field in $d > 2$ maximally sliced AdS-Schwarzschild
black hole backgrounds, and hence the observations we make below also
hold in higher dimensions ($d > 2$).

For the two-sided BTZ black hole in $2+1$ dimensions, an alternate
expression for the two-point function can be obtained by leveraging
the fact that the two-sided BTZ geometry is a quotient of AdS$_3$ by a
discrete group of translations and using the method of images
\cite{Lifschytz:1993eb,Ichinose:1994rg,Keski-Vakkuri:1998gmz}. The
expression is in terms of Kruskal-Szekeres coordinates and is given by
\begin{align}\label{scalar2ptimages}
   &G(\bar{t}, x, \varphi; \bar{t}', x', \varphi') =\langle \text{HH}| \phi(\bar{t}, x, \varphi) \phi(\bar{t}', x', \varphi') |\text{HH}\rangle\nonumber\\
  &\quad= \sum_{n \in \mathbf{Z}} \frac{1}{\sqrt{z_n^2 - 1}} \left(z_n + \sqrt{z_n^2 -1}\right)^{1 - \Delta_+}\, ,
\end{align}
where
\begin{multline}
z_n = \frac{1}{(1 + \eta^2 UV)(1 + \eta^2 U'V')} \times \\ \times \bigg( (1 - \eta^2 UV)(1 - \eta^2 U'V') \cosh \big(\ell\eta (\varphi - \varphi' + 2n\pi)\big) \\ + 2 \eta^2(VU' + UV') \bigg)\, .
\end{multline}
While the two expressions \eqref{scalar2pt} and
\eqref{scalar2ptimages} look very different, they are equivalent upon
doing the smooth coordinate transformation between the wormhole and
Kruskal-Szekeres coordinates.

\subsection{Two-point function in Region I in BTZ
  coordinates}\label{RegionIBTZ}

The two-point function \eqref{scalar2pt} (equivalently,
\eqref{scalar2ptimages}) with the insertion points in Region I can be
rewritten in terms of the BTZ coordinates $(t, r, \varphi)$ by using
the coordinate transformation from wormhole coordinates
$(\bar{t}, x, \varphi)$ to $(t, r, \varphi)$. Attention must be paid
to the fact that this coordinate transformation is singular at the
horizons since the BTZ coordinates are valid only in Region I. The
correlation function turns out to be
\begin{align}
  &G({t}, r, \varphi; {t}', r', \varphi') = \nonumber\\
  &\quad= \sum_{n \in \mathbf{Z}} \frac{1}{\sqrt{z_n^2 - 1}} \left(z_n + \sqrt{z_n^2 -1}\right)^{1 - \Delta_+}\, ,
\end{align}
with
\begin{align}
  & z_n(t,r,\varphi; t',r',\varphi')\nonumber \\
  &= \frac{1}{R_h^2}\bigg(r r'  \cosh \big(\ell\eta (\varphi - \varphi' + 2n\pi)\big)\nonumber \\ &\quad- \sqrt{(r^2 - R_h^2)(r'^2 - R_h^2)}\cosh \big(\eta (t - t')\big) \bigg)\, .
\end{align}
The above correlation function satisfies the KMS property
\begin{equation}
  G({t}, r, \varphi; {t}', r', \varphi') = G(t'-\i\beta, r', \varphi';{t}, r, \varphi)\ ,
\end{equation}
where $\beta = 2\pi / \eta$. Thus, the correlators in Region I in the
singular BTZ coordinates are that of a thermal system in Region I with
temperature $\eta / 2\pi$. The thermal density matrix of the system is
obtained by tracing out the degrees of freedom in the Hartle-Hawking
state that arise from Region II. Curiously, this tracing out is
automatically performed by the singular coordinate transformation from
wormhole coordinates to BTZ coordinates in Region I.

\subsection{The two-point Wightman functions for specific insertion
  points}\label{various2pt}

We now present the values of the two-point Wightman functions for
various insertion points of the scalar field operators. Since the
expression \eqref{scalar2ptimages} is valid everywhere in the
maximally extended BTZ black hole geometry, we will use it to study
correlation functions with insertions throughout the extended
spacetime, as a function of the wormhole coordinates $\bar{t}$ and
$x$. 
We separately consider timelike and spacelike separated insertions below.

\subsubsection{Timelike separated insertions}\label{ACsec}

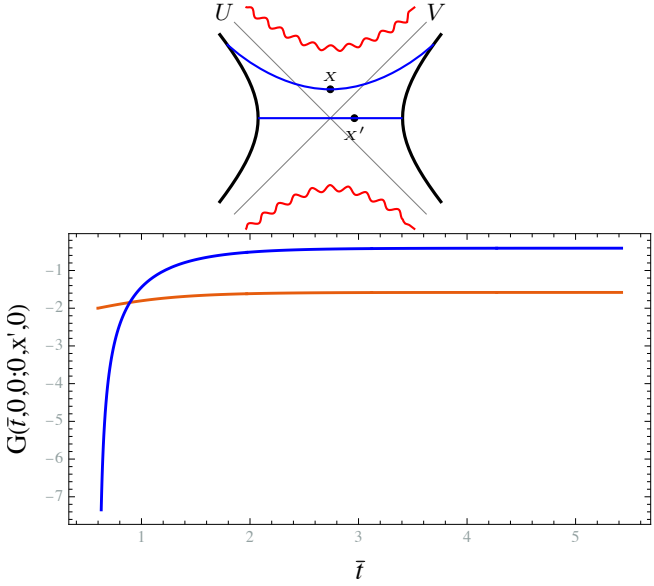
\begin{figure}[]
	\centering 
	\begin{tikzpicture}[scale=0.64]
		\draw[black, very thick] plot[domain=-1.:1.] ({1.5*cosh(\x)},{1.5*sinh(\x)});
		\draw[black, very thick] plot[domain=-1.:1.] ({-1.5*cosh(\x)},{1.5*sinh(\x)});
		\draw[decorate, decoration={snake, amplitude=0.4mm, segment length=2.4mm}, thick, red] plot[domain=-1.:1.] ({1.5*sinh(\x)},{1.5*cosh(\x)});
		\draw[decorate, decoration={snake, amplitude=0.4mm, segment length=2.4mm}, thick, red] plot[domain=-1.:1.] ({1.5*sinh(\x)},{-1.5*cosh(\x)});
		\draw[gray] (-2,-2) -- (2,2);
		\draw[gray] (-2,2) -- (2,-2);
		\filldraw[black] ({0},{0.6}) circle (2pt) node[anchor=south]{{\tiny $X$}};
		\filldraw[black] (0.5, 0) circle (2pt) node[anchor=north]{{\tiny $X'$}};
		\draw[blue, thick] (-1.5,0) -- (1.5, 0);
		\draw[blue, thick] plot[smooth, tension=1] coordinates { ({-1.5*cosh(-0.9)},{1.5*sinh(0.9)}) ({0},{0.6}) ({1.5*cosh(-0.9)},{1.5*sinh(0.9)}) };
		\node[black] at (2.2,2.2) {$V$};
		\node[black] at (-2.2,2.2) {$U$};
	\end{tikzpicture}\hfill
	\includegraphics[width=\linewidth]{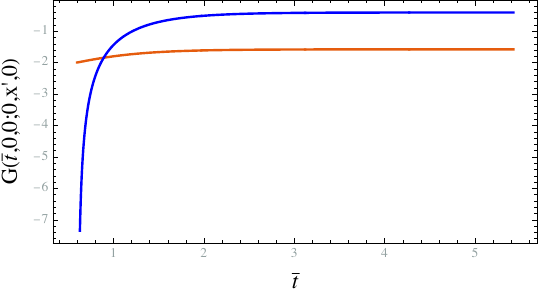}
	\caption{\label{fig:Gtxp05} The function \eqref{w1-ac} for timelike separated insertions is plotted as a function of time $\bar{t}$ with $x'=0.5 R_h$ held fixed. The imaginary part (blue) has the UV divergence at small times but decays at large times. On the other hand the real part (orange) saturates to a finite non-zero value at late times, showing that information is not lost even as $\bar{t}\to\infty$.  The two insertions are also shown schematically in the Kruskal-Szekeres diagram in the top figure.}
\end{figure}

We study the two-point correlator with one insertion in the exterior region I and another in the black hole interior in region F,
\begin{align}\label{w1-ac}
	&G(\bar{t},0,0;0,x',0) = \nonumber \\
	&\langle \text{HH}| \phi(\bar{t}, x=0, \varphi=0) \phi(\bar{t}'=0, x'>0, \varphi'=0) |\text{HH}\rangle .
\end{align}
For small $\bar t$, the two points can be spacelike separated, however for times large enough the separation is always timelike. This setup then provides a proxy for a correlator measured by an observer falling into the horizon. Keeping $x'>0$ fixed, we plot this correlator as a function of the time $\bar{t}$ in Figure \ref{fig:Gtxp05}. Clearly, this correlator does not decay to zero; instead (the real part) saturates to a non-zero value while the imaginary part decays to zero. 
This seems to be the consequence of bulk unitarity.

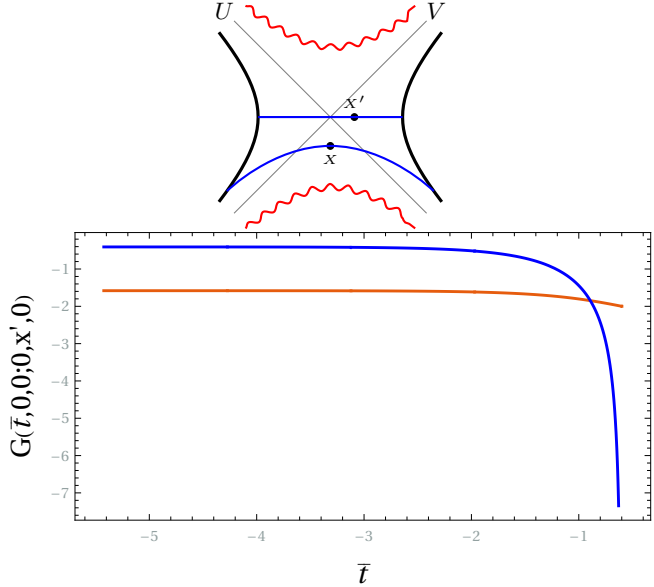
\begin{figure}[]
	\centering 
	\begin{tikzpicture}[scale=0.64]
		\draw[black, very thick] plot[domain=-1.:1.] ({1.5*cosh(\x)},{1.5*sinh(\x)});
		\draw[black, very thick] plot[domain=-1.:1.] ({-1.5*cosh(\x)},{1.5*sinh(\x)});
		\draw[decorate, decoration={snake, amplitude=0.4mm, segment length=2.4mm}, thick, red] plot[domain=-1.:1.] ({1.5*sinh(\x)},{1.5*cosh(\x)});
		\draw[decorate, decoration={snake, amplitude=0.4mm, segment length=2.4mm}, thick, red] plot[domain=-1.:1.] ({1.5*sinh(\x)},{-1.5*cosh(\x)});
		\draw[gray] (-2,-2) -- (2,2);
		\draw[gray] (-2,2) -- (2,-2);
		\filldraw[black] ({0},{-0.6}) circle (2pt) node[anchor=north]{{\tiny $X$}};
		\filldraw[black] (0.5, 0) circle (2pt) node[anchor=south]{{\tiny $X'$}};
		\draw[blue, thick] (-1.5,0) -- (1.5, 0);
		\draw[blue, thick] plot[smooth, tension=1] coordinates { ({-1.5*cosh(-0.9)},{-1.5*sinh(0.9)}) ({0},{-0.6}) ({1.5*cosh(-0.9)},{-1.5*sinh(0.9)}) };
		\node[black] at (2.2,2.2) {$V$};
		\node[black] at (-2.2,2.2) {$U$};
	\end{tikzpicture}\hfill
	\includegraphics[width=\linewidth]{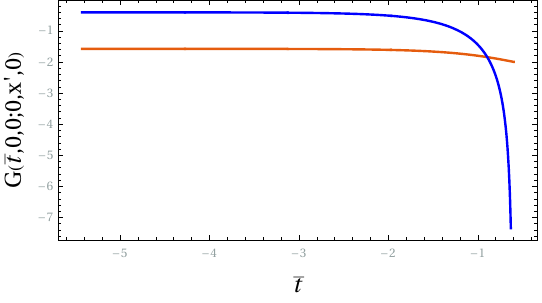}
	\caption{\label{fig:Gtxp05-TR} The function \eqref{w1-ac} for timelike separated insertions is plotted for $\bar{t}<0$ with $x'=0.5 R_h$ held fixed. The imaginary part (blue) has the UV divergence at small times but decays at large times. On the other hand the real part (orange) saturates to a finite non-zero value at large negative times, which follows from the time reflection property. The two insertions are also shown schematically in the Kruskal-Szekeres in the top figure.}
\end{figure}

We also note that for an insertion in region P ($\bar t<0$) on the $x=0,\varphi=0$ line, the two-point function in \eqref{w1-ac} will saturate to the same finite non-zero value for large negative times as well, as shown in figure \ref{fig:Gtxp05-TR}. This is a consequence of the fact that the background classical solution is invariant under $\bar t \to -\bar t$, and the scalar field transforms as $\phi (\bar t, x, \varphi) \to \phi (-\bar t,x, \varphi)$. We would not expect this to happen if our time evolution was dissipative. 

As a special case, we put $x'=0$ together with $\bar{t}'=0$, which implies that $U'=V'=0$, i.e., this insertion is exactly at the bifurcate horizon. In this case the correlator $G(\bar{t},0,0;0,0,0)$
also saturates to a finite non-zero value. 

\subsubsection{Spacelike separated insertions}

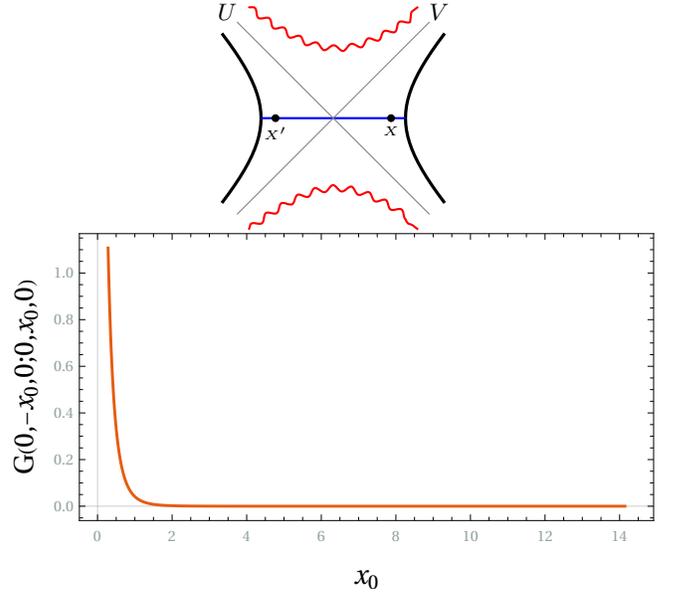
\begin{figure}[]
	\centering 
	\begin{tikzpicture}[scale=0.64]
		\draw[black, very thick] plot[domain=-1.:1.] ({1.5*cosh(\x)},{1.5*sinh(\x)});
		\draw[black, very thick] plot[domain=-1.:1.] ({-1.5*cosh(\x)},{1.5*sinh(\x)});
		\draw[decorate, decoration={snake, amplitude=0.4mm, segment length=2.4mm}, thick, red] plot[domain=-1.:1.] ({1.5*sinh(\x)},{1.5*cosh(\x)});
		\draw[decorate, decoration={snake, amplitude=0.4mm, segment length=2.4mm}, thick, red] plot[domain=-1.:1.] ({1.5*sinh(\x)},{-1.5*cosh(\x)});
		\draw[blue, thick] plot[domain=-1.5:1.5] ({\x},{0});
		\draw[gray] (-2,-2) -- (2,2);
		\draw[gray] (-2,2) -- (2,-2);
		\filldraw[black] (-1.2,0) circle (2pt) node[anchor=north]{{\tiny $X'$}};
		\filldraw[black] (1.2,0) circle (2pt) node[anchor=north]{{\tiny $X$}};
		\node[black] at (2.2,2.2) {$V$};
		\node[black] at (-2.2,2.2) {$U$};
	\end{tikzpicture}\hfill
	\includegraphics[width=\linewidth]{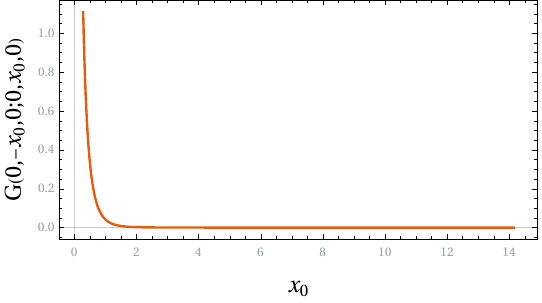}
	\caption{\label{fig:W2} The Wightman correlator \eqref{w2} plotted as a function of the spatial separation $x_0$ on the $\bar{t}=0$ slice. The real part clearly decays. The imaginary part simply vanishes and is therefore not shown. For this plot, we have taken $\Delta_+=3$. The insertions are also shown schematically in the Kruskal-Szekeres diagram in the top figure.}
\end{figure}

When the two insertions in regions I and II each, the separation is always spacelike. First we consider the two insertions on the $\bar{t}=0$ slice and study
\begin{align}\label{w2}
	&G(0, x_0, 0; 0, -x_0, 0) = \nonumber \\
	&\langle \text{HH}| \phi(\bar{t}=0, x=x_0, \varphi=0) \phi(\bar{t}'=0, x'=-x_0, \varphi'=0) |\text{HH}\rangle
\end{align}
as a function of the separation $x_0$. This correlation function
decays with increasing $x_0$ as can be seen in Figure
\ref{fig:W2}. This is analogous to the decay of the massive scalar
field two-point function on a constant time slice in Minkowski
spacetime as the separation between the points increases.

\begin{figure}[]
	\centering 
	\begin{tikzpicture}[scale=0.64]
		\draw[black, very thick] plot[domain=-1.:1.] ({1.5*cosh(\x)},{1.5*sinh(\x)});
		\draw[black, very thick] plot[domain=-1.:1.] ({-1.5*cosh(\x)},{1.5*sinh(\x)});
		\draw[decorate, decoration={snake, amplitude=0.4mm, segment length=2.4mm}, thick, red] plot[domain=-1.:1.] ({1.5*sinh(\x)},{1.5*cosh(\x)});
		\draw[decorate, decoration={snake, amplitude=0.4mm, segment length=2.4mm}, thick, red] plot[domain=-1.:1.] ({1.5*sinh(\x)},{-1.5*cosh(\x)});
		\draw[gray] (-2,-2) -- (2,2);
		\draw[gray] (-2,2) -- (2,-2);
		\filldraw[black] ({1.6},{1.1}) circle (2pt) node[anchor=south]{{\tiny $X$}};
		\filldraw[black] (-1.3, 0) circle (2pt) node[anchor=north]{{\tiny $X'$}};
		\draw[blue, thick] (-1.5,0) -- (1.5, 0);
		\draw[blue, thick] plot[smooth, tension=1] coordinates { ({-1.5*cosh(-0.9)},{1.5*sinh(0.9)}) ({0},{0.6}) ({1.5*cosh(-0.9)},{1.5*sinh(0.9)}) };
		\node[black] at (2.2,2.2) {$V$};
		\node[black] at (-2.2,2.2) {$U$};
	\end{tikzpicture}\hfill
	\includegraphics[width=\linewidth]{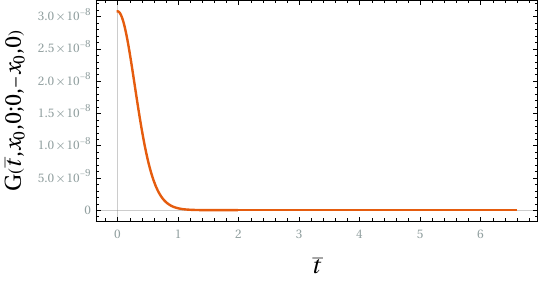}
	\caption{\label{fig:W3} The Wightman correlator \eqref{w3} plotted as a function of the time $\bar{t}$, keeping $x_0=10 R_h$ fixed. The real part clearly decays. The imaginary part simply vanishes and is therefore not shown. For this plot, we have taken $\Delta_+=3$. The two insertions are also shown schematically in the Kruskal-Szekeres diagram in the top figure.}
\end{figure}

We then also consider insertions near the two asymptotic boundaries in regions I and II, but at different time slices,
\begin{align}\label{w3}
	&G(\bar{t}, x_0, 0;0, -x_0, 0) = \nonumber \\
	&\langle \text{HH}| \phi(\bar{t}, x=x_0, \varphi=0) \phi(\bar{t}'=0, x'=-x_0, \varphi'=0) |\text{HH}\rangle .
\end{align}
With $x_0$ held fixed, we find that this correlation function decays as $\bar{t}$ increases, as is evident from Figure \ref{fig:W3}.

%
%

\subsection{The boundary limit of the correlators}

In this subsection, we evaluate the two point function with one scalar field insertion near the right boundary and one near the left boundary by considering $x = \varepsilon^{-1}$ and $x' = -\varepsilon^{-1}$, in the limit that the cutoff $\varepsilon$ is taken to zero.

Since the dependence of $\varepsilon$ comes through the dependence of
$U$ and $V$ in the correlation function \eqref{scalar2pt} or
\eqref{scalar2ptimages}, we have to look at the $\varepsilon \to 0$
limit of the expansions of $U$ and $V$ in region I,
\begin{subequations}\label{UV-expansion}
	\begin{align}
		U  &= -\eta^{-1} \e^{-\eta\bar{t}}  \Big( 1 - R_h\varepsilon + \frac{1}{2} (R_h\varepsilon)^2 \nonumber \\
		&\quad-\frac{-3 - 2 \frac{T(\bar{t})}{T_\infty} + 3\sqrt{1 - \frac{T(\bar{t})^2}{T_\infty^2}}}{12} (R_h\varepsilon)^3 + \mathcal{O}(\varepsilon^4) \Big)\, , \\
		V  &= \eta^{-1} \e^{\eta\bar{t}}  \Big( 1 - R_h\varepsilon + \frac{1}{2} (R_h\varepsilon)^2 \nonumber \\
		&\quad+\frac{-3 + 2 \frac{T(\bar{t})}{T_\infty} + 3\sqrt{1 - \frac{T(\bar{t})^2}{T_\infty^2}}}{12} (R_h\varepsilon)^3 + \mathcal{O}(\varepsilon^4) \Big)\, .
	\end{align}
\end{subequations}
It is easy to see that in the limit $\varepsilon \to 0$, the
Kruskal-Szekeres coordinates become
\begin{equation}
  \lim_{\varepsilon \to 0} U(\bar{t}, \varepsilon^{-1}) = -\eta^{-1} \e^{-\eta \bar{t}}\ ,\quad   \lim_{\varepsilon \to 0} V(\bar{t}, \varepsilon^{-1}) = \eta^{-1} \e^{\eta \bar{t}}\ .
\end{equation}

In the BTZ black hole background, we can substitute the above values
for $U$, $V$, and the appropriate values for $U'$, $V'$ (which are in region II), in the expression \eqref{scalar2ptimages} to get
\begin{align}\label{scalarRL}
	&\lim_{\varepsilon \to 0}\langle \text{HH}| \phi(\bar{t},\varepsilon^{-1}, \varphi) \phi(\bar{t}',-\varepsilon^{-1}, \varphi') |\text{HH}\rangle = (\varepsilon^{\Delta_+})^2 \times \nonumber\\
	& \times \sum_{n \in \mathbf{Z}}\Big(\cosh \big(\eta(\bar{t} - \bar{t}')\big) + \cosh \big(\ell \eta (\varphi - \varphi' + 2n\pi)\big)\Big)^{-\Delta_+} ,
\end{align}
where recall that $\Delta_+ = 1 + \sqrt{1 + \ell^2 m^2}$ is the
conformal dimension of the generalized free field operator $\mc{O}$
dual to the bulk scalar field $\phi$.

The two-point function of the generalized free field $\mc{O}$ in the
dual CFT is obtained by using the usual extrapolate dictionary \eqref{phiextrapol}:
\begin{equation}\label{phiextrapol}
  \phi(\bar{t}, x, \varphi) \to \left\{\def\arraystretch{1.5}\begin{array}{cc} x^{-\Delta_+} \mc{O}({t}, \varphi) & x \to +\infty\, , \\ (-x)^{-\Delta_+} \tl{\mc{O}}(\tl{t},\varphi) & x \to -\infty\, , \end{array}\right.
\end{equation} 
The Hartle-Hawking state is dual to the thermofield double state in
the product of the two CFTs. Then we get, in the product of the dual
CFTs on the left and right boundaries,
\begin{align}\label{CFTRL}
	&\langle \text{TFD}| \mc{O}({t}, \varphi) \tl{\mc{O}}({t}', \varphi') |\text{TFD}\rangle  \nonumber\\
	& = \sum_{n \in \mathbf{Z}}\Big(\cosh \big(\eta({t} + {t}')\big) + \cosh \big(\ell \eta (\varphi - \varphi' + 2n\pi)\big)\Big)^{-\Delta_+} ,
\end{align}
which agrees with the expression computed in
\cite{Maldacena:2001kr}.\footnote{See also \cite{Mandal:2014wfa} for
  computation of the left-right boundary correlator using the bulk
  geodesic approximation in the large scalar field mass limit.} This
expression clearly decays at large times. The correlation function
\eqref{CFTRL} above can also be obtained from the extrapolate limit of
the bulk correlator \eqref{w3} with non-zero $t'$. Then the decay 
is not surprising from the bulk point of view, since we already
expected it to decay to zero (see figure \ref{fig:W3}). 

Using the bulk reconstruction formula \eqref{BRWH}, the two-point Wightman correlator in \eqref{w1-ac}, with one insertion in Region I and another in Region F, can be written as\footnote{Note that any correlator with insertions in region $R_0$ can expressed in terms of boundary operators. We consider this correlator as a specific example.}
\begin{widetext}
	\begin{align}\label{bulkbdry}
		&\langle\text{HH}| \phi(\bar{t},0,0) \phi(0,x'>0,0) |\text{HH}\rangle \nonumber\\
		&= \int_0^{2\pi} \ud\varphi' \ud\varphi'' \int_{-\infty}^\infty \ud t' \ud t'' \Big( \mathcal{K}(\bar{t},0,0;t',\varphi') \mathcal{K}(0,x',0; \tilde{t}'', \tilde{\varphi}'') \langle\text{HH}| \mathcal{O}(t',\varphi') \mathcal{O}(\tilde{t}'', \tilde{\varphi}'') |\text{HH}\rangle \nonumber \\
		&\qquad\qquad\qquad\qquad\qquad\qquad + \tilde{\mathcal{K}}(\bar{t},0,0;t',\varphi') \mathcal{K}(0,x',0; \tilde{t}'', \tilde{\varphi}'') \langle\text{HH}| \tilde{\mathcal{O}}(t',\varphi') \mathcal{O}(\tilde{t}'', \tilde{\varphi}'') |\text{HH}\rangle \Big)\ .
	\end{align}
\end{widetext}
For the insertion in Region I, the second term involving $\tilde{O}$ vanishes. The bulk correlation function is thus a integral transform of boundary correlators. While the extrapolate limit of bulk correlation functions decay to zero, the correlation function deep in the bulk need not decay to zero due to the complicated form of the $\mc{K}$ and $\tl{\mc{K}}$ kernels in \eqref{bulkbdry}. This is precisely what we have observed in the sample calculations in this section.

\begin{acknowledgments}
  We would like to thank Rajesh Gopakumar, Hong Liu, Suresh Govindarajan, Alok Laddha, Gautam Mandal, Suvrat Raju, Ashoke Sen, 
  Ronak  Soni, Sandip Trivedi, Shiraz Minwalla, Edward Witten, and especially Raghu Mahajan and Kyriakos Papadodimas, for discussions and comments.
  We also acknowledge the organizers and participants of the `Workshop on Canonical Gravity', held at Chennai Mathematical Institute, for discussions.
  S.~R.~W. would like to thank the CERN Theory Division for
  hospitality, and the Infosys Foundation Homi Bhabha Chair at
  ICTS-TIFR for its support.
\end{acknowledgments}

\appendix

\section{Derivation of the maximally sliced two-sided BTZ black hole
  solution} \label{msbtz-app}

In this appendix, we outline the solution of Einstein's equations in
the maximal slicing gauge in $2+1$ dimensions which gives the
maximally sliced two-sided BTZ black hole. A very similar procedure
gives the maximally sliced two sided AdS-Schwarzschild black hole in
$d + 1$ dimensions for $d > 2$. We closely follow
\cite{Estabrook:1973ue} in which the maximal slicing of the flat
Schwarzschild black hole in $3+1$ dimensions was derived \footnote{See
  also the textbook \cite{Baumgarte:2010ndz} for a detailed
  discussion}.

\subsection{The maximal slicing solution}
The general ADM ansatz for a non-rotating metric is
\begin{align}\label{ds2-ansatzapp}
  \ud s^2 &= -\alpha(r,\bar t)^2 \ud\bar t^2 + A(r,\bar t) \left(\ud r + \frac{\beta(r,\bar t)}{A(r,\bar t)} \ud\bar t \right)^2 + r^2 \ud \varphi^2\, .
\end{align}
The spatial metric is
\begin{equation}
  \ud s^2 = g_{ij} \ud x^i \ud x^j = A(\bar{t}, r) \ud r^2 + r^2 \ud \varphi^2\, .
\end{equation}
The Ricci scalar of $g_{ij}$ is
\begin{equation}
  R = \frac{\partial_r A}{r A^2}\, .
\end{equation}
We use the notation $\dot{}$ to denote a derivative with respect to
the time $\bar{t}$. The extrinsic curvature $K_{ij}$ is defined as
\begin{equation}
  K_{ij} = \frac{1}{2N} (\dot{g}_{ij} - 2 D_{(i} N_{j)})\, ,
\end{equation}
where $N$ is the lapse and $N^i$ is the shift. Here, $N = \alpha$,
$N^r = \beta / A$, $N^\varphi = 0$. We then have
$N_r = g_{rr} N^r = \beta$. The non-zero components of the extrinsic
curvature for the ansatz are \footnote{It is convenient to use a
  Mathematica package such as diffgeo.m by Matthew Headrick.}
\begin{align}\label{Kvalapp}
  K_{rr} &= \frac{1}{2\alpha} \left(\partial_{\bar{t}}A + \beta\frac{\partial_r A}{A} - 2\partial_r\beta \right)\, , \quad	K_{\varphi\varphi} = -\frac{r\beta}{A\alpha}\, .
\end{align}
The maximal slicing gauge condition is
\begin{equation}
    K = g^{ij} K_{ij} = 0\, ,
\end{equation}
which, with the above ansatz plugged in, becomes
\begin{equation}\label{Kzeroapp}
  \partial_{\bar t} A - \beta \partial_r \log\left(\frac{r^2\beta^2}{A}\right) = 0\, .
\end{equation}
Let us look at the momentum constraints
$\mc{H}^j = -2 D_i K^{ij} = 0$. The momentum constraint
$\mc{H}^\varphi$ is automatically satisfied due to the angular
symmetry. The radial momentum constraint $\mc{H}^r$ becomes the simple
equation
\begin{equation}
  \partial_r K_{\varphi\varphi} = 0\, ,\quad\text{with solution} \quad K_{\varphi\varphi} = -T(\bar{t})\, ,
\end{equation}
where $T$ is independent of the spatial coordinates but can be a
function of time $\bar{t}$.  Using the expression for
$K_{\varphi\varphi}$ from \eqref{Kvalapp}, we then get
\begin{equation}
  \frac{r \beta}{A\alpha} = T(\bar{t})\, .
\end{equation}
We next solve the Hamiltonian constraint
$\mc{H}_\perp = K^{ij} K_{ij} - K^2 - R + 2\Lambda = 0$ which becomes
the following first order equation for $A$:
\begin{equation}
  2 \frac{T(\bar{t})^2}{r^4} - \frac{\partial_r A}{r A^2} + 2\Lambda = 0\, ,
\end{equation}
The solution is
\begin{equation}\label{ASol-app}
  A(r) = \left(\frac{r^2}{\ell^2} - \frac{M}{2\pi} + \frac{T(\bar{t})^2}{r^2}\right)^{-1}\, ,
\end{equation}
where $M$ is the integration constant. It turns out that $M$ will be
ADM mass of the solution, and it can be shown that it is constant
under time evolution using the $\dot{K}_{rr}$ Einstein equation.

The maximal slicing condition \eqref{Kzeroapp} can now be solved for
the lapse $\alpha$ to get
\begin{equation}\label{alphaSol}
  \alpha(r,\bar{t}) = A(r,\bar{t})^{-1/2} \left(c - \dot{T} \int_\infty^r\frac{\ud\rho}{\rho} A(\rho,\bar{t})^{3/2} \right)\, .
\end{equation}
The integration constant is $c$, and is imposed as a boundary
condition on $\alpha$ as $r\to \infty$:
\begin{equation}
  \lim_{r\to\infty} \alpha = c \frac{r}{\ell}\, ,
\end{equation}
which is the familiar asymptotic AdS boundary condition for the
lapse. The usual AAdS boundary conditions set $c=1$ so that the time
$\bar{t}$ agrees with the time $t$ of the asymptotic AdS metric near
the boundary.
\begin{equation}
  \text{AAdS boundary condition}:\quad c = 1\, .
\end{equation}
Note that $T$ as a function of $\bar t$ is still undetermined.


Thus, the solution to the Einstein's equations in maximal slicing
gauge is
\begin{align}\label{ds2-Sol-app}
  \ud s^2 &= -\alpha(r,\bar t)^2 \ud\bar t^2 + A(r,\bar t) \left(\ud r + \frac{\alpha(r,\bar t) T(\bar t)}{r} \ud\bar t \right)^2\nonumber\\
          &\quad + r^2 \ud \varphi^2\ ,\nonumber \\
  A(r,\bar t) &=\left(\frac{r^2}{\ell^2} -\frac{M}{2\pi} +\frac{T(\bar t)^2}{r^2}\right)^{-1}\ ,\nonumber  \\
  \alpha(r,\bar t) &= A(r,\bar t)^{-1/2} \left(1 - \dot{T}(\bar t) \int_\infty^r\frac{\ud\rho}{\rho} A(\rho,\bar{t})^{3/2}  \right)\ . 
\end{align}
The above solution has coordinate singularities since the component
$g_{rr}$ blows up at the roots of the equation
$r^4 - \frac{M}{2\pi}\ell^2 r^2 + \ell^2 T^2 = 0$ which appears in the
denominator of $A(r,\bar{t})$, see \eqref{ds2-Sol-app}. The roots of
the equation are the following two repeated roots
\begin{equation}\label{rpmdefapp}
  R_\pm(T) = \ell\sqrt{\frac{M}{4\pi}} \left[1 \pm \sqrt{1-\left(\frac{4\pi T(\bar t)}{M\ell}\right)^2}\right]^{1/2}\, ,
\end{equation}
Thus, there are coordinate singularities at $r = R_\pm(T)$. From the expressions above, it is clear that the range of the function $T(\bar{t})$ is restricted to
\begin{equation}\label{Tdefapp}
  -T_\infty \leq T \leq T_\infty\, ,\quad\text{with}\quad T_\infty = \frac{\ell M}{4\pi}\, .
\end{equation}
It is also clear that $R_\pm(T)$ satisfy
\begin{equation}
R_-(T) \leq R_+(T)\, ,
\end{equation}
with the equality satisfied only when
$T = T_\infty = \ell M / 4\pi$:
\begin{equation}
  R_+(T) = R_-(T) = \ell\sqrt{\frac{M}{4\pi}}\quad\text{when}\quad T = \pm T_\infty\, .
\end{equation}
Below, we consider the diffeomorphism of the above solution to the
two-sided BTZ black hole. We shall see that the above solution covers
only a part of the two-sided BTZ black hole which includes the right
boundary. Coming in from the boundary $r = \infty$, the above metric
stops being valid at the first coordinate singularity $r = R_+(T)$. We
call this the \emph{right areal chart}.

Beyond this, it turns out that we need another chart to cover
a similar region that starts from the left boundary of the two-sided
BTZ black hole and comes up to the radius $R_+(T)$. To describe this
solution, we use a different areal radial coordinate $\tl{r}$ (which
coincides with the areal radial coordinate of the BTZ solution in
Region II, see Figure \ref{BTZKruskal}). The solution in the maximal
slicing gauge is the same as earlier, with obvious changes in
notation:
\begin{align}\label{ds2-Sol-left-app}
  \ud s^2 &= -\tl\alpha(\tl{r},\bar t)^2 \ud\bar t^2 + \tl{A}(\tl{r},\bar t) \left(\ud \tl{r} + \frac{\tl\alpha(\tl{r},\bar t) \tl{T}(\bar t)}{\tl{r}} \ud\bar t \right)^2\nonumber\\
          &\qquad+ \tl{r}^2 \ud \varphi^2\, , \nonumber\\
  {A}(\tl{r},\bar t) &=\left(\frac{\tl{r}^2}{\ell^2} -\frac{M}{2\pi} +\frac{\tl{T}(\bar t)^2}{\tl{r}^2}\right)^{-1}\, , \nonumber\\
  \tl\alpha(\tl{r},\bar t) &= {A}(\tl{r},\bar t)^{-1/2} \left(1 - \dot{\tl{T}}(\bar t) \int_\infty^{\tl{r}}\frac{\ud\rho}{\rho} A(\tl{r}, \bar{t})^{3/2} \right)\, . 
\end{align}
We call this solution the \emph{left areal chart}.


\subsection{Diffeomorphism to the two-sided BTZ black hole}

In this subsection, we obtain the diffeomorphism from the maximal
slicing solution obtained previously to the two-sided BTZ black
hole. Our strategy is to look at the diffeomorphism in each of the
four Regions of the two-sided BTZ black hole separately and then put
them together. The right areal chart solution \eqref{ds2-Sol-app} will
be compared with the solution in Region I and it can be continued into
Regions F and P till the radius $r = R_+(T)$. Similarly, Region II and
remaining parts of Regions F and P will be covered by the left areal
chart solution \eqref{ds2-Sol-left-app}.

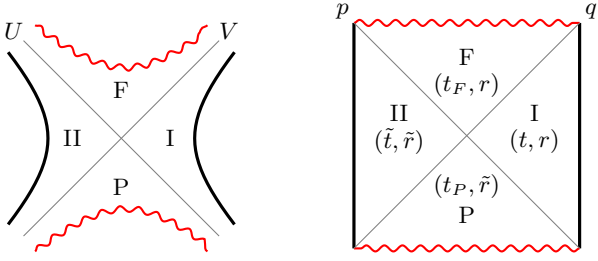
\begin{figure}
  \centering
  \begin{tikzpicture}[scale=0.65]
    \draw[black, very thick] plot[domain=-1.:1.] ({1.5*cosh(\x)},{1.5*sinh(\x)});
    \draw[black, very thick] plot[domain=-1.:1.] ({-1.5*cosh(\x)},{1.5*sinh(\x)});
    \draw[decorate, decoration={snake, amplitude=0.4mm, segment length=2.4mm}, thick, red] plot[domain=-1.:1.] ({1.5*sinh(\x)},{1.5*cosh(\x)});
    \draw[decorate, decoration={snake, amplitude=0.4mm, segment length=2.4mm}, thick, red] plot[domain=-1.:1.] ({1.5*sinh(\x)},{-1.5*cosh(\x)});
    \draw[gray] (-2,-2) -- (2,2);
    \draw[gray] (-2,2) -- (2,-2);
    \node[black] at (2.2,2.2) {$V$};
    \node[black] at (-2.2,2.2) {$U$};
    \node[black] at (1,0) {I};
    \node[black] at (0,1) {F};
    \node[black] at (-1,0) {II};
    \node[black] at (0,-1) {P};
  \end{tikzpicture}
  \qquad\quad \begin{tikzpicture}[scale=0.75]
    \draw[black, very thick] (-2,-2) -- (-2,2);
    \draw[black, very thick] (2,-2) -- (2,2);
    \draw[decorate, decoration={snake, amplitude=0.4mm, segment length=2.4mm}, thick, red] (-2,2) -- (2,2);
    \draw[decorate, decoration={snake, amplitude=0.4mm, segment length=2.4mm}, thick, red] (2,-2) -- (-2,-2);
    \draw[gray] (-2,-2) -- (2,2);
    \draw[gray] (-2,2) -- (2,-2);
    \node[black] at (2.2,2.2) {$q$};
    \node[black] at (-2.2,2.2) {$p$};
    \node[black] at (1.2,0.4) {I};
    \node[black] at (1.2,-0.1) {$(t,r)$};
    \node[black] at (0,1.4) {F};
    \node[black] at (0,0.9) {$(t_F,r)$};
    \node[black] at (-1.2,0.4) {II};
    \node[black] at (-1.2,-0.1) {$(\tl{t},\tl{r})$};
    \node[black] at (0,-1.4) {P};
    \node[black] at (0,-0.9) {$(t_P,\tl{r})$};
  \end{tikzpicture}
  \caption{\label{BTZKruskalapp} The Kruskal (left) and Penrose (right)
    diagrams for the two-sided BTZ black hole solution. The two solid
    black lines are the asymptotic AdS boundaries, the diagonal gray
    lines are the event horizons and the wiggly red lines are the
    singularities. The horizons divide the two-sided black hole into
    four regions labelled I, II, F and P, each of which have a
    static Schwarzschild-like coordinate system which is displayed in
    the Penrose diagram.}
\end{figure}

Recall the standard BTZ metric in Region I:
\begin{equation}\label{BTZmetapp}
  \ud s^2 = -f(r)^2 \ud t^2 + f(r)^{-1} \ud r^2 + r^2 \ud \varphi^2\, , \quad f(r) = \frac{r^2}{\ell^2} - \frac{M}{2\pi}\, ,
\end{equation}
with the Killing event horizon located at $r = R_h$ with
\begin{equation}\label{horapp}
  R_h = \ell \sqrt{\frac{M}{2\pi}}\, .
\end{equation}
See Figure \ref{BTZKruskalapp} for a pictorial description of the
various regions of the two-sided BTZ black hole.

The simplest diffeomorphism is to map the radial coordinate $r$ and
angle $\varphi$ in the above metric to $r$ and $\varphi$ in the the
maximally sliced solution \eqref{ds2-Sol-app}. The only non-trivial
function to compute then is $t(\bar{t}, r)$, i.e., the BTZ coordinate
time $t$ as a function of $\bar{t}$ and $r$. Plugging this into
\eqref{BTZmetapp} and comparing with \eqref{ds2-ansatzapp}, we get
\begin{align}\label{BTZdiffapp}
  \partial_{\bar t}t &= \epsilon\alpha\sqrt{A}\, ,\quad \partial_r t = -\epsilon\frac{T \sqrt{A}}{r f(r)}\, ,
\end{align}
where $\epsilon$ is a sign. In Region I for $t>0$, since $t$ increases
as $r$ decreases along a maximal slice, we must choose
$\epsilon=+1$. Similarly, $\epsilon = -1$ for the $t < 0$ part of
Region I. These signs can be inferred visually by tracking the change
in $r$ and $t$ along a maximal slice in Figure \ref{fig:WH1F}.

Suppose we consider a maximal slice that starts at the boundary in the
$t > 0$ part of Region I. Then, the first equation in
\eqref{BTZdiffapp} gives
\begin{equation}\label{ttbarbdryapp}
  \lim_{r\to\infty} t = \bar{t} + t_0\, ,
\end{equation}
where $t_0$ is an arbitrary constant which sets the origin of time on
the boundary in Region I. We take this constant $t_0$ to be zero since
we would like to identify the origin of $t$ and
$\bar{t}$. Integrating the second equation gives the
diffeomorphism $t(r,\bar{t})$ in Region I:
\begin{equation}\label{BTZ-time-app}
  t(r, \bar t)  = \bar{t} + T(\bar t) \int_r^{\infty} \ud\rho\frac{A(\rho,\bar{t})^{1/2}}{\rho f(\rho)}\, .
\end{equation}

Next, we study the diffeomorphism in Region F. The BTZ metric in
Region F is
\begin{equation}\label{BTZmetappF}
  \ud s^2 = -f(r)^2 \ud t_F^2 + f(r)^{-1} \ud r^2 + r^2 \ud \varphi^2\ ,\quad r< \ell \sqrt{\frac{M}{2\pi}}\ ,
\end{equation}
with $r < R_h$. The diffeomorphism is obtained by solving the equations
\begin{align}\label{BTZdiffappF}
  \partial_{\bar t}t_F &= \epsilon\alpha\sqrt{A}\ ,\quad \partial_r t_F = -\epsilon\frac{T \sqrt{A}}{r f(r)}\ ,
\end{align}
where the sign $\epsilon$ must be chosen to be $+1$ for the $t_F > 0$
part of Region F and $-1$ in the $t_F < 0$ part of Region F.

Note that the derivative $\partial t_F / \partial r$ blows up at the
roots $R_\pm(T)$ \eqref{rpmdefapp}. It is easy to see that these roots
satisfy
\begin{equation}
  R_-(T) \leq R_+(T) \leq R_h\, ,
\end{equation}
where recall that $R_h$ is the horizon radius \eqref{horapp}. Thus,
both $R_\pm$ are located in Region F with $r < R_h$. As one comes in
from the horizon $r = R_h$ on a constant $\bar{t}$ slice, the
divergence of $\partial t_F / \partial r$ at $r = R_+$ is first
attained. The divergence of $\partial t_F / \partial r$ indicates that
the constant $\bar t$ slice becomes tangent to the constant $r$
hyperbola at $r=R_+$. Beyond this the areal radial coordinate ceases
to be a good coordinate since the tangent condition implies that
continuing beyond this point increases $r$ again to values greater
than $R_+$.

Integrating the second equation in \eqref{BTZdiffappF} for points on
the maximal slice that are in the $t_F > 0$ part of Region F, we get
\begin{equation}\label{BTZ-time-appF}
  t_F(r, \bar t) - {t}_F(R_+(T),\bar{t}) = - T(\bar t) \int^r_{R_+(T)} \ud\rho\frac{A(\rho,\bar{t})^{1/2}}{\rho f(\rho)}\, ,
\end{equation}
where we have chosen the lower limit of integration to be $R_+(T)$
where the solution has a coordinate singularity. The above
diffeomorphism in Region F can be extended to Region I as well by
taking $r > \ell \sqrt{\frac{M}{2\pi}}$ in \eqref{BTZ-time-appF}; the
integration range will now have a simple pole at the horizon which has
to be dealt with the usual Cauchy principal-value prescription. After
this, it is easy to see that it satisfies the second equation in
\eqref{BTZdiffapp}. So, the diffeomorphism in Region I is
\begin{equation}\label{BTZ-time-app-full}
  t(r, \bar t) - {t}_F(R_+(T),\bar{t}) = - T(\bar t) \dashint^r_{R_+(T)}  \ud\rho\frac{A(\rho,\bar{t})^{1/2}}{\rho f(\rho)}\, .
\end{equation}
Note that the reference time $\hat{t}_F(R_+,\bar{t})$ is not yet
fixed. We fix this later after we obtain maximal slices in Region II
as well below.

The function $T(\bar{t})$ is implicitly determined by using the
boundary condition \eqref{ttbarbdryapp} on $t$ as $r \to \infty$:
\begin{equation}\label{bart-Sol-app}
   \bar t = {t}_F(R_+(T),\bar{t}) - T(\bar t) \dashint^\infty_{R_+(T)}  \ud\rho\frac{A(\rho,\bar{t})^{1/2}}{\rho f(\rho)}\, .
\end{equation}

We now describe the maximal slicing solution that is diffeomorphic to
a region of the fully extended BTZ black hole that includes only the
boundary in Region II, which we call the left boundary. Recall the BTZ
metric in Region II:
\begin{equation}\label{BTZmetappII}
  \ud s^2 = -f(\tl{r})^2 \ud \tl{t}^2 + f(\tl{r})^{-1} \ud \tl{r}^2 + \tl{r}^2 \ud\varphi^2\, ,\quad \tl{r}> \ell \sqrt{\frac{M}{2\pi}}\, .
\end{equation}
Note that $\tl{t}$ is chosen to increase in the opposite sense to the
that of the time $t$ in Region I. Thus, the $\bar{t} > 0$ maximal
slice actually starts from the boundary in the $t > 0$ part of Region
I, enters the $t_F > 0$ part of Region F at the future horizon
$U = 0$, $V > 0$, goes to the $t_F < 0$ part of Region F and exits
Region F at the past horizon $U > 0$, $V = 0$ and reaches the boundary
in Region II in the $\tl{t} < 0$ part.

To obtain maximal slices in the $\tl{t} < 0$ region, we use the left
areal chart maximal slicing solution \eqref{ds2-Sol-left-app}. The
diffeomorphism to the BTZ metric in Region II is obtained by solving
the equations
\begin{align}
  \partial_{\bar t}\tl{t} &= \epsilon\tl\alpha\sqrt{{A(\tl{r},\bar{t})}}\, ,\quad \partial_{\tl{r}} \tl{t} = -\epsilon\frac{\tl{T }\sqrt{{A}(\tl{r},\bar{t})}}{\tl{r} f(\tl{r})}\, ,
\end{align}
where the sign $\epsilon$ must be chosen to be $-1$ for the
$\tl{t} < 0$ part of Region II and $+1$ in the $\tl{t} > 0$ part of
Region II. The first equation with $\epsilon = -1$ gives the boundary
condition
\begin{equation}
  \lim_{\tl{r} \to \infty} \tl{t} = - \bar{t} - \tl{t}_0\, .
\end{equation}
Following the same procedure as earlier, the coordinate transformation
to BTZ coordinates in the $\tl{t} < 0$ part of Region II is
\begin{equation}\label{BTZ-time-left-app}
  \tl{t}(\tl{r},\bar{t}) - {t}_F(R_+(\tl{T}),\bar{t}) = \tl{T} \dashint^{\tl{r}}_{R_+(\tl{T})} \ud\rho\frac{A(\rho, \bar{t})^{1/2}}{\rho f(\rho)}\ ,
\end{equation}
where ${t}_F({R}_+(\tl{T}))$ is the $t_F$ coordinate of the point
$\tl{r} = R_+(\tl{T})$ in Region F where the metric
\eqref{ds2-Sol-left-app} has a coordinate singularity. The
$r\to\infty$ limit of \eqref{BTZ-time-left-app} gives the following
implicit expression for $\tl{T}(\bar{t})$:
\begin{align}
  - \bar t &= {t}_F(R_+(\tl{T}),\bar{t}) + \tl{T}(\bar t) \dashint^\infty_{R_+(\tl{T})}   \ud\rho\frac{A(\rho,\bar{t})^{1/2}}{\rho f(\rho)}\, . \label{bart-Sol-left-app}
\end{align}

We match the maximal slices from the left and right charts. Continuity
in the radial coordinate forces the reference points ${t}_F(R_+(T))$
and ${t}_F(R_+(\tl{T}))$ to be the same. This in turn requires the
functions $\tl{T}(\bar t)$ and $T(\bar t)$ to be the same, i.e.,
$\tl{T}(\bar t) = T(\bar t)$. Since these slices are tangent to the
same constant $r$ hyperbola, they are guaranteed to be smooth at
$r=R_+(T(\bar t))$. Adding \eqref{bart-Sol-app} and
\eqref{bart-Sol-left-app} we get
\begin{equation}
	{t}_F(R_+(T),\bar{t}) = 0\, .
\end{equation}
Taking the difference of \eqref{bart-Sol-app} and
\eqref{bart-Sol-left-app} gives
\begin{equation}\label{Tdef-sym-app}
  \bar t = -T \dashint^\infty_{R_+(T)} \ud\rho \frac{A(\rho,\bar{t})^{1/2}}{\rho f(\rho)}\, .
\end{equation}
This equation gives an implicit equation for $T$ in terms of
$\bar{t}$.

Thus, the two areal radial coordinate charts together cover a region
$R_0$ of the two-sided BTZ black hole which includes both AAdS
boundaries. The region $R_0$ is depicted in Figure
\ref{fig:WH1Regions}. It is demarcated by two constant $r$ hyperbolas,
one in Region F and one in Region P. These correspond to the
$\bar{t} \to \pm \infty$ spatial slices respectively.

One way to obtain the location of these \emph{limit slices} is as
follows. Recall that the maximum value of $|T|$ that is allowed by the
expression for $R_\pm$ in \eqref{rpmdefapp} is
\begin{equation}\label{Tdefapp1}
  -T_\infty \leq  T \leq T_\infty\, ,\quad T_\infty = \frac{\ell M}{4\pi}\, .
\end{equation}
At this value of $T$, the two roots $R_+$ and $R_-$ coincide and
become
\begin{equation}
  R_+(T = T_\infty) = R_-(T = T_\infty) = R_\infty \equiv \ell \sqrt{\frac{M}{4\pi}}\, .
\end{equation}
From the implicit equation for $T$ \eqref{Tdef-sym-app}, it is clear
that when $T = \pm T_\infty$, there is a pole in the integrand at
$\rho = R_\infty$ due to the $A(\rho,\bar{t})^{1/2}$ factor. This pole
in the integrand is non-removable since it occurs at the end of the
integration range. This leads to $\bar{t} \to \pm \infty$ when
$T = \pm T_\infty$. Thus, the limit slices $\bar{t} \to \pm \infty$
are located at
\begin{equation}
  r = R_\infty = \ell\sqrt{\frac{M}{4\pi}}\, .
\end{equation}
Since $R_\infty < R_h$, these limit slices are hyperbolas which
completely lie in Regions F and P. In Kruskal-Szekeres coordinates,
these limit slices are the two branches of the hyperbola
\begin{equation}
  UV = \eta^2\frac{R_h - R_\infty}{R_h + R_\infty} = \eta^2\frac{\sqrt{2} - 1}{\sqrt{2} + 1}\, .
\end{equation}
These are displayed in Figure \ref{fig:WH1Regions} as the orange
curves in Regions F and P.


\subsection{Diffeomorphism to Kruskal coordinates} \label{UVDiff}

Recall from Section \ref{BTZreview} that the Kruskal coordinates were
defined in terms of the Region I coordinates as
$U = -\eta^{-1}\e^{-\eta(t - r_*)}$ and $V = \eta \e^{\eta(t+ r_*)}$,
where $r_*$ is the tortoise coordinate
\begin{equation}
  r_* = \int_\infty^r\frac{\ell^2 \ud \rho}{\rho^2 - R_h^2} = \frac{1}{2\eta} \log \frac{r - R_h}{r + R_h}\, . 
\end{equation}
Using the expression for the diffeomorphism $t(r,\bar{t})$
\eqref{BTZ-time-app} for the BTZ time coordinate $t$ from the right
chart $(\bar{t}, r)$, we get the following diffeomorphism from the
right areal chart $(\bar{t}, r)$ in Region I to $(U,V)$:
\begin{multline}
  V(\bar{t}, r) = 
  \\ \eta^{-1} \e^{\eta \bar{t} - R_h \int^\infty_r \frac{\rho^2\ud\rho}{\left(\ell T + \sqrt{(\rho^2 - R_+^2)(\rho^2 - R_-^2)} \right) \sqrt{(\rho^2 - R_+^2)(\rho^2 - R_-^2)}}}\, ,
\end{multline}
\begin{multline}
  U(\bar{t}, r) = 
\\ -\eta^{-1}\e^{\eta\bar{t} - R_h \int^\infty_r \frac{\rho^2\ud\rho}{\left(\ell T - \sqrt{(\rho^2 - R_+^2)(\rho^2 - R_-^2)} \right) \sqrt{(\rho^2 - R_+^2)(\rho^2 - R_-^2)}}}\, .
\end{multline}
When $T > 0$, i.e., when $\bar{t} > 0$, the integral in the expression
for $U$ has a pole at $\rho = R_h$ which we can see by using
$R_+(T)^2 + R_-(T)^2 = R_h^2$ for all $T$ and
$R_+^2 R_-^2 = \ell^2 T^2$. When the lower limit $r$ of the integral
is set to $r = R_h$, the integral diverges, so that $U$ is driven to
zero at the future horizon. For $r < R_h$, the integral can be
evaluated using the principal value prescription. The additional
$\i \pi/\eta$ which comes from the contour integral around the pole
flips the sign of $U$ when $r < R_h$. The expression for $v$ is
regular at $r = R_h$ for $T > 0$, and hence it can be continued into
Region F without encountering divergences.

When $T < 0$, i.e., when $\bar{t} < 0$, the integrand for $v$ has a
pole at $\rho = R_h$ and the integral for $v$ diverges to $-\infty$
when $r = R_h$. This drives $V = \eta^{-1} \e^{\eta v}$ to zero at the
past horizon. To continue into Region P from Region I, the integral
for $r < R_h$ has to be evaluated with the principal value
prescription. The imaginary part $\i \pi /\eta$ from the contour
integral around the pole flips the sign of $V$ as one crosses the past
horizon.

Thus, the diffeomorphism for the Kruskal-Szekeres coordinates $U, V$
from the right areal radial coordinate chart $(\bar{t}, r)$ which
covers Region I fully including the right AAdS boundary, and half each
of Region F and Region P is
\begin{multline}
  V(\bar{t}, r) =
  \\ \pm \eta^{-1} \e^{\eta \bar{t} - R_h \dashint^\infty_r \frac{\rho^2\ud\rho}{(\ell T + \sqrt{(\rho^2 - R_+^2)(\rho^2 - R_-^2)})\sqrt{(\rho^2 - R_+^2)(\rho^2 - R_-^2)}}}\, ,
\end{multline}
\begin{multline}
  U(\bar{t}, r) =
\\              \pm\eta^{-1}\e^{\eta\bar{t} - R_h \dashint^\infty_r \frac{\rho^2\ud\rho}{(\ell T - \sqrt{(\rho^2 - R_+^2)(\rho^2 - R_-^2)})\sqrt{(\rho^2 - R_+^2)(\rho^2 - R_-^2)}}}\, ,
\end{multline}
where the signs for Region I are $(-,+)$, Region F are $(+,+)$, and
for Region P are $(+,-)$. A similar diffeomorphism can be written down
for the left areal radial coordinate chart $(\bar{t}, \tl{r})$ with
$\tl{r} > R_+(T)$.

A few constant $\bar{t}$ slices are plotted in Figure
\ref{fig:WH1F}.


\subsection{The wormhole coordinate}

While we have matched the left and right charts, the metric component
$g_{rr} = A(r)$ still diverges at the locus of matching points. We can
remedy this by defining a new coordinate by
\begin{equation}
	x^2 = r^2 - R_+(T(\bar t))^2\, .
\end{equation}
This coordinate transformation is valid everywhere except the final
limit slice where
$r = R_+(T_\infty) = R_\infty = \ell \sqrt{M / 4\pi}$. In this case,
the entire final slice lies at $x = 0$ and $x$ fails to be a good
spatial coordinate on this final slice.

In terms of these coordinates ($\bar{t},x,\varphi$), the metric is
given by
\begin{equation}\label{WormholeM-A}
	\ud s^2 = -N^2 \ud\bar t^2 + g_{xx} \left(\ud x + N^x \ud\bar t \right)^2 + \left(x^2 + R_+(\bar{t})^2\right) \ud \varphi^2\, ,
\end{equation}
with
\begin{subequations}
	\begin{align}
		N(\bar{t},x) &=
                               \frac{x \sqrt{x^2 + R_+^2 - R_-^2}}{\ell\sqrt{x^2 + R_+^2}} \times \nonumber\\
          &\quad\times \left(1 + \dot{T} \ell^3 \int_x^\infty \frac{\ud y}{y^2} \frac{\sqrt{y^2 + R_+^2}}{\left(y^2 +R_+^2 -R_-^2\right)^{3/2}} \right)\, ,\label{WHlapse-A} \\ 
		N^x(\bar{t},x) &= \frac{1}{x}\left(N(\bar{t},x) T +  R_+ \dot R_+ \right)\, , \label{WHshift-A} \\
		g_{xx}(\bar t, x) &= \frac{\ell^2}{x^2 + R_+^2 - R_-^2}\, . \label{WHgxx-A}
	\end{align}
\end{subequations}
All components of the above spacetime metric are smooth and regular
everywhere except at the limit slices $\bar{t} \to \pm \infty$ where
the entire slice lies at $x = 0$.

First, we can see that metric component $g_{xx}$ is regular everywhere
since the denominator is positive everywhere except the limit slice
where $R_+ = R_- = R_\infty$ and $x = 0$.

Similarly, the lapse and shift are regular and smooth everywhere. For
instance, the Taylor expansion near $x = 0$ is
\begin{align}
	\lim_{x\to 0^+} N &= \frac{\dot{T}\ell^2}{R_+^2 - R_-^2} + \dot{T}\ell^2 \frac{R_+^2 + R_-^2}{R_+^2(R_+^2 - R_-^2)} x^2 + \mathcal{O}(x^4)\, \label{Nxzero}\ .
\end{align}
Thus we have constructed a coordinate chart ($\bar{t},x,\varphi$) with
metric \eqref{WormholeM-A}, whose constant time slices are maximal, go
from the left asymptotic boundary to the right asymptotic boundary
while smoothly cutting across the two horizons.

The Kruskal-Szekeres coordinates in Region I are related to wormhole
coordinates by
\begin{align}
	&U(\bar{t},x) = \nonumber \\
	& -\eta^{-1} \e^{-\eta \bar{t} - R_h \dashint_x^\infty \frac{\ud y}{(y^2 - R_-^2)\sqrt{y^2 + R_+^2}} \left(\frac{\ell T}{\sqrt{y^2 + R_+(\bar{t})^2 - R_-(\bar{t})^2}} + y\right)} , \\
	&V(\bar{t},x) = \nonumber \\
	& \eta^{-1} \e^{\eta \bar{t} + R_h \dashint_x^\infty \frac{\ud y}{(y^2 - R_-^2)\sqrt{y^2 + R_+^2}} \left(\frac{\ell T}{\sqrt{y^2 + R_+(\bar{t})^2 - R_-(\bar{t})^2}} - y\right)} .
\end{align}
In Region F they are related by
\begin{align}
	&U(\bar{t},x) = \nonumber \\
	& \eta^{-1} \e^{-\eta \bar{t} - R_h \dashint_x^\infty \frac{\ud y}{(y^2 - R_-^2)\sqrt{y^2 + R_+^2}} \left(\frac{\ell T}{\sqrt{y^2 + R_+(\bar{t})^2 - R_-(\bar{t})^2}} + y\right)} , \\
	&V(\bar{t},x) = \nonumber \\
	& \eta^{-1} \e^{\eta \bar{t} + R_h \dashint_x^\infty \frac{\ud y}{(y^2 - R_-^2)\sqrt{y^2 + R_+^2}} \left(\frac{\ell T}{\sqrt{y^2 + R_+(\bar{t})^2 - R_-(\bar{t})^2}} - y\right)} .
\end{align}
Similarly one can write down the diffeomorphisms in regions II and P.

\subsection{A second family of maximal slices}\label{sec:2ndFam}

\begin{figure}[tbp!]
	\centering 
	\includegraphics[width=0.45\linewidth]{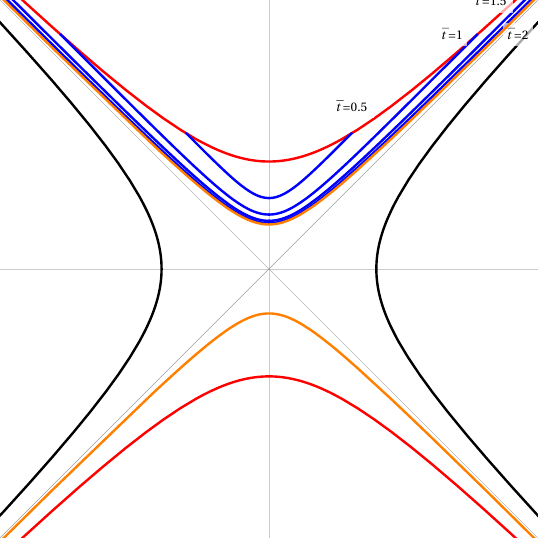}\hfill
	\includegraphics[width=0.45\linewidth]{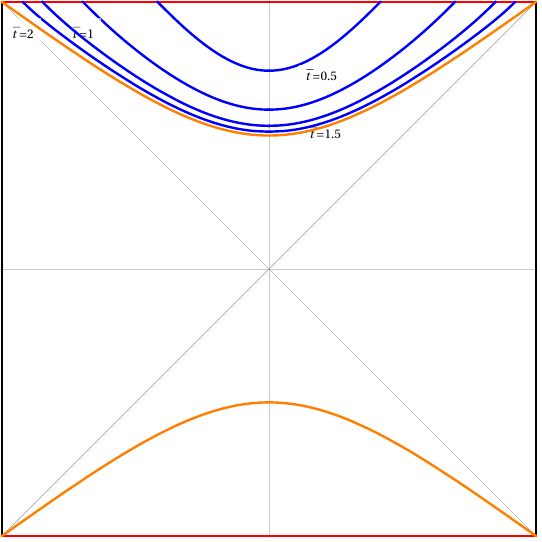}
	\caption{\label{fig:WH2F} Constant $\bar t$ slices of the second family are drawn in Kruskal-Szekeres diagram on the left and the Penrose diagram on the right. The red curves are the singularities at $r=0$ while the black curves are the asymptotic boundaries. The maximal slices are drawn in blue. These slices start at the singularity, grow outwards and end at the final slice at 		$R_\infty=\ell\sqrt{\frac{M}{4\pi}}$ (orange curve) as $\bar t\to\infty$, thus cover region $R_1$. Similarly one can cover the region $R_{-1}$ by a similar set of slices of the second family.}
\end{figure}

There exist another family of maximal slices that lie completely inside the horizon. These spatial slices are sausage-like: they are cylindrical and cap off at the two ends at $r = 0$. We define the globally defined `sausage' coordinate (analogous to the wormhole coordinate $x$ for the first family)
\begin{equation}
	z(r,\bar{t}) = \left\{\begin{array}{cc} \sqrt{R_-(\bar{t})^2 - r^2} &\quad \text{right chart} \\ - \sqrt{R_-(\bar{t})^2 - \tl{r}^2} &\quad \text{left chart} \end{array}\right. .
\end{equation}
Since $z \in [-R_-(T), R_-(T)]$, the range of the sausage coordinate $z$ is time dependent. The metric in these coordinates is given by
\begin{equation}\label{WormholeM-2F}
	\ud s^2 = -N^2 \ud \bar{t}^2 + g_{z z} \left(\ud z + N^{z} \ud \bar{t} \right)^2 + \left( R_-(\bar{t})^2 - z^2 \right) \ud \varphi^2\, ,
\end{equation}
with
\begin{subequations}
	\begin{align}
		N(\bar{t}, z) =& \frac{z \sqrt{z^2 + R_+^2 - R_-^2}}{\ell \sqrt{R_-^2 - z^2}} \Bigg(1+ \dot{T} \ell^3 \times \nonumber\\ 
		&\times \int_{ \sqrt{R_-^2 - z^2} }^\infty \frac{\ud\rho}{\rho} \left(\frac{\rho^2}{\ell^2} -\frac{M}{2\pi} +\frac{T(\bar{t})^2}{\rho^2}\right)^{-3/2} \Bigg)\, ,\label{WHlapse-2F} \\ 
		N^{z}(\bar{t}, z) =& - \frac{1}{z} \left(N(\bar{t}, z) T + R_- \dot R_- \right)\, , \label{WHshift-2F} \\
		g_{z z}(\bar{t}, z) =& \frac{\ell^2}{z^2 + R_+^2 - R_-^2}\, . \label{WHgxx-2F}
	\end{align}
\end{subequations}
The coordinate transformation to the BTZ coordinates
$(t_F, r, \varphi)$ in Region F is given by
\begin{subequations}
	\begin{align}
		r^2 &= R_-(T(\bar t))^2 - z^2\, , \label{r-x-rel-2F} \\
		t_F &= -T(\bar t) \int^{ \sqrt{R_-^2 - z^2} }_{R_-(T)} \frac{\ud\rho}{\rho \left(\frac{\rho^2}{\ell^2} -\frac{M}{2\pi} \right) \sqrt{\frac{\rho^2}{\ell^2} -\frac{M}{2\pi} +\frac{T^2}{\rho^2}}} \, . 
		\label{t-x-rel-2F}
	\end{align}
\end{subequations}
These second family of slices are plotted in Figure \ref{fig:WH2F}. They start from $r=0$ and cover the interior of the black hole up to a final limit slice at constant radial value $R_\infty=\ell\sqrt{M/4\pi}$, i.e., the region $R_1$ in Figure \ref{fig:WH1Regions}. The first family of maximal slices reached the same final slice but from the exterior (see Figure \ref{fig:WH1F}).  Similarly one can cover the region $R_{-1}$ by a similar set of slices of the second family. Together the two families foliate the entire black hole spacetime.


\section{Probe scalar field is gauge invariant}\label{pertgaugefix}

Since we are interested in treating the scalar field as a perturbation around a pure gravity solution of the Einstein's equations, we would like to establish its gauge invariance in this situation. This analysis is done in arbitrary spacetime dimension $d+1$.

Recall that the gravitational Hamiltonian is given by (here $\kappa^2=16\pi G_N$ is the gravitational constant)
\begin{equation}\label{grham}
	H[N, N_i] = \int_\Sigma \ud^d x\,(N \mc{H}_\perp + N_i \mc{H}^i) + H_\partial\, ,
\end{equation}
where $\mc{H}_\perp$ and $\mc{H}^i$ are respectively the Hamiltonian and momentum constraints
\begin{align}\label{grconst}
	\mc{H}_\perp
	&= \frac{\kappa^2}{\sqrt{g}}\Big(\pi^{ij} \pi_{ij} - \frac{1}{d-1}\pi^2\Big) - \frac{1}{\kappa^2}\sqrt{g}(R - 2\Lambda) + \rho\, ,\nonumber\\
	\mc{H}^i &= -2 D_j \pi^{ij} + S^i\, , \nonumber \\
	H_\partial &= - \frac{2}{\kappa^2}\, N\sqrt{\sigma} k + 2 {r}_i N_j \pi^{ij} - \mc{B}_{\rm ct}\, .
\end{align}
Here, $\pi = g_{ij}\pi^{ij}$ is the trace of $\pi^{ij}$ and the
quantities $\rho$ and $S^i$ are the scalar field energy density and
momentum density respectively,
\begin{align}
	\rho &= \frac{\pi_\phi^2}{2\sqrt{g}} +  \sqrt{g}\big(\tfrac{1}{2}g^{ij}\partial_i\phi \partial_j\phi + V(\phi)\big)\ , \nonumber\\
	S^i &= g^{ij} \pi_\phi \partial_j\phi\ ,
\end{align}
with $V(\phi)$ being the scalar field potential.

Note that the scalar field appears in the Hamiltonian at
$\mc{O}(\kappa^0)$ whereas the gravity part starts at
$\mc{O}(\kappa^{-2})$. Since we want all terms corresponding to the
pure gravity background to contribute at the same order of $\kappa$,
we write down perturbations as
\begin{align}\label{pertexpr}
	&g_{ij} = \hat{g}_{ij} + \kappa h_{ij}\, ,\quad \pi^{ij} = \frac{1}{\kappa^2}(\hat{P}^{ij} + \kappa p^{ij})\, ,\nonumber\\
	&\quad N_i = \hat{N}_i + \kappa \beta_i\, ,\quad N = \hat{N}(1 + \kappa \alpha)\, ,
\end{align}
where the hatted quantities correspond to the pure gravity background
solution and the perturbations are $h_{ij}$, $p^{ij}$, $\beta_i$ and
$\alpha$. The scalar field is zero in the background and hence is
purely a perturbation.

The Hamiltonian dynamics can now be analyzed order by order in
$\kappa$. First, we describe the dynamics of the pure gravity
background by solving the constraints and gauge conditions for the
background and analyzing the Hamiltonian dynamics on the reduced phase
space involving only the gravitational variables. This was discussed
in detail in the companion paper \cite{KPWI}. Skipping over the
details of the calculation, we find that the dynamics of the
fluctuations are then governed by the quadratic Hamiltonian
\begin{align}\label{H2}
	H_2 &= \int_\Sigma \ud^dx \Big(\alpha \mc{C}_\perp(h_{ij},p^{ij}) + \beta_i \mc{C}^i(h_{ij},p^{ij})\nonumber\\
	&\quad \ + \mc{H}_{\rm grav}(h_{ij},p^{ij}) + \mc{H}_{\rm s}(\phi,\pi_\phi)\Big) + \text{boundary terms}\, ,
\end{align}
where $\mc{C}_\perp$ and $\mc{C}^i$ are the linearized Hamiltonian and
momentum constraints that depend (in addition to the pure gravity
background) only on the gravity fluctuations,
\begin{align}\label{linconst}
	\mc{C}_\perp =& \hat{N}\big(2 \hat{P}^{ij} p_{ij} + 2 h_{ik} \hat{P}^k{}_j \hat{P}^{ij} - h\hat{P}^{ij} \hat{P}_{ij} \nonumber\\
	&- \hat{g}(\hat{D}^a\hat{D}^b h_{ab} - \hat{D}^2 h - h^{ac}\hat{R}_{ac})\big)\, ,\nonumber\\
	\mc{C}^l =& 2 \hat{D}_i p^{il} + \hat{P}^{ij} (\hat{D}_i h^l{}_j + \hat{D}_j h^l{}_i - \hat{D}^l h_{ij})\, .
\end{align}
$\mc{H}_{\rm grav}(h_{ij},p^{ij})$ and $\mc{H}_{\rm s}(\phi,\pi_\phi)$ are the dynamical, non-constraint, quadratic Hamiltonians for the metric and scalar degrees of freedom respectively,
\begin{align}\label{H2pert}
	&\mc{H}_{\rm grav}(h_{ij}, p^{ij}) \nonumber\\
	=&  \frac{\hat{N}}{\sqrt{\hat{g}}}\Big\{-\tfrac{1}{2} h\big(2 \hat{P}^{ij} p_{ij} + 2 h_{ik} \hat{P}^k{}_j \hat{P}^{ij}\big) \nonumber \\
	&+ \tfrac{1}{4}(h^{ij} h_{ij} + \tfrac{1}{2} h^2)\hat{P}^{ij} \hat{P}_{ij}  \nonumber\\
	& + p^{ij} p_{ij} + 2 p^{ij} h_{ik} \hat{P}^k{}_j + \hat{P}^{ij} \hat{P}^{kl} h_{ik} h_{jl} + 2 \hat{P}^{ij} h_{ik} p^k{}_j\Big\} \nonumber\\
	&- \hat{N} \sqrt{\hat{g}} \Big\{\tfrac{1}{4}(\tfrac{1}{2}h^2 - h_{ij} h^{ij})(\hat{R} - 2\Lambda)\nonumber\\
	& + \tfrac{1}{2}h(\hat{D}^a\hat{D}^b h_{ab} - \hat{D}^2 h - h^{ac}\hat{R}_{ac})\nonumber\\
	&+ \tfrac{3}{4}\hat{D}^a h^{bc} \hat{D}_a h_{bc} +  h^{ab} \hat{D}^2 h_{ab} -  \hat{D}_b h^{bc} \hat{D}^a h_{ac} -  h^{bc} \hat{D}_b \hat{D}^a h_{ac}\nonumber\\
	&+  h^{ab} \hat{D}_a \hat{D}_b h +  \hat{D}_a h^{ac} \hat{D}_c h - \tfrac{1}{2} \hat{D}_b h^{ac} \hat{D}_a h^b{}_c - \tfrac{1}{4} \hat{D}^a h \hat{D}_a h \nonumber\\
	& - h^{ac} \hat{D}_b \hat{D}_a h^b{}_c  +  h^{ae} h_{e}{}^c \hat{R}_{ac}\Big\} \nonumber\\
	& - (\hat{N}^l p^{ij} - \hat{N}_k h^{kl}\hat{P}^{ij})C_{ijl} \, ,
\end{align}
\begin{align}
	\mc{H}_{\rm s}(\phi,\pi_\phi) =& \hat{N} \Big(\tfrac{1}{2\sqrt{\hat{g}}}\pi_\phi^2 + \sqrt{\hat{g}}(\tfrac{1}{2} \hat{g}^{ij} \partial_i\phi\partial_j\phi + V(\phi))\Big) \nonumber \\
	&- \hat{N}_i \hat{g}^{ij} \pi_\phi \partial_j\phi \, .
\end{align}  

The important feature of the quadratic Hamiltonian $H_2$ \eqref{H2} is that the gravitational perturbations $h_{ij}, p^{ij}$ and scalar fluctuations $\phi, \pi_\phi$ do not mix. In $d=2$, since there are no local fluctuating gravitational degrees of freedom, the quadratic Hamiltonian $H_2$ \eqref{H2} involves only the scalar perturbations. Moreover, the linearized constraints do not contain the scalar field perturbations and hence do not act on the scalar field. Thus, if one has succeeded in completely fixing the diffeomorphism freedom on the background pure gravity solution, then \emph{the scalar field is completely gauge invariant to this order in perturbation theory.}

\section{Solutions of the Klein-Gordon equation in the BTZ spacetime}\label{BTZKGsol}

\subsection{Separation of variables}

The $2+1$ components of the BTZ metric \eqref{BTZmet} are
\begin{equation}
	N(r) = \sqrt{f(r)}\, ,\ N_i = 0\, ,\ g_{rr} = \frac{1}{f(r)}\, ,\ g_{\varphi\varphi} = r^2\, ,\ g_{r\varphi} = 0\, ,
\end{equation}
which gives $\sqrt{g} = r f(r)^{-1/2}$. The wave equation takes the
form
\begin{equation}
	-f^{-1} \ddot\phi + r^{-1}\partial_r (r f \partial_r \phi) +  r^{-1}\partial_\varphi (r^{-1}\partial_\varphi\phi) - m^2 \phi = 0\ .
\end{equation} 
Since $\partial_t$ and $\partial_\varphi$ are Killing vectors, we can
separate variables and write a particular mode of the scalar field as
\begin{equation}\label{capFdef}
	F_{\omega q}(t,r,\varphi) \equiv  \e^{-\i\omega t} \e^{-\i q \varphi}\, f_{\omega q}(r)\, ,
\end{equation}
where $\omega$, $q$ are the frequency and angular momentum along the
circle respectively.
Changing variables to the dimensionless $\rho = r^2 / R_h^2$ and
defining\begin{align}
	f_{\omega q}(\rho) =& (\rho - 1)^\alpha \rho^\gamma \chi(\rho)\, , \\
	\text{with} \quad \alpha =& -\i\frac{\omega }{2 \eta}\, ,\quad \gamma = \i\frac{q \ell}{2 R_h} \, ,
\end{align}
we see that $\chi$ satisfies the hypergeometric equation:
\begin{align}\label{chieqn}
	&\rho (1-\rho) \chi''(\rho) + (c - (a+b+1)\rho)\chi'(\rho) - ab \chi(\rho) = 0\, ,
\end{align}
where we have defined
\begin{align}\label{abcdefapp}
	a =& \alpha + \gamma + \tfrac{1}{2}\Delta_+\, , \quad b = \alpha + \gamma + \tfrac{1}{2}\Delta_-\, , \nonumber \\
	c =& 2\gamma +1\, ,\qquad \Delta_\pm = 1 \pm \sqrt{1 + \ell^2 m^2}\, .
\end{align}
Note that $\Delta_+$ is the conformal dimension of the operator dual
to $\phi$ in the dual CFT. The hypergeometric equation has regular
singular points at $\rho = 0$, $\rho=1$ and $\rho = \infty$,
corresponding to the origin (or inner horizon), the horizon (or the
outer horizon), and infinity. There are solutions readily available
(as a convergent series via the Frobenius method) in the neighbourhood
of each of these points. This greatly facilitates further
analysis. For instance, in the neighbourhood of $z = 0$,
\begin{equation}
	{}_2F_1(a_1,a_2;b_1;z) = \sum_{n = 0}^\infty \frac{(a_1)_n (a_2)_n}{(b_1)_n} \frac{z^n}{n!}\, ,
\end{equation}
with
\begin{equation}
	(a)_n = a (a+1)\cdots (a+n-1)\, ,
\end{equation}
so that ${}_2F_1(a_1,a_2;b_1;0) = 1$. 
Below, we list the solutions $\chi(\rho)$ that are regular in the
neighbourhood of the three singular points $\rho = \infty, 1, 0$. We
use the notation of DLMF \cite[\href{http://dlmf.nist.gov/15.10}{\S
	15.10}]{NIST:DLMF}. The solutions regular at $\rho=1$ are
\begin{align}\label{w34}
	w_3(\rho) &= {}_2F_1\left(a,b;a+b-c+1;1-\rho\right)\, ,\nonumber\\
	w_4(\rho) &= (1-\rho)^{c-a-b} {}_2F_1\left(c-a,c-b;c-a-b+1;1-\rho\right) ,
\end{align}
the solutions regular at $\rho = \infty$ are
\begin{align}\label{w56}
	w_5(\rho) &=  \e^{\i \pi a}  \rho^{-a}\ {}_2F_1\left( a, a-c+1; a-b+1; \frac{1}{\rho}\right) ,\nonumber\\
	w_6(\rho) &= \e^{\i\pi b}  \rho^{-b}\ {}_2F_1\left(b,b-c+1; b-a+1; \frac{1}{\rho}\right) ,
\end{align}
and the solutions regular at $\rho = 0$ are
\begin{align}\label{w12}
	w_1(\rho) &= {}_2F_1(a,b;c;\rho)\, ,\nonumber\\
	w_2(\rho) &= \rho^{1-c} {}_2F_1\left(a-c+1,b-c+1;2-c;\rho\right)\, .
\end{align}

\subsection{Solution in Region I}

Let us look at the solution in the exterior region I. Recalling that
$f_{\omega q}(\rho) = (\rho-1)^\alpha \rho^\gamma \chi(\rho)$, the
general solution in the neighbourhood of $\rho = \infty$ is a linear
combination of $(\rho-1)^\alpha \rho^\gamma w_5$ and
$(\rho-1)^\alpha \rho^\gamma w_6$ in \eqref{w56}:
\begin{align}\label{gensolnI}
	& f_{\omega q}(\rho) \nonumber \\
	&= C_{\omega q}\, \e^{-\i\pi a} (\rho-1)^\alpha \rho^\gamma w_5(\rho) + C'_{\omega q}\,\e^{-\i\pi b} (\rho-1)^\alpha \rho^\gamma w_6(\rho) \nonumber\\
	&= C_{\omega q} (\rho-1)^\alpha \rho^{\gamma-a}\ {}_2F_1\left(a,a-c+1;a-b+1;\frac{1}{\rho}\right)\nonumber\\
	&\ \ + C'_{\omega q}\, (\rho-1)^\alpha \rho^{\gamma-b}\ {}_2F_1\left(b,b-c+1;b-a+1;\frac{1}{\rho}\right) ,
\end{align}
for some constants $C_{\omega q}$ and $C'_{\omega q}$.
We find that the normalizable component of
$f_{\omega q}(\rho)$ w.r.t.~the Klein-Gordon inner product is
proportional to $w_5$ and is given by
\begin{align}\label{solnIinf}
	f_{\omega q}(\rho) &= C_{\omega q} (\rho-1)^\alpha \rho^{\gamma-a} {}_2F_1\left(a,a-c+1; a-b+1; \frac{1}{\rho}\right).
\end{align}
The constant $C_{\omega q}$ can be fixed by demanding that
$F_{\omega q} =  \e^{-\i\omega t} \e^{-\i
	q\varphi} f_{\omega q}$ satisfy the canonical Klein-Gordon inner
product
\begin{equation}\label{FKGip}
	(F_{\omega'q'}, F_{\omega q})_{\rm KG} = 4\pi\omega \delta(\omega-\omega') \delta_{qq'}\, .
\end{equation}
We do this in Appendix \ref{KGnormapp} below and obtain
\begin{equation}
	C_{\omega q} = \frac{1}{N_{\omega q}\sqrt{2\pi R_h}}\, ,\ N_{\omega q} = \left|\frac{\Gamma(a-b+1) \, \Gamma\left(c-a-b\right)}{\Gamma(1-b) \Gamma(c-b)}\right| .
\end{equation}
The mode expansion for the scalar field in Region I reads
\begin{equation}\label{phiBTZIapp}
	\phi(t,r,\varphi) = \sum_{q\in \mathbb{Z}} \int_0^\infty \frac{\ud\omega}{\sqrt{4\pi\omega}}\, a_{\omega q} F_{\omega q}(t,r,\varphi) + \text{c.c.}\ .
\end{equation}
We then have
\begin{equation}
	a_{\omega q} = \frac{1}{\sqrt{4\pi\omega}}(F_{\omega q}, \phi)_{\rm KG}\, .
\end{equation}
Using that $F_{\omega q}$ satisfy the canonical Klein-Gordon inner
product \eqref{FKGip}, we get the commutation relations
\begin{equation}
	[a_{\omega q}, a^\dag_{\omega'q'}] = \delta(\omega-\omega')\delta_{qq'} ,\ \ \ [a_{\omega q}, a_{\omega'q'}] =   [a^\dag_{\omega q}, a^\dag_{\omega'q'}] = 0\, .
\end{equation}
\paragraph{Solutions regular at the horizon.} The mode functions
\eqref{solnIinf} have a good Taylor series near $\rho = \infty$. To
obtain an expression near the horizon for the mode function in Region
I, we have to first use a connection formula to write $w_5(\rho)$ in
terms of hypergeometric functions that have a good expansion near
$\rho = 1$:
\begin{align}\label{w534}
	w_5(\rho) =& \e^{\i \pi a} \frac{\Gamma(a-b+1)\Gamma(c-a-b)}{\Gamma(1-b)\Gamma(c-b)} w_3(\rho) \nonumber \\
	&+ \e^{\i \pi (c-b)} \frac{\Gamma(a-b+1)\Gamma(a+b-c)}{\Gamma(a) \Gamma(a-c+1)}w_4(\rho)\, .
\end{align}
Plugging in the values of $a,b,c$ \eqref{abcdefapp} in the coefficients,
we see that, when $\omega$ and $q$ are real,
\begin{align}
	&\frac{\Gamma(a-b+1)\Gamma(c-a-b)}{\Gamma(1-b)\Gamma(c-b)} = \left(\frac{\Gamma(a-b+1)\Gamma(a+b-c)}{\Gamma(a) \Gamma(a-c+1)}\right)^* \nonumber \\ 
	&= \frac{\Gamma(\Delta_+)\,\Gamma\left( \i \omega / \eta\right)}{\Gamma\left(\frac{1}{2} (\Delta_+ + \frac{\i \omega}{\eta} - \frac{\i q \ell}{R_h})\right) \Gamma\left(\frac{1}{2}(\Delta_+ + \frac{\i \omega}{\eta} +\frac{\i q\ell}{R_h})\right)}\, .
\end{align}
Let
\begin{equation}\label{comegaqdef}
	\frac{\Gamma(a-b+1)\Gamma(c-a-b)}{\Gamma(1-b)\Gamma(c-b)} = N_{\omega q} \e^{\i\delta_{\omega q}}\ ,
\end{equation}
with $N_{\omega q}$ the magnitude and $\delta_{\omega q}$ the phase of
the left-hand side. We have already encountered $N_{\omega q}$ in the
normalization of the mode function \eqref{solnIinf}. We can then write
\begin{align}\label{solnIhor}
	f_{\omega q} =& \frac{\e^{-\i\pi a}}{N_{\omega q} \sqrt{2\pi R_h}} (\rho-1)^\alpha \rho^\gamma\, w_5(\rho)\ ,\nonumber\\
	=& \frac{1}{\sqrt{2\pi R_h}} \Big( \e^{\i\delta_{\omega q}} (\rho-1)^\alpha \rho^\gamma w_3(\rho)  \nonumber \\
	&+ \e^{\i \pi (c-b-a)} \e^{-\i\delta_{\omega q}} (\rho-1)^\alpha \rho^\gamma w_4(\rho)\Big)\, ,
\end{align}
which can be written more explicitly using \eqref{w34} as \footnote{To obtain this expression, we have absorbed the phase $\e^{-\i\pi 2\alpha}$ into
	$(1-\rho)^{-2\alpha}$ to obtain $(\rho-1)^{-2\alpha}$:
	\begin{equation*}
		\e^{-\i\pi 2\alpha} (1-\rho)^{-2\alpha}  = \e^{-2\alpha \log \big((1-\rho) \e^{-\i\pi}\big)} = (\rho-1)^{-2\alpha}\, .
	\end{equation*}
}
\begin{align}\label{Foriginalapp}
	& f_{\omega q} = \frac{\rho^\gamma}{\sqrt{2\pi R_h}} \Big( \e^{\i\delta_{\omega q}} (\rho-1)^\alpha {}_2F_1(a,b;a+b-c+1;1-\rho)\nonumber\\
	&\ + \e^{-\i\delta_{\omega q}} (\rho-1)^{-\alpha} {}_2F_1(c-a,c-b;c-a-b+1;1-\rho)\Big) .
\end{align}
It will be useful to write the full mode function $F_{\omega q}$ which
also includes the $t$ and $\varphi$ dependence succinctly as
\begin{equation}\label{reghorI}
	F_{\omega q} = \e^{-\i\omega t - \i q\varphi} f_{\omega q} = \frac{1}{\sqrt{2\pi R_h}} (F^1_{\omega q} + F^2_{\omega q})\, ,
\end{equation}
with
\begin{align}\label{F12def}
	F^1_{\omega q} &= \e^{-\i\omega t - \i q\varphi} \e^{\i\delta_{\omega q}} (\rho-1)^\alpha \rho^\gamma\times \nonumber\\
	&\quad\quad \times {}_2F_1(a,b;a+b-c+1;1-\rho)\, ,\nonumber\\
	F^2_{\omega q} &= \e^{-\i\omega t - \i q\varphi} \e^{-\i\delta_{\omega q}} (\rho-1)^{-\alpha} \rho^\gamma \times \nonumber\\
	&\quad\quad \times {}_2F_1(c-a,c-b;c-a-b+1;1-\rho)\, ,
\end{align}
Near the horizon $\rho = 1$ in Region I, we have
$\rho - 1 \sim 4 \e^{2\eta r_*}$ (with $r_*$ the tortoise coordinate
in Region I) so that the above functions behave as
\begin{align}\label{nearhorI}
	F^1_{\omega q} &\sim \e^{-\i q\varphi} \e^{\i\delta_{\omega q}} (2\eta V)^{-\i\omega /\eta}\, , \nonumber \\
	F^2_{\omega q} &\sim \e^{-\i q\varphi} \e^{-\i\delta_{\omega q}} (-2\eta U)^{\i\omega /\eta}\, .
\end{align}
Thus, the scalar field behaves as
\begin{align}\label{scalarnearhorI}
	\phi(U,V,\varphi) \approx & \frac{1}{\sqrt{2\pi R_h}}\sum_{q \in \mathbb{Z}} \int_0^\infty \frac{\ud \omega}{\sqrt{4\pi\omega}}\,a_{\omega q} \e^{-\i q\varphi} \times \nonumber \\
	&(\e^{\i\delta_{\omega q}} (2\eta V)^{-\i\omega / \eta} + \e^{-\i \delta_{\omega q}} (-2\eta U)^{\i\omega /\eta}) + \text{c.c.}\, .
\end{align}

\subsection{Solution in region II}

We write the ansatz for a mode of the scalar field in Region II as
\begin{equation}
	\tl{F}_{\omega q}(\tl{t},\tl{r},\varphi) = \e^{\i\omega \tl{t}} \e^{\i q \varphi}\, \tl{f}_{\omega q}(\tl{r})\ .
\end{equation}
The solution for $\tilde{f}_{\omega q}$ that is normalizable has the
same functional form as \eqref{solnIinf} above, but with the choices
\begin{align}\label{abcdefII}
	&\tl\alpha = \alpha^* = \frac{\i\omega}{2\eta}\, ,\ \tl\gamma = \gamma^* = -\frac{\i q \ell}{2 R_h}\, ,\ \Delta_\pm = 1 \pm \sqrt{1 + \ell^2 m^2}\, , \nonumber \\
	&\tl{c} = 2\tl\gamma +1\, ,\quad\,\ \tl{a} = \tl\alpha + \tl\gamma + \tfrac{1}{2}\Delta_+\, , \,\ \tl{b} = \tl\alpha + \tl\gamma + \tfrac{1}{2}\Delta_-\, .
\end{align}
The mode function in region II is then
\begin{align}
	\tl{f}_{\omega q}(\tl\rho) 
	&= {C}_{\omega q}\, (\tl\rho-1)^{\tl\alpha} \tl\rho^{\tl\gamma -\tl{a}}\ {}_2F_1\left(\tl{a},\tl{a}-\tl{c}+1; \tl{a}-\tl{b}+1; \frac{1}{\tl\rho}\right) .
\end{align}
$\tl\alpha$ and $\tl\gamma$ are chosen as above so that
$\tl{f}_{\omega q}^*$ is exactly the same function of $\rho$ as
$f_{\omega q}$ in Region I \eqref{solnIinf}. This choice is useful
when we define the Hartle-Hawking modes.

The scalar field mode expansion in Region II is then
\begin{equation}\label{phiBTZII}
	\phi(\tl{t},\tl{r},\varphi) = \sum_{q\in \mathbb{Z}} \int_0^\infty \frac{\ud\omega}{\sqrt{4\pi\omega}}\, \tl{a}_{\omega q} \tl{F}_{\omega q}(\tl{t},\tl{r},\varphi) + \text{c.c.}\, ,
\end{equation}
with the canonical commutation relations
\begin{equation}
	[\tl{a}_{\omega q}, \tl{a}^\dag_{\omega'q'}] = \delta(\omega-\omega')\, ,\quad [\tl{a}_{\omega q}, \tl{a}_{\omega'q'}] = [\tl{a}^\dag_{\omega q}, \tl{a}^\dag_{\omega'q'}] = 0\, .
\end{equation}
We note the following property of $\tl{F}_{\omega q}$ under complex
conjugation:
\begin{equation}\label{Ftlconj}
	\tl{F}^*_{-\omega,-q} = \tl{F}_{\omega q} = F_{-\omega,-q}\ .
\end{equation}
We also record mode functions that are regular near the horizon
$\tl\rho = 1$. From the relations \eqref{Ftlconj}, it follows that
\begin{equation}\label{reghorII}
	\tl{F}_{\omega q} = \frac{1}{\sqrt{2\pi R_h}} (F^{1*}_{\omega q} + F^{2*}_{\omega q})\, .
\end{equation}
The near-horizon $\tl\rho \to 1$ behaviour of the scalar field is then
\begin{align}\label{scalarnearhorII}
	\phi(U,V,\varphi) \approx & \frac{1}{\sqrt{2\pi R_h}}\sum_{q \in \mathbb{Z}} \int_0^\infty \frac{\ud \omega}{\sqrt{4\pi\omega}}\,\tl{a}_{\omega q} \e^{\i q\varphi} \times \nonumber \\
	&(\e^{-\i\delta_{\omega q}} (-2\eta V)^{\i\omega / \eta} + \e^{\i \delta_{\omega q}} (2\eta U)^{-\i\omega /\eta}) + \text{c.c.}\, .
\end{align}

\subsection{Solution in Region F}
The ansatz in Region F is
\begin{equation}\label{scalardefF}
	\e^{-\i\omega t_F} \e^{-\i q \varphi}\, f^F_{\omega q}(r)\, .
\end{equation}
As usual, we write $f^F(r) = (1 - \rho)^\alpha \rho^\gamma \chi(\rho)$
so that $\chi$ satisfies the hypergeometric equation
\eqref{chieqn}. There is no normalizability condition imposed on the
solution in Region F. Hence, both linearly independent solutions to
the Klein-Gordon equation are admitted. We write them in terms of
hypergeometric functions $w_3$, $w_4$ \eqref{w34} which are defined in
a neighbourhood of the horizon $\rho = 1$.
The mode expansion of the scalar field is then
\begin{align}
	\phi(t_F,r,\varphi) =& \frac{1}{\sqrt{2\pi R_h}}\sum_{q \in \mathbb{Z}} \int_0^\infty \frac{\ud\omega}{\sqrt{4\pi\omega}} \Big(C_{\omega q}^{\rm F} G^1(t_F,r,\varphi) \nonumber \\
	&+ C_{\omega q}^{\rm F}{}' G^2(t_F,r,\varphi) + \text{c.c.}\Big) \, ,
\end{align}
where $C^{\rm F}_{\omega q}$ and $C_{\omega q}^{\rm F}{}'$ are some
constants, and $G^1_{\omega q}$ and $G^2_{\omega q}$ are defined as
\begin{align}\label{G12def}
	G^1_{\omega q} 
	=& \e^{\i\delta_{\omega q}}\,\e^{-\i \omega t_F -\i q\varphi}  (1-\rho)^{\alpha} \rho^{\gamma} \times\, \nonumber \\
	& \times {}_2F_1\left(a,b;a+b-c+1;1-\rho\right)\, ,\nonumber\\
	G^2_{\omega q} 
	=& \e^{-\i\delta_{\omega q}}\,\e^{-\i \omega t_F-\i q\varphi} (1-\rho)^{-\alpha} \rho^{\gamma} \times \nonumber \\
	& \times {}_2F_1\left(c-a,c-b;c-a-b+1;1-\rho\right)\, .
\end{align}
We fix the constants $C^{\rm F}_{\omega q}$ and
$C_{\omega q}^{\rm F}{}'$ by matching with the mode expansions in
Regions I and II at the horizons.

The near horizon behaviour of the $G^i$ can be obtained by writing
$1 - \rho \sim 4 \e^{2\eta r_{*F}}$. Then,
\begin{align}\label{nearhorF}
	G^1_{\omega q} &\sim \e^{-\i q\varphi} \e^{\i\delta_{\omega q}} (2\eta V)^{-\i\omega /\eta}\, ,\nonumber \\
	G^2_{\omega q} &\sim \e^{-\i q\varphi} \e^{-\i\delta_{\omega q}} (2\eta U)^{\i\omega /\eta}\, ,
\end{align}
which implies

\begin{align}\label{scalarnearhorF}
	&\phi(U,V,\varphi) \sim \frac{1}{\sqrt{2\pi R_h}}\sum_{q \in \mathbb{Z}} \int_0^\infty \frac{\ud\omega}{\sqrt{4\pi\omega}} \times \nonumber \\&\times 
	\Big[ \e^{-\i q \varphi}\big(C_{\omega q}^{\rm F} \e^{\i\delta_{\omega q}} (2\eta V)^{-\i\omega/\eta}  + C_{\omega q}^{\rm F}{}' \e^{-\i\delta_{\omega q}} (2\eta U)^{\i\omega/\eta}  \big) \nonumber \\
	&\ + \e^{\i q \varphi}\big(C_{\omega q}^{\rm F *} \e^{-\i\delta_{\omega q}} (2\eta V)^{\i\omega/\eta}  + C_{\omega q}^{\rm F'*} \e^{\i\delta_{\omega q}} (2\eta U)^{-\i\omega/\eta} \big)\Big]\, .
\end{align}
This expression should match the expression from Region I at the
future horizon $U = 0$ and that from Region II at the past horizon
$V = 0$ so that the scalar field is continuous across the horizons
\cite{Papadodimas:2012aq}. As $U \to 0$, the $U$ dependent part
oscillates progressively faster and can be set to zero in all
expressions (similarly for the $V \to 0$ case). The near-horizon
expression for the scalar field in Region I is
\begin{align}\label{scalarnearhorI1}
	&\phi(t,r,\varphi) \approx \frac{1}{\sqrt{2\pi R_h}}\sum_{q \in \mathbb{Z}} \int_0^\infty \frac{\ud \omega}{\sqrt{4\pi\omega}} \times \nonumber \\&\times 
	\Big[ a_{\omega q} \e^{-\i q\varphi} \big(\e^{\i\delta_{\omega q}} (2\eta V)^{-\i\omega / \eta} + \e^{-\i \delta_{\omega q}} (-2\eta U)^{\i\omega /\eta}\big)\nonumber\\
	&+ a^\dag_{\omega q} \e^{\i q\varphi} \big(\e^{-\i\delta_{\omega q}} (2\eta V)^{\i\omega / \eta} + \e^{\i \delta_{\omega q}} (-2\eta U)^{-\i\omega /\eta}\big)\Big]\, .
\end{align}
Similarly, the near-horizon expression for the scalar field in Region
II is
\begin{align}\label{scalarnearhorII1}
	&\phi(\tl{t},\tl{r},\varphi) \approx \frac{1}{\sqrt{2\pi R_h}}\sum_{q \in \mathbb{Z}} \int_0^\infty \frac{\ud \omega}{\sqrt{4\pi\omega}} \times \nonumber \\&\times 
	\Big[\tl{a}_{\omega q} \e^{\i q\varphi} \big(\e^{-\i\delta_{\omega q}} (-2\eta V)^{\i\omega / \eta} + \e^{\i \delta_{\omega q}} (2\eta U)^{-\i\omega /\eta}\big) \nonumber\\
	& + \tl{a}^\dag_{\omega q} \e^{-\i q\varphi} \big(\e^{\i\delta_{\omega q}} (-2\eta V)^{-\i\omega / \eta} + \e^{-\i \delta_{\omega q}} (2\eta U)^{\i\omega /\eta}\big)\Big]\, .
\end{align}
Matching \eqref{scalarnearhorF} with \eqref{scalarnearhorI1} and
\eqref{scalarnearhorII1} at $U = 0$ and $V = 0$ respectively, we get
\begin{equation}
	C^{\rm F}_{\omega q} = a_{\omega q}\, ,\quad C^{F'}_{\omega q} = \tl{a}^{\dag}_{\omega q}\, .
\end{equation}
Thus, the mode expansion in Region F is
\begin{align}
	\phi(t_F,r,\varphi) =& \frac{1}{\sqrt{2\pi R_h}}\sum_{q \in \mathbb{Z}}\int_0^\infty \frac{\ud \omega}{\sqrt{4\pi\omega}} \times \nonumber \\
	& \times  \Big( a_{\omega q}  G^1_{\omega q} +  \tl{a}^\dag_{\omega q}   G^2_{\omega q} + \text{c.c.}\Big)\, .
\end{align}

\subsection{Solution in Region P}

The ansatz in Region P for the scalar field is
\begin{equation}\label{scalardefP}
  \e^{\i\omega t_P} \e^{\i q \varphi}\, f^P_{\omega q}(\tl{r})\ ,
\end{equation}
with the solutions
\begin{align}
  & f^P_{\omega q}(\tl{r}) \nonumber \\
  &= C^{\rm P}_{\omega q}\, (1-\tl\rho)^{\tl\alpha} \tl\rho^{\tl\gamma} w_3(\tl\rho) + C_{\omega q}^{\rm P}{}'\, (1-\tl\rho)^{\tl\alpha} \tl\rho^{\tl\gamma} w_4(\tl\rho)\, ,\nonumber\\
  &=  C^{\rm P}_{\omega q} (1-\tl\rho)^{\tl\alpha} \tl\rho^{\tl\gamma}\ {}_2F_1\left(\tl{a},\tl{b};\tl{a}+\tl{b}-\tl{c}+1;1-\tl\rho\right) \nonumber\\
  &+ C_{\omega q}^{\rm P}{}'\, (1-\tl\rho)^{-\tl\alpha} \rho^{\tl\gamma} \ {}_2F_1\left(\tl{c}-\tl{a},\tl{c}-\tl{b};\tl{c}-\tl{a}-\tl{b}+1;1-\tl\rho\right) ,
\end{align}
for some constants $C^{\rm P}_{\omega q}$ and
$C_{\omega q}^{\rm P}{}'$. Note that the functions above are simply
$G^{1*}$ and $G^{2*}$ respectively so that the scalar field in Region
P has the mode expansion
\begin{align}
  \phi(t_P,\tl{r},\varphi) =& \frac{1}{\sqrt{2\pi R_h}}\sum_{q \in \mathbb{Z}}\int_0^\infty \frac{\ud \omega}{\sqrt{4\pi\omega}} \Big( C^P_{\omega q} G^{1*}_{\omega q}(t_P, \tl{r}, \varphi) \nonumber \\ &+ C^{P'}_{\omega q}   G^{2*}_{\omega q}(t_P, \tl{r}, \varphi) + \text{c.c.}\Big)\ .
\end{align}
The near horizon expression is
\begin{align}\label{scalarnearhorP}
  &\phi(t_P,\tl{r},\varphi) \sim \frac{1}{\sqrt{2\pi R_h}}\sum_{q \in \mathbb{Z}}\int_0^\infty \frac{\ud \omega}{\sqrt{4\pi\omega}} \times \nonumber \\
  &\times \Big[ \e^{\i q\varphi} \big( C^P_{\omega q} \e^{-\i\delta_{\omega q}} (-2\eta V)^{\i\omega/\eta} +  C^{P'}_{\omega q} \e^{\i\delta_{\omega q}} (-2\eta U)^{-\i\omega/\eta}\big) \nonumber\\
  & + \e^{-\i q\varphi} \big( C^{P*}_{\omega q} \e^{\i\delta_{\omega q}} (-2\eta V)^{-\i\omega/\eta} +  C^{P'*}_{\omega q} \e^{-\i\delta_{\omega q}} (-2\eta U)^{\i\omega/\eta}\big)\Big] .
\end{align}
At the future horizon $U = 0$, \eqref{scalarnearhorP} has to be
matched with the expression \eqref{scalarnearhorII1} in Region II,
whereas at the past horizon $V = 0$, it has to be matched with the
expression \eqref{scalarnearhorI1} in Region I. We get
\begin{equation}
  C^P_{\omega q} = \tl{a}_{\omega q}\ ,\quad C^{P'}_{\omega q} = a^\dag_{\omega q}\ ,
\end{equation}
so that the mode expansion in Region P is
\begin{align}
  \phi(t_P,\tl{r},\varphi) =& \frac{1}{ \sqrt{2\pi R_h}}\sum_{q \in \mathbb{Z}}\int_0^\infty \frac{\ud \omega}{\sqrt{4\pi\omega}} \times \nonumber \\
  &\times \Big( \tl{a}_{\omega q} G^{1*}_{\omega q} +  {a}^\dag_{\omega q}   G^{2*}_{\omega q} + \text{c.c.}\Big)\, .
\end{align}

\subsection{Analytic continuation to regions F and P}\label{ACtoFP}

\begin{figure}
	\centering
	\begin{tikzpicture}[scale=1]
		\draw[black] (-2,0) -- (0,0);
		\draw[black] (0,-2) -- (0,2);
		\draw[black,thick,decorate,decoration={snake, amplitude=1.5pt, segment length=4pt}] (0,0) -- (2,0);
		\draw [->,thick] (-1,0) arc (180:350:1);
		\draw[black] (1.5,1.5) -- (1.5,2);
		\draw[black] (1.5,1.5) -- (2,1.5);
		\node[black] at (1.8,1.8) {$U$};
	\end{tikzpicture}
	\caption{\label{analytic} Analytic continuation in the
		lower-half $U$-plane.}
\end{figure}
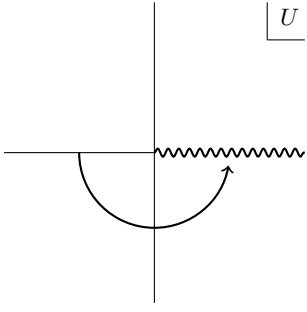

In the previous subsections, we solved the Klein-Gordon equation
separately in each Region and then matched the expressions across the
horizons. These mode functions were analytic in the lower-half $t$
plane (where $t$ has to be replaced by its counterpart in each
Region). One can also consider functions which are analytic in the
lower-half $U$ or $V$ planes. This consideration is motivated by the
simple observation, due to Unruh \cite{Unruh:1976db}, that
translations in the $V$ coordinate are a symmetry of the metric on the
future horizon $U = 0$ (and similarly, translations in $U$ on the
future horizon $V = 0$). One can then naturally write down mode
functions $\e^{-\i \omega V}$ which are analytic in the lower-half $V$
plane. Following Unruh, let us try to obtain appropriate analytic
continuations of the mode functions $F^i_{\omega q}$ such that they are analytic in the lower-half $U$ and $V$ planes. 

The mode functions $F^i_{\omega q}$ (defined in \eqref{F12def}) in Region I are valid for $U < 0$,
$V > 0$, and are certainly not analytic in the lower-half $U$ and $V$
planes. However, one can make them analytic in the lower-half $U$
plane by appropriately defining their continuation to $U > 0$, across
the future horizon $U = 0$ (and similarly, across the past horizon
$V = 0$). Recall from \eqref{nearhorI} that
$F^1_{\omega q} \sim V^{-\i\omega / \eta}$. Hence, this continues
uneventfully across $U = 0$. The function
$F^2_{\omega q} \sim U^{\i\omega / \eta}$ requires the introduction of
a branch cut in the complex $U$ plane to be well-defined. Let us take
the branch cut along the positive $U$ axis and define
\begin{equation}
	\text{For $U > 0$,}\quad \underbrace{U}_{\text{in F}} = \e^{\i\pi} \underbrace{(-U)}_{\text{in I}}\ .
\end{equation}
This corresponds to a rotation by $\pi$ in the counter-clockwise
direction in the lower-half $U$ plane, see Figure \ref{analytic}. This
ensure that $U^{\i\omega / \eta}$ is analytic in the lower-half
plane. How do we implement this for the full function $F^2_{\omega q}$
and not just its near-horizon expression?
\begin{figure}
	\centering
	\begin{tikzpicture}[scale=0.66]
		\draw[black,thick] (3,3) -- (3,-3);
		\draw[black,thick] (-3,3) -- (-3,-3);
		\draw[black,thick] (-3,3) -- (3,3);
		\draw[black,thick] (-3,-3) -- (3,-3);
		\draw[gray, dashed] (-3,-3) -- (3,3);
		\draw[gray, dashed] (-3,3) -- (3,-3);
		\node[black] at (1.8,0) {\footnotesize$F^1_{\omega q} + F^2_{\omega q}$};
		\node[black] at (0,2) {\footnotesize$G^1_{\omega q} + \e^{-\pi\omega / \eta} G^2_{\omega q}$};
		\node[black] at (-1.8,0) {\footnotesize 0};
		\node[black] at (0,-2) {\footnotesize$\e^{-\pi\omega/\eta}G^1_{\omega q} + G^2_{\omega q}$};
	\end{tikzpicture}
	\hfill
	\begin{tikzpicture}[scale=0.66]
		\draw[black,thick] (3,3) -- (3,-3);
		\draw[black,thick] (-3,3) -- (-3,-3);
		\draw[black,thick] (-3,3) -- (3,3);
		\draw[black,thick] (-3,-3) -- (3,-3);
		\draw[gray, dashed] (-3,-3) -- (3,3);
		\draw[gray, dashed] (-3,3) -- (3,-3);
		\node[black] at (1.8,0) {\footnotesize 0};
		\node[black] at (0,2) {\footnotesize$\e^{\pi\omega/\eta} G^{1}_{\omega q} + G^{2}_{\omega q}$};
		\node[black] at (-1.8,0) {\footnotesize$F^{1}_{\omega q} + F^{2}_{\omega q}$};
		\node[black] at (0,-2) {\footnotesize$G^{1}_{\omega q} + \e^{\pi\omega/\eta} G^{2}_{\omega q}$};
	\end{tikzpicture}
	\caption{\label{analyticcontBTZ} The above are Penrose
		diagrams for the fully extended BTZ black
		hole. \textbf{Left}: The analytic continuation of Region I
		mode $F_{\omega q} = F^1_{\omega q} + F^2_{\omega q}$ to
		rest of the fully extended BTZ
		spacetime. \textbf{Right}: The analytic continuation of
		Region II conjugate modes
		$\tl{F}^*_{\omega q} = F^{1}_{\omega q} + F^{2}_{\omega q}$. }
\end{figure}
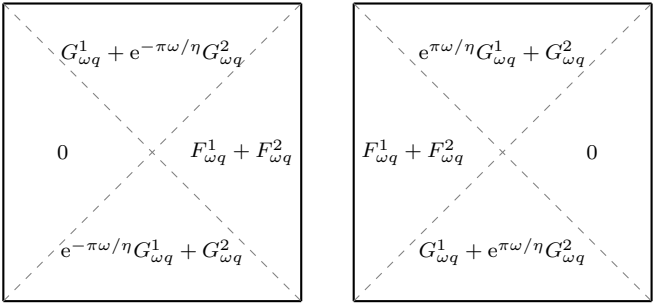
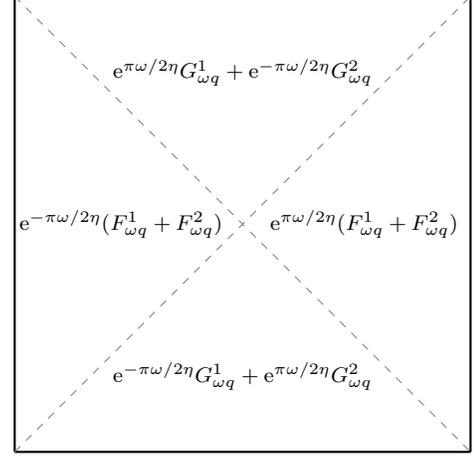
\begin{figure}[!t]
	\centering
	\begin{tikzpicture}[scale=1.01]
		\draw[black,thick] (3,3) -- (3,-3);
		\draw[black,thick] (-3,3) -- (-3,-3);
		\draw[black,thick] (-3,3) -- (3,3);
		\draw[black,thick] (-3,-3) -- (3,-3);
		\draw[gray, dashed] (-3,-3) -- (3,3);
		\draw[gray, dashed] (-3,3) -- (3,-3);
		\node[black] at (1.6,0) {\footnotesize$\e^{\pi\omega /2 \eta}(F^1_{\omega q} + F^2_{\omega q})$};
		\node[black] at (0,2) {\footnotesize$\e^{\pi\omega /2 \eta}G^1_{\omega q} + \e^{-\pi\omega /2 \eta} G^2_{\omega q}$};
		\node[black] at (-1.6,0) {\footnotesize $\e^{-\pi\omega /2 \eta}(F^1_{\omega q} + F^2_{\omega q})$};
		\node[black] at (0,-2) {\footnotesize$\e^{-\pi\omega/2\eta}G^1_{\omega q} + \e^{\pi\omega/2\eta} G^2_{\omega q}$};
	\end{tikzpicture}
	\caption{\label{BTZHHmodesfig} The definition of the Hartle-Hawking
		modes $h_{\omega q}$ obtained by the Unruh procedure.}
\end{figure}
Recall the definition of the Kruskal $U$ in Regions I and F:
\begin{equation}
	\text{I}:\ U = -\eta^{-1} \e^{-\eta(t-r_*)}\, ,\quad   \text{F}:\  U = \eta^{-1} \e^{-\eta(t_F-r_{*F})}\, .
\end{equation}
To get formulas in Region F, we replace $t - r_*$ by
$t - r_* -\i\pi/\eta$ in all formulas in Region I. But since the
local coordinates in Region F are called $t_F$ and $r_F$, we relabel
$t$ and $r_*$ as $t_F$ and $r_F$ respectively, after the
replacement. Effectively, we have
\begin{equation}
	t - r_* \to    t_F - r_{*F} - \i\pi / \eta\ ,\quad t + r_* \to t_F + r_{*F}\ ,
\end{equation}
that is, in continuing from Region I $\to$ F:
\begin{align}
	& t \to t_F - \i\pi/2\eta,\ r_{*} \to r_{*F} + \i\pi/2\eta  
	\Rightarrow \rho - 1 \to \e^{\i\pi} (1-\rho) .
\end{align}
Next, the continuation across the past horizon $V = 0$ involves taking
$V$ to $-V$ through the lower half complex $V$ plane while keeping $U$
fixed:
\begin{equation}
	\text{For $V < 0$,}\quad \underbrace{V}_{\text{in P}} = \e^{-\i\pi} \underbrace{(-V)}_{\text{in I}}  = - \e^{-\i\pi}  \eta^{-1} \e^{\eta(t+r_{*})}\ ,
\end{equation}
which continuing from Region I $\to$ P, corresponds to
\begin{align}
	& t \to t_P - \i\pi/2\eta,\ r_{*} \to r_{*P} - \i\pi/2\eta 
	\Rightarrow \rho-1 \to \e^{-\i\pi} (1-\tl\rho)\, .
\end{align}
Thus, we apply the above rules to $F^i_{\omega q}$ to obtain the
appropriate analytic continuation to Regions F and P. We then get

\begin{widetext}
\begin{align}
\text{Region I to F:}\qquad  F^1_{\omega q}(t,r,\varphi) &\to G^1_{\omega q}(t_F,r,\varphi)\, ,\quad   F^2_{\omega q}(t,r,\varphi) \to \e^{-\pi\omega /\eta} G^2_{\omega q}(t_F,r,\varphi)\, ,\nonumber\\
\text{Region I to P:}\qquad  F^1_{\omega q}(t,r,\varphi) &\to \e^{-\pi\omega /\eta} G^1_{\omega q}(t_P,\tl{r},\varphi)\, ,\quad   F^2_{\omega q}(t,r,\varphi) \to G^2_{\omega q}(t_P,\tl{r},\varphi)\, .
\end{align}
The analytic continuation of the $F^i_{\omega q}$ are then
\begin{equation}\label{BTZFiac}
  F^1_{\omega q} =  \left\{\def\arraystretch{1.5} \begin{array}{cl} F^1_{\omega q} \quad & \text{Region I} \\  G^1_{\omega q} \quad & \text{Region F} \\ \e^{-\pi\omega/\eta} G^1_{\omega q} \quad & \text{Region P} \\ 0 \quad & \text{Region II} \ .\end{array}\right.\qquad   F^2_{\omega q} =  \left\{\def\arraystretch{1.5} \begin{array}{cl} F^2_{\omega q} \quad & \text{Region I} \\ e^{-\pi\omega/\eta} G^2_{\omega q} \quad & \text{Region F} \\  G^2_{\omega q} \quad & \text{Region P} \\ 0 \quad & \text{Region II} \ .\end{array}\right.
\end{equation}
This is summarized on the left of Figure \ref{analyticcontBTZ}.

\textbf{Note:} Most importantly, the function presented
\eqref{BTZFiac} is analytic for negative values of $\omega$ as well,
since the procedure above did not involve positivity of $\omega$.

Similarly, we can work out the continuation rules for Region II to
Regions F and P:
\begin{align}\label{IItoFP}
	&\text{Region II $\to$ Region F:}\quad  \tl{t} \to t_F + \i\pi/2\eta\ ,\quad \tl{r}_{*} \to r_{*F} + \i\pi/2\eta\quad\Rightarrow\quad  \tl\rho-1 \to \e^{\i\pi} (1-\rho)\ ,\nonumber\\
	&\text{Region II $\to$ Region P:}\quad  \tl{t} \to t_P + \i\pi/2\eta\ ,\quad \tl{r}_{*} \to r_{*P} - \i\pi/2\eta\quad\Rightarrow\quad  \tl\rho-1 \to \e^{-\i\pi} (1-\tl\rho)\ .
\end{align}
The Region II modes \eqref{reghorII} involve $F^{i*}$. Under the rules
\eqref{IItoFP}, the analytic continuation to Regions F and P are
\begin{align}
\text{Region II to F}:\quad  F^{1*}_{\omega q}(\tl{t},\tl{r},\varphi) &\to \e^{-\pi\omega/\eta}\, G^{1*}_{\omega q}(t_F,r,\varphi)\, ,\quad 
  F^{2*}_{\omega q}(\tl{t},\tl{r},\varphi) \to G^{2*}_{\omega q}(t_F,r,\varphi)\, ,\nonumber\\
\text{Region II to P}:\quad  F^{1*}_{\omega q}(\tl{t},\tl{r},\varphi) &\to G^{1*}_{\omega q}(t_P,\tl{r},\varphi)\, ,\quad 
  F^{2*}_{\omega q}(\tl{t},\tl{r},\varphi) \to \e^{-\pi\omega/\eta}\,  G^{2*}_{\omega q}(t_P,\tl{r},\varphi)\, .
\end{align}
We also need the analytic continuation of the conjugate mode functions
in Region II. These simply involve the $F^i_{\omega q}$. Applying the
rules \eqref{IItoFP}, we get
\begin{align}
\text{Region II to F}:\quad  F^{1}_{\omega q}(\tl{t},\tl{r},\varphi) &\to \e^{\pi\omega/\eta}\, G^{1}_{\omega q}(t_F,r,\varphi)\ ,\quad  F^{2}_{\omega q}(\tl{t},\tl{r},\varphi) \to G^{2}_{\omega q}(t_F,r,\varphi)\, ,\nonumber\\
\text{Region II to P}:\quad  F^{1}_{\omega q}(\tl{t},\tl{r},\varphi) &\to G^{1}_{\omega q}(t_P,\tl{r},\varphi)\ ,\quad  F^{2}_{\omega q}(\tl{t},\tl{r},\varphi) \to \e^{\pi\omega/\eta}\,  G^{2}_{\omega q}(t_P,\tl{r},\varphi)\, .
\end{align}
\end{widetext}
The analytic continuations of the Region II conjugate modes are
summarized on the right in Figure \ref{analyticcontBTZ}. Observe that
the analytic continuation of the rescaled Region I mode function
$\e^{\pi\omega/2\eta} F_{\omega q} = \e^{\pi\omega/2\eta} (F^1 + F^2)$
and the rescaled Region II mode function
$\e^{-\pi\omega/2\eta} \tl{F}^*_{\omega q} = \e^{-\pi\omega/2\eta}
(F^1 + F^2)$ agree on Region F and Region P.  Using these analytic
continuations, we can define a function on the fully extended BTZ
black hole that is analytic in the lower half $U$, $V$ planes as
\begin{align}\label{BTZHHmodemain}
  &h_{\omega}(U,V,\varphi) = \frac{1}{\sqrt{4\pi\omega}\sqrt{2\sinh (\pi\omega/\eta)}} \times\nonumber\\
  & \times \left\{\def\arraystretch{1.5} \begin{array}{cl} \e^{\pi\omega/2\eta} (F^1_{\omega q} + F^2_{\omega q}) \quad & \text{Region I} \\ \e^{\pi\omega/2\eta} G^1_{\omega q} + \e^{-\pi\omega/2\eta} G^2_{\omega q} \quad & \text{Region F} \\ \e^{-\pi\omega/2\eta} G^1_{\omega q} + \e^{\pi\omega/2\eta} G^2_{\omega q} \quad & \text{Region P} \\ \e^{-\pi\omega/2\eta} (F^1_{\omega q} + F^2_{\omega q}) \quad & \text{Region II} \, .\end{array}\right.
\end{align}
Note that these modes are analytic in the lower half $U$ and $V$
planes for both positive and negative $\omega$. Thus, $\omega$ is a
label running over all real numbers for these modes, and only when we
include $\omega$ of both signs do we get a complete set of modes.

\subsection{The Hartle-Hawking modes and Bogoliubov coefficients}

The scalar field can then be expanded in terms of these Hartle-Hawking
modes:
\begin{equation}\label{phiBTZall}
  \phi(U,V,\varphi) = \frac{1}{\sqrt{2\pi R_h}}\sum_{q \in \mathbb{Z}}\int_{-\infty}^\infty \ud \omega\, c_{\omega q} h_{\omega q} + \text{c.c.}\, .
\end{equation}
For future use, we define $\tl{c}_{\omega q} = c_{-\omega,-q}$ for
$\omega > 0$. Restricting to Region I, we get
\begin{align}
	\phi(U,V,\varphi)\big|_{\rm I} =& \sum_{q \in \mathbb{Z}}\int_{0}^\infty \frac{\ud \omega}{\sqrt{4\pi\omega}\sqrt{2\sinh \frac{\pi\omega}{\eta}}}\, \times \nonumber \\
	&\times \Big((c_{\omega q} \e^{\pi\omega/2\eta} + \tl{c}^\dag_{\omega q} \e^{-\pi\omega/2\eta}) F_{\omega q} \nonumber\\
  	& + (\tl{c}_{\omega q} \e^{-\pi\omega/2\eta} + c^\dag_{\omega q} \e^{\pi\omega/2\eta}) F^*_{\omega q}\Big)\, .
\end{align}
Comparing with the mode expansion in Region I \eqref{phiBTZIapp}, we get
the following relations between the operators $a_{\omega q}$ and
$c_{\omega q}$, $\tl{c}_{\omega q}$:
\begin{align}\label{acctl}
	a_{\omega q} =& \frac{c_{\omega q} \e^{\pi\omega/2\eta} + \tl{c}^\dag_{\omega q} \e^{-\pi\omega/2\eta}}{\sqrt{2\sinh (\pi\omega/\eta)}}\, ,\nonumber \\ 
	a^\dag_{\omega q} =& \frac{c^\dag_{\omega q} \e^{\pi\omega/2\eta} + \tl{c}_{\omega q} \e^{-\pi\omega/2\eta}}{{\sqrt{2\sinh (\pi\omega/\eta)}}}\, .
\end{align}
Similarly, restricting the global mode expansion \eqref{phiBTZall} to
Region II, and comparing with the mode expansion in Region II
\eqref{phiBTZII}, we get
\begin{align}\label{tlacctl}
	\tl{a}_{\omega q} =& \frac{\tl{c}_{\omega q} \e^{\pi\omega/2\eta} + {c}^\dag_{\omega q} \e^{-\pi\omega/2\eta}}{\sqrt{2\sinh (\pi\omega/\eta)}}\, ,\nonumber \\
	 \tl{a}^\dag_{\omega q} =& \frac{\tl{c}^\dag_{\omega q} \e^{\pi\omega/2\eta} + c_{\omega q} \e^{-\pi\omega/2\eta}}{\sqrt{2\sinh (\pi\omega/\eta)}}\, .
\end{align}
We can invert the relations \eqref{acctl} and \eqref{tlacctl} together to get the Bogoliubov transformations
\begin{align}\label{cctldef}
	c_{\omega q} =& \frac{a_{\omega q} \e^{\pi\omega/2\eta} - \tl{a}^\dag_{\omega q} \e^{-\pi\omega/2\eta}}{\sqrt{2\sinh (\pi\omega/\eta)}}\, ,\nonumber \\
	\tl{c}_{\omega q} =& \frac{\tl{a}_{\omega q} \e^{\pi\omega/2\eta} - a^\dag_{\omega q} \e^{-\pi\omega/2\eta}}{\sqrt{2\sinh (\pi\omega/\eta)}}\, ,
\end{align}
and their conjugates. The commutation relations for $c_{\omega q}$,
$\tl{c}_{\omega q}$ for $\omega > 0$ can be obtained from those of the ${a}_{\omega q}$, $\tl{a}_{\omega q}$:
\begin{align}
	[{c}_{\omega q}, {c}^\dag_{\omega'q'}] =& [\tl{c}_{\omega q}, \tl{c}^\dag_{\omega'q'}] = \delta(\omega-\omega')\delta_{qq'}\, ,\nonumber \\  
	[{c}_{\omega q}, \tl{c}_{\omega'q'}] =& [{c}_{\omega q}, {c}_{\omega'q'}] = [\tl{c}_{\omega q}, \tl{c}_{\omega'q'}] = 0\, .
\end{align}

\subsection{Klein-Gordon normalization of the exterior mode functions}\label{KGnormapp}

The Klein-Gordon inner product between two solutions $\phi_1$ and
$\phi_2$ of the Klein-Gordon equation is defined by
\begin{equation}
	(\phi_1,\phi_2)_{\rm KG} =  \i \int_\Sigma \ud^{d-1}x \sqrt{g} n^\mu (\phi_1^* D_\mu \phi_2 - D_\mu \phi^*_1 \, \phi_2)\ , 
\end{equation}
where $\Sigma$ is a constant time slice and $n^\mu$ is the future directed unit vector normal to it.
In a static spacetime with metric $ -N^2 \ud t^2 + g_{ij} \ud x^i \ud x^j$, the normal to the constant time $t$ slice in the coordinate basis
$(\partial/ \partial t, \partial / \partial {x^i})$ is given by $n^\mu \partial_\mu = N^{-1}\partial_t$.
Thus, the inner product becomes
\begin{equation}
	(\phi_1,\phi_2)_{\rm KG} =  \i \int_\Sigma  \ud^{d-1}x \, \sqrt{g} N^{-1} (\phi_1^* \dot \phi_2 - \dot \phi^*_1 \phi_2)\ .
\end{equation}
For Region I of the BTZ black hole, any constant $t$ slice is
described by the range of coordinates
$ R_h < r < \infty\ ,\quad 0 \leq \varphi < 2\pi$. Recall that after writing $F_{\omega q}(t,r,\varphi) = \e^{-\i\omega t}\e^{-\i q \varphi} {f}_{\omega q}(r)$, the wave equation for $f_{\omega q}(r)$ in terms of the variable $\rho = r^2 / R_h^2$ reads
\begin{equation}\label{hypergeomrho}
	\frac{\ud}{\ud \rho} \big(\rho(\rho-1) \frac{\ud f_{\omega q}}{\ud \rho}\big) = \left(\frac{-1}{4(\rho-1)} \frac{\omega^2}{\eta^2} + \frac{q^2\ell^2}{4\rho R_h^2} + \frac{\ell^2 m^2}{4} \right) f_{\omega q} .
\end{equation}
The Klein-Gordon norm between the modes $F_{\omega q}, F_{\omega' q'}$ is
\begin{align}\label{KGexpr}
  &(F_{\omega' q'},F_{\omega q})_{\rm KG} \nonumber \\
  &= 2\pi(\omega + \omega') \e^{\i(\omega' - \omega)t} \delta_{q q'}\int_{R_h}^\infty \ud r\, r f^{-1}\, f^*_{\omega' q} f_{\omega q}\, ,\nonumber\\
  &= \pi\ell^2(\omega + \omega') \e^{\i(\omega' - \omega)t} \delta_{q q'}\int_{1}^\infty \frac{\ud \rho}{\rho-1}\, f^*_{\omega' q} f_{\omega q}\, ,\nonumber\\
  &= -\pi\ell^2\eta(\omega + \omega') \e^{\i(\omega' - \omega) t} \delta_{q q'}\int_{-\infty}^0 \ud r_*\coth {\eta r_*} f^*_{\omega' q} f_{\omega q}\, .
\end{align}
Let us analyze the integrals near the end-points which are the horizon
at $\rho = 1$ and the boundary at $\rho = \infty$. 

\paragraph{Near the boundary.} The two linearly independent solutions
of the equation \eqref{hypergeomrho} can be written as
\begin{align}\label{fpm}
  f^\pm_{\omega q}(\rho) =& C^\pm_{\omega q} (\rho-1)^{\alpha} \rho^{-\alpha-\frac{1}{2}\Delta_\pm} \times \nonumber \\
  &\times {}_2F_1\left(\alpha+\gamma + \tfrac{1}{2}\Delta_\pm, \alpha-\gamma + \tfrac{1}{2}\Delta_\pm;\Delta_\pm;\frac{1}{\rho}\right) ,
\end{align}
where $C^\pm$ are undetermined normalization constants. We see that
the integrand near $\rho = \infty$ behaves as $\rho^{-\Delta_\pm - 1}$
for the two solutions so that the integral behaves as
$\rho^{-\Delta_\pm}$. From the expression
$\Delta_\pm = 1\pm \sqrt{1 + \ell^2 m^2}$, it is clear that, for
$m^2 > 0$, the normalizable solution is $f^+_{\omega q}$. For the
range $-\ell^{-2} < m^2 < 0$, both solutions are normalizable but one
set of normalizable functions may be more appropriate, depending 
on the context.

\paragraph{Near the horizon.} The integrand has a pole at the horizon
$\rho = 1$ and the integral seems to be singular near the
horizon. However, the expression for the inner product in terms of the
tortoise coordinate $r_*$ (the last line of \eqref{KGexpr}) displays
no singularity at the horizon as $r_* \to -\infty$. Near the horizon, the
mode function $f^+_{\omega q}$ becomes a plane wave in terms
of the tortoise coordinate:
\begin{align}\label{fpnearhor}
	f^+_{\omega q} 
	&= 2\, C^+_{\omega q} N_{\omega q} \cos\left(\delta_{\omega q}- \omega r_* - \frac{\omega}{\eta}\log 2\right) + \mathcal{O}(\e^{2\eta r_*})\, .
\end{align}      
Since $\coth \eta r_* \to -1$ as $r_* \to -\infty$, the integral near
the horizon gives a delta function $\delta(\omega - \omega')$. We see
below that the Klein-Gordon inner product is precisely this delta
function and hence the mode functions $F_{\omega q}$ are delta
function normalized.

\paragraph{Evaluating the Klein-Gordon inner product} We show that the
integral over the radial coordinate in \eqref{KGexpr} is in fact a sum
of boundary terms on the spatial slice. From the equation
\eqref{hypergeomrho} satisfied by $f_{\omega q}$, we find that
\begin{widetext}
\begin{align}
	f^*_{\omega' q}\frac{\ud}{\ud \rho} \left(\rho(\rho-1) \frac{\ud f_{\omega q}}{\ud \rho}\right) - f_{\omega q}\frac{\ud}{\ud \rho} \left( \rho(\rho-1) \frac{\ud f^*_{\omega' q}}{\ud \rho}\right) = - \frac{1}{4(\rho-1)} \frac{(\omega^2-\omega'^2)}{\eta^2} f^*_{\omega' q}f_{\omega q}\, .
\end{align}
Integrating both sides over $\rho$, we get
\begin{equation}
	\int_1^\infty \ud \rho\, f^*_{\omega' q}\frac{\ud}{\ud \rho} \left(\rho(\rho-1) \frac{\ud f_{\omega q}}{\ud \rho}\right) - \int_1^\infty \ud \rho\,   f_{\omega q}\frac{\ud}{\ud \rho} \left(\rho(\rho-1) \frac{\ud f^*_{\omega' q}}{\ud \rho}\right) = -\frac{(\omega^2-\omega'^2)}{4 \eta^2} \int_1^\infty\frac{ \ud \rho}{\rho-1} f^*_{\omega' q}f_{\omega q}\, .
\end{equation}
Integrating by parts in both terms on the left-hand side, we see that
the volume terms cancel each other, and there is only a surface
contribution left:
\begin{align}
	\left[  \rho(\rho-1)\left(f^*_{\omega' q}\frac{\ud f_{\omega q}}{\ud \rho} -   f_{\omega q} \frac{\ud f^*_{\omega' q}}{\ud \rho}\right)\right]_1^\infty = - \frac{(\omega^2-\omega'^2)}{4 \eta^2} \int_1^\infty \frac{\ud \rho}{\rho-1}    f^*_{\omega' q}f_{\omega q}\, .
\end{align}
The right hand side is proportional to the Klein-Gordon inner product
\eqref{KGexpr} of $F_{\omega' q}$ and $F_{\omega q}$:
\begin{align}\label{KGBoundary}
	\left[  \rho(\rho-1)\left(f^*_{\omega' q}\frac{\ud f_{\omega q}}{\ud \rho} -   f_{\omega q} \frac{\ud f^*_{\omega' q}}{\ud \rho}\right)\right]_{1}^{\infty} =  \frac{(\omega'-\omega)}{4\pi \eta^2\ell^2}  \e^{-\i(\omega'-\omega) t} (F_{\omega'q},F_{\omega q})_{\rm KG}\ .
\end{align}
Let us first study the boundary term at $\rho = \infty$. 
As $\rho \to \infty$, the solutions $f^\pm_{\omega q}(\rho)$ in
\eqref{fpm} behave as
\begin{align}\label{falloffapp}
  f^\pm_{\omega q}(\rho) &= C^\pm_{\omega q} \rho^{-\Delta_\pm / 2}\left(1 + \frac{4(\alpha^2 - \gamma^2) + \Delta^2_\pm}{4\Delta_\pm} \frac{1}{\rho} + \mc{O}(\rho^{-2})\right)\ , \\
	\frac{\ud f^\pm_{\omega q}}{\ud \rho} &= C^\pm_{\omega q} \rho^{-\Delta_\pm / 2 -1}\left(-\frac{\Delta_\pm}{2} + \frac{(4(\alpha^2 - \gamma^2) + \Delta^2_\pm)(2+\Delta_\pm)}{8\Delta_\pm} \frac{1}{\rho} + \mc{O}(\rho^{-2})\right)\ .
\end{align}
Plugging this into the left hand side at $\rho = \infty$, the leading
contribution cancels, while the subleading contribution is
\begin{equation}
	\left[ \rho (\rho-1)\left(f^{\pm *}_{\omega' q}\frac{\ud f^\pm_{\omega q}}{\ud \rho} - f^\pm_{\omega q} \frac{\ud f^{\pm*}_{\omega' q}}{\ud \rho}\right)\right]_{\rho \to \infty} = \quad C^{\pm*}_{\omega' q} C^\pm_{\omega q} \frac{(2+\Delta_\pm) (\omega^2 - \omega'^2)}{8\Delta_\pm \eta^2} \rho^{-\Delta_\pm}\Big|_{\rho \to \infty}\, .
\end{equation}
Thus, we indeed see that, when $m^2 > 0$, $f^+_{\omega q}$ is
normalizable since $\Delta_+ > 0$ whereas $f^-_{\omega q}$ is not
normalizable since $\Delta_- < 0$. The upper limit in
\eqref{KGBoundary} simply vanishes for the normalizable solution.

Recall the behaviour \eqref{fpnearhor} of $f^+_{\omega q}$ near the
horizon $\rho = 1$. This gives
\begin{align}
	f^{+ *}_{\omega' q}\, \rho (\rho-1) \frac{\ud f^+_{\omega q}}{\ud \rho} = C^{+*}_{\omega' q} \, N_{\omega' q} \, C^+_{\omega q} \, N_{\omega q} \,  \frac{\omega}{\eta} \Bigg[& \sin\left(\delta_{\omega q} + \delta_{\omega' q} - (\omega + \omega') r_* - \frac{\omega + \omega'}{\eta}\log 2 \right) \nonumber \\
	& + \sin\left(\delta_{\omega q} - \delta_{\omega' q} - (\omega - \omega') r_* - \frac{\omega - \omega'}{\eta}\log 2 \right) \Bigg] + \mathcal{O}(\e^{2\eta r_*}) \ . 
\end{align}
Using, $\lim_{r_*\to -\infty} \sin(-\omega r_*)/\omega = \pi \delta(\omega)$, we get
\begin{align}
  \frac{1}{(\omega - \omega')} \lim_{r_*\to -\infty} f^{+ *}_{\omega' q}\, \rho (\rho-1) \frac{\ud f^+_{\omega q}}{\ud \rho} = |C^{+}_{\omega' q}|^2 \, N_{\omega q}^2 \frac{\pi\omega}{\eta} \left(\delta(\omega-\omega') + \delta(\omega+\omega')\right)\, .
\end{align}
Since $\omega, \omega' \geq 0$, the second delta function never
clicks. Therefore, we have
\begin{equation}
	\frac{1}{(\omega - \omega')} \left[ \rho (\rho-1)\left(f^{+ *}_{\omega' q}\frac{\ud f^+_{\omega q}}{\ud \rho} - f^+_{\omega q} \frac{\ud f^{+*}_{\omega' q}}{\ud \rho}\right)\right]_{\rho \to 1^+} = |C^{+}_{\omega' q}|^2 \, N_{\omega q}^2\, \frac{\pi}{\eta}\, 2\omega \delta(\omega-\omega')\, ,
\end{equation}
from which we find the Klein-Gordon inner product
\begin{equation}
	(F_{\omega'q},F_{\omega q})_{\rm KG} =  |C^{+}_{\omega' q}|^2 \, N_{\omega q}^2 4\pi^2 R_h 2\omega \delta(\omega-\omega')\, .
\end{equation}
Similarly, we find
\begin{equation}
	(F^*_{\omega'q},F^*_{\omega q})_{\rm KG} =  -|C^{+}_{\omega' q}|^2 \, N_{\omega q}^2 4\pi^2 R_h 2\omega \delta(\omega-\omega')\, ,\quad 	(F^*_{\omega'q},F_{\omega q})_{\rm KG} = 0\, .
\end{equation}
Setting
\begin{equation}
  C^{+}_{\omega q} = \frac{1}{N_{\omega q} \sqrt{2\pi R_h}}\, ,
\end{equation}
we have
\begin{equation}\label{normsolI}
	f^+_{\omega q}(\rho) = \frac{1}{N_{\omega q} \sqrt{2\pi R_h}} (\rho-1)^{\alpha} \rho^{-\alpha-\frac{1}{2}\Delta_+}\ {}_2F_1\left(\alpha+\gamma+\tfrac{1}{2}\Delta_+,\alpha-\gamma+\tfrac{1}{2}\Delta_+;\Delta_+;\frac{1}{\rho}\right)\, ,
\end{equation}
so that the modes
$F_{\omega q} = \e^{-\i\omega t} \e^{-\i q \varphi} f_{\omega q}$
satisfy the canonical Klein-Gordon inner products
\begin{equation}
	(F_{\omega' q'}, F_{\omega q})_{\rm KG} = 4\pi\omega \delta(\omega-\omega') \delta_{qq'}\ ,\quad 	(F^*_{\omega' q'}, F^*_{\omega q})_{\rm KG} = -4\pi\omega \delta(\omega-\omega') \delta_{qq'}\ ,\quad (F^*_{\omega' q'}, F_{\omega q})_{\rm KG} = 0\ . 
\end{equation}

\subsection{Klein-Gordon inner product of the Hartle-Hawking modes
  and the smeared modes}\label{KGsmeared}

Recall the Klein-Gordon inner product
\begin{equation}
	(\phi_1,\phi_2)_{\rm KG} =  \i \int_\Sigma \ud^{d-1}x \sqrt{g} n^\mu (\phi_1^* \partial_\mu \phi_2 - \partial_\mu \phi^*_1 \, \phi_2)\, .
\end{equation}
For a given $d-1$ dimensional slice $\Sigma$ in the spacetime geometry
and given functions $\phi_1$ and $\phi_2$ are fixed, different
coordinate systems give the same answer for the inner product. This is
because $n^\mu\partial_\mu$ is covariant and so is
$\ud^{d-1}x \sqrt{g}$. We demonstrate this in the case of the
time-symmetric slice in the fully extended BTZ black hole.

The time-symmetric slice is described in BTZ coordinates as the union
of the $t = 0$ slice in Region I and $\tl{t} = 0$ slice in Region
II. In wormhole coordinates, it is the $\bar{t} = 0$ slice and in
Kruskal coordinates, it is the $\tau = \frac{\eta}{2}(U + V) = 0$
slice. We summarize the various quantities required for the
Klein-Gordon norm in Table \ref{tableKG}. The quantities in different
coordinate systems can be obtained from each other by the appropriate
diffeomorphisms.
\begin{table}[!htbp]
	\centering
	\begin{tabular}{ |c|c|c|c| }
		\hline
		& BTZ & Kruskal & Wormhole \\
		\hline\hline
		Slice & $\{t = 0\} \cup \{\tl{t} = 0\}$ & $\tau = 0$ \phantom{$\displaystyle \frac{A}{B}$} & $\bar{t} = 0$ \\
		\hline
		Spatial range & $\{0 < r < \infty\} \cup \{0 < \tl{r} < \infty\}$ & $-1 < \chi < 1$\phantom{$\displaystyle \frac{A}{B}$} & $-\infty < x < \infty$ \\
		\hline
		$n^\mu \partial_\mu$ & $f(r)^{-1/2} \partial_t - f(\tl{r})^{-1/2} \partial_{\tl{t}}$ & $\displaystyle \frac{1 - \chi^2}{2\ell} \partial_\tau$ & $N^{-1} \partial_{\bar{t}}$\\
		\hline
		$\ud^2x\,\sqrt{g}$ & $r f(r)^{-1/2} \ud r \ud\varphi$,\ $\tl{r} f(\tl{r})^{-1/2} \ud \tl{r} \ud\varphi$ & $\displaystyle 2\ell R_h \frac{1 + \chi^2}{(1 - \chi^2)^2} \ud\chi\ud\varphi$ & $\ell \ud x \ud\varphi$\\
		\hline
	\end{tabular}
	\caption{The various quantities required for the Klein-Gordon inner product}\label{tableKG}
\end{table}
Suppose we want to compute the Klein-Gordon inner product of the
smeared modes $\hat{g}_{nq}(\bar{t},x,\varphi)$ and
$\hat{g}_{n'q'}(\bar{t},x,\varphi)$ on the $\bar{t} = 0$ slice. We
have the formula
\begin{equation}
	(\hat{g}_{nq}, \hat{g}_{n'q'})_{\rm KG} = \int_\Sigma \ud x\ud \varphi\, \ell N(\bar{t}=0, x)^{-1} \left(\hat{g}^*_{n q} \frac{\partial  \hat{g}_{n'q'}}{\partial{\bar{t}}} - \frac{\partial \hat{g}^*_{n q}}{\partial {\bar{t}}} \hat{g}_{n'q'}\right) .
\end{equation}
We can evaluate the above integral directly in wormhole coordinates,
but that would be tedious. Instead, we first go back to their
expression in terms of Hartle-Hawking mode functions:
\begin{equation}
	\hat{g}(\bar{t}(U,V), x(U,V), \varphi) = g_{nq}(U, V, \varphi) = \int_{-\infty}^\infty \frac{\ud\omega}{\eta} h_{\omega q}(U,V,\varphi) \psi_n(\omega /\eta)\, .
\end{equation}
Plugging this into the Klein-Gordon inner product and also changing to
Kruskal coordinates, we get
\begin{align}
	(\hat{g}_{nq}, \hat{g}_{n'q'})_{\rm KG} = \int_{-\infty}^\infty \frac{\ud\omega}{\eta} \psi_n(\omega / \eta) \int_{-\infty}^\infty \frac{\ud\omega'}{\eta} \psi_{n'}(\omega' / \eta) \int_\Sigma \ud \chi \ud \varphi\,  R_h \frac{1 - \chi^2}{1 + \chi^2}  \left(h^*_{\omega q} \frac{\partial  h_{\omega'q'}}{\partial {\tau}} -  \frac{\partial {h}^*_{\omega q}}{\partial {\tau}} h_{\omega'q'}\right) .
\end{align}
Now, once again, we can map back to the original BTZ coordinates on
each of the $t = 0$ and $\tl{t} = 0$ slices and use the expression for
the Hartle-Hawking mode functions in terms of the original mode
functions $F_{\omega q}$ in Regions I and II. This gives
\begin{multline}
	(\hat{g}_{nq}, \hat{g}_{n'q'})_{\rm KG} = 2\pi R_h\int_{-\infty}^\infty \frac{\ud\omega}{\eta} \psi_n(\omega / \eta) \int_{-\infty}^\infty \frac{\ud\omega'}{\eta} \psi_{n'}(\omega' / \eta) \frac{1}{4\pi\sqrt{\omega\omega' 2\sinh(\pi\omega/\eta)2\sinh(\pi\omega'/\eta)}} \times \\ \times \Bigg[\e^{\pi(\omega + \omega')/2\eta}   \int_{\Sigma_R} \ud r \ud \varphi\,  r f(r)^{-1} \left(F^*_{\omega q} \frac{\partial  F_{\omega'q'}}{\partial {t}} -  \frac{\partial {F}^*_{\omega q}}{\partial {t}} F_{\omega'q'}\right) \\ - \e^{-\pi(\omega + \omega')/2\eta}  \int_{\Sigma_L} \ud \tl{r} \ud \varphi\,  \tl{r} f(\tl{r})^{-1} \left(F^*_{\omega q} \frac{\partial  F_{\omega'q'}}{\partial \tl{t}} -  \frac{\partial {F}^*_{\omega q}}{\partial \tl{t}} F_{\omega'q'}\right) \Bigg] \, .
\end{multline}
The integrands have a pole at the horizon due to the $f(r)^{-1}$
factors. However, recall from our discussion in Appendix
\ref{KGnormapp} after \eqref{KGexpr} that, when we transform to
tortoise coordinates the integral is well-defined and leads to a delta
function normalization for the $F_{\omega q}$.

The integrals over $\Sigma_R$ and $\Sigma_L$ are the Klein-Gordon
inner products of $F_{\omega q}$, $F_{\omega'q'}$ in their original
regions of quantization and have been computed previously:
\begin{equation}
	(F_{\omega' q'}, F_{\omega q})_{\rm KG} = 4\pi\omega \delta(\omega-\omega') \delta_{qq'}\, ,\quad 	(F^*_{\omega' q'}, F^*_{\omega q})_{\rm KG} = -4\pi\omega \delta(\omega-\omega') \delta_{qq'}\, ,\quad (F^*_{\omega' q'}, F_{\omega q})_{\rm KG} = 0\, . 
\end{equation}
Plugging this into the previous expression, we get
\begin{multline}
	(\hat{g}_{nq}, \hat{g}_{n'q'})_{\rm KG} = 2\pi R_h\int_{-\infty}^\infty \frac{\ud\omega}{\eta} \psi_n(\omega / \eta) \int_{-\infty}^\infty \frac{\ud\omega'}{\eta} \psi_{n'}(\omega' / \eta) \frac{1}{4\pi\sqrt{\omega\omega' 2\sinh(\pi\omega/\eta)2\sinh(\pi\omega'/\eta)}} \times \\ \times \Bigg[\e^{\pi(\omega + \omega')/2\eta} 4\pi\omega \delta(\omega - \omega')\delta_{qq'} - \e^{-\pi(\omega + \omega')/2\eta} 4\pi\omega \delta(\omega - \omega')\delta_{qq'} \Bigg] \, .
\end{multline}
Carrying out the $\omega$, $\omega'$ integrals, we get
\begin{align}
	(\hat{g}_{nq}, \hat{g}_{n'q'})_{\rm KG} = \frac{2\pi R_h}{\eta}\int_{-\infty}^\infty \frac{\ud\omega}{\eta} \psi_n(\omega / \eta) \psi_{n'}(\omega / \eta) = 2\pi \ell^2 \delta_{nn'}\delta_{qq'}\, .
\end{align}
We can compute the other inner products analogously and obtain
\begin{align}
	&(\hat{g}_{n q}, \hat{g}_{n' q'})_{\rm KG} = 2\pi \ell^2 \delta_{nn'}\delta_{qq'}\, ,\quad   (\hat{g}^*_{n q}, \hat{g}_{n' q'})_{\rm KG} = 0\, ,\quad (\hat{g}^*_{n q}, \hat{g}^*_{n' q'})_{\rm KG} = -2\pi \ell^2 \delta_{nn'}\delta_{qq'}\, .
\end{align}
Finally, the Hartle-Hawking modes' inner products can be computed along the same lines:
\begin{equation}
	(h_{\omega' q'}, h_{\omega q})_{\rm KG} = 2\pi R_h \delta(\omega-\omega') \delta_{qq'}\ ,\quad 	(h^*_{\omega' q'}, h^*_{\omega q})_{\rm KG} = -2\pi R_h \delta(\omega-\omega') \delta_{qq'}\, ,\quad (h^*_{\omega' q'}, h_{\omega q})_{\rm KG} = 0\, . 
\end{equation}
\end{widetext}

\section{Bulk reconstruction of local scalar field operator in wormhole coordinates} \label{BRinWC}

Recall the mode expansion in region I
\begin{align}
	\phi(t,r,\varphi) = \sum_{q\in \mathbb{Z}} \int_0^\infty \frac{\ud\omega}{\sqrt{4\pi\omega}} \Big( a_{\omega q}  F_{\omega q}(t,r,\varphi) + \text{c.c.} \Big)\, ,
\end{align}
with $F_{\omega q}(t,r,\varphi)= \e^{-\i\omega t} \e^{-\i q\varphi} f_\omega(r)$. This allows us to invert this equation, namely,
\begin{equation}
	a_{\omega q} f_\omega(r) = \frac{1}{4\pi^2} \int_0^{2\pi} \ud\varphi'  \int_{-\infty}^\infty \ud t' \e^{\i\omega t'} \e^{\i q\varphi'} \phi(t',r,\varphi') .
\end{equation}
On taking the extrapolate limit, i.e., $\lim_{r\to\infty} r^{\Delta_+} \phi(t',r,\varphi') = \mathcal{O}(t',\varphi')$, we get an expression for the right CFT oscillator
\begin{align}\label{aHKLL}
	a_{\omega q} =& \frac{N_{\omega q} \sqrt{2\pi R_h}}{4\pi^2 R_h^{\Delta_+}} \int_0^{2\pi} \ud\varphi'  \int_{-\infty}^\infty \ud t' \e^{\i\omega t'} \e^{\i q\varphi'} \mathcal{O}(t',\varphi') . 
\end{align}
Similar expression exists for the scalar field in region II. Recall
\begin{align}
	\phi(\tilde{t}, \tilde{r}, \tilde{\varphi}) = \sum_{q\in \mathbb{Z}} \int_0^\infty \frac{\ud\omega}{\sqrt{4\pi\omega}} \Big( \tilde{a}_{\omega q}  \tilde{F}_{\omega q}( \tilde{t}, \tilde{r}, \tilde{\varphi}) + \text{c.c.} \Big)\, ,
\end{align}
with $\tilde{F}_{\omega q}( \tilde{t}, \tilde{r}, \tilde{\varphi}) =  \e^{\i\omega \tilde{t}} \e^{\i q\tilde{\varphi}} f_\omega(r)$. This gives
\begin{align}\label{atHKLL}
	\tilde{a}_{\omega q} =& \frac{N_{\omega q} \sqrt{2\pi R_h}}{4\pi^2 R_h^{\Delta_+}} \int_0^{2\pi} \ud\tilde{\varphi}' \int_{-\infty}^\infty \ud \tilde{t}' \e^{-\i\omega \tilde{t}'} \e^{-\i q\tilde{\varphi}'} \tilde{\mathcal{O}}(\tilde{t}', \tilde{\varphi}') . 
\end{align}
Now let us go back to the discrete scalar field expansion in terms of the smooth mode functions,
\begin{equation}
	\phi(\bar{t},x,\varphi) = \frac{1}{\sqrt{2\pi} \ell}\sum_{q \in \mathbb{Z}}\sum_{n=0}^\infty \big(e_{nq}\, \hat{g}_{n q}(\bar{t},x,\varphi) + \text{c.c.}\Big)\, ,
\end{equation}
and recall that the $e_{n q}$ oscillators are combinations of $a_{\omega q}$ and $\tilde{a}_{\omega q}$:
\begin{widetext}
	\begin{align}
		&e_{nq} = \int_{0}^{\infty} \frac{\ud\omega}{\sqrt{\eta}} \left( c_{\omega q} \psi_n(\omega/\eta) + c_{-\omega q} \psi_n(-\omega/\eta) \right) = \int_{0}^{\infty} \frac{\ud\omega}{\sqrt{\eta}} \left( c_{\omega q} \psi_n(\omega/\eta) + \tl{c}_{\omega, -q} \psi_n(-\omega/\eta) \right) \nonumber \\
        &= \int_{0}^{\infty} \frac{\ud\omega}{\sqrt{2\eta \sinh (\pi\omega/\eta)}} \left[ (a_{\omega q} \e^{\pi\omega/2\eta} - \tl{a}^\dag_{\omega q} \e^{-\pi\omega/2\eta}) \psi_n(\omega/\eta) + (\tl{a}_{\omega, -q} \e^{\pi\omega/2\eta} - a^\dag_{\omega, -q} \e^{-\pi\omega/2\eta}) \psi_n(-\omega/\eta) \right]\nonumber \\
		&= \int_{0}^{\infty} \frac{\ud\omega}{\sqrt{2\eta \sinh (\pi\omega/\eta)}} \left[ (a_{\omega q} \e^{\pi\omega/2\eta} \psi_n(\omega/\eta) - a^\dag_{\omega, -q} \e^{-\pi\omega/2\eta} \psi_n(-\omega/\eta) )  + (\tl{a}_{\omega, -q} \e^{\pi\omega/2\eta} \psi_n(-\omega/\eta) - \tl{a}^\dag_{\omega q} \e^{-\pi\omega/2\eta} \psi_n(\omega/\eta)) \right].
	\end{align}
	Plugging in the individual oscillator expressions \eqref{aHKLL} and \eqref{atHKLL}, we get a bulk expression for the scalar field everywhere in the wormhole coordinates (including the part of interior regions F and P contained in $R_0$), in terms of the left and right CFT operators
	\begin{align}
		\phi(\bar t,x,\varphi) = \int_0^{2\pi} \ud\varphi' \int_{-\infty}^\infty  \ud t' \mathcal{K}(\bar t,x,\varphi;t',\varphi') \mathcal{O}(t',\varphi') + \int_0^{2\pi} \ud\tilde{\varphi}' \int_{-\infty}^\infty \ud \tilde{t}' \tilde{\mathcal{K}}(\bar t,x,\varphi; \tilde{t}', \tilde{\varphi}') \tilde{\mathcal{O}}( \tilde{t}', \tilde{\varphi}') \, ,
	\end{align}
	with
	\begin{align}\label{kxktx}
		\mathcal{K}(\bar t,x,\varphi;t',\varphi') =& \frac{N_{\omega q} \sqrt{R_h}}{4\pi^2 \ell R_h^{\Delta_+}} \sum_{n,q} \hat{g}_{nq}(\bar t,x,\varphi) \times \nonumber \\
		&\times \int_{0}^{\infty} \frac{\ud\omega}{\sqrt{2\eta \sinh (\pi\omega/\eta)}} \left( \e^{\pi\omega/2\eta} \psi_n(\omega/\eta) \e^{\i\omega t'+\i q\varphi'} - \e^{-\pi\omega/2\eta} \psi_n(-\omega/\eta) \e^{-\i\omega t'+\i q\varphi'} \right) + \text{c.c.}\, , \\
		\tilde{\mathcal{K}}(\bar t,x,\varphi; \tilde{t}', \tilde{\varphi}') =& \frac{N_{\omega q} \sqrt{R_h}}{4\pi^2 \ell R_h^{\Delta_+}} \sum_{n,q} \hat{g}_{nq}(\bar t,x,\varphi) \times \nonumber \\
		&\times \int_{0}^{\infty} \frac{\ud\omega}{\sqrt{2\eta \sinh (\pi\omega/\eta)}} \left( \e^{\pi\omega/2\eta} \psi_n(-\omega/\eta) \e^{-\i\omega \tilde{t}'+\i q\tilde{\varphi}'} - \e^{-\pi\omega/2\eta} \psi_n(\omega/\eta) \e^{\i\omega \tilde{t}'+\i q\tilde{\varphi}'} \right) + \text{c.c.}\, .
	\end{align}
\end{widetext}
Note that the bulk spacetime dependence is coming entirely from the the mode functions $\hat{g}_{nq}(\bar t,x,\varphi)$. Consequently these boundary-to-bulk kernels satisfy the Klein-Gordon equation \eqref{KGwave} in the wormhole coordinates.

\bibliography{refs}

\begin{thebibliography}{81}%
\makeatletter
\providecommand \@ifxundefined [1]{%
 \@ifx{#1\undefined}
}%
\providecommand \@ifnum [1]{%
 \ifnum #1\expandafter \@firstoftwo
 \else \expandafter \@secondoftwo
 \fi
}%
\providecommand \@ifx [1]{%
 \ifx #1\expandafter \@firstoftwo
 \else \expandafter \@secondoftwo
 \fi
}%
\providecommand \natexlab [1]{#1}%
\providecommand \enquote  [1]{``#1''}%
\providecommand \bibnamefont  [1]{#1}%
\providecommand \bibfnamefont [1]{#1}%
\providecommand \citenamefont [1]{#1}%
\providecommand \href@noop [0]{\@secondoftwo}%
\providecommand \href [0]{\begingroup \@sanitize@url \@href}%
\providecommand \@href[1]{\@@startlink{#1}\@@href}%
\providecommand \@@href[1]{\endgroup#1\@@endlink}%
\providecommand \@sanitize@url [0]{\catcode `\\12\catcode `\$12\catcode
  `\&12\catcode `\#12\catcode `\^12\catcode `\_12\catcode `\%12\relax}%
\providecommand \@@startlink[1]{}%
\providecommand \@@endlink[0]{}%
\providecommand \url  [0]{\begingroup\@sanitize@url \@url }%
\providecommand \@url [1]{\endgroup\@href {#1}{\urlprefix }}%
\providecommand \urlprefix  [0]{URL }%
\providecommand \Eprint [0]{\href }%
\providecommand \doibase [0]{https://doi.org/}%
\providecommand \selectlanguage [0]{\@gobble}%
\providecommand \bibinfo  [0]{\@secondoftwo}%
\providecommand \bibfield  [0]{\@secondoftwo}%
\providecommand \translation [1]{[#1]}%
\providecommand \BibitemOpen [0]{}%
\providecommand \bibitemStop [0]{}%
\providecommand \bibitemNoStop [0]{.\EOS\space}%
\providecommand \EOS [0]{\spacefactor3000\relax}%
\providecommand \BibitemShut  [1]{\csname bibitem#1\endcsname}%
\let\auto@bib@innerbib\@empty
\bibitem [{\citenamefont {Hawking}(1975)}]{Hawking:1975vcx}%
  \BibitemOpen
  \bibfield  {author} {\bibinfo {author} {\bibfnamefont {S.~W.}\ \bibnamefont
  {Hawking}},\ }\href {https://doi.org/10.1007/BF02345020} {\bibfield
  {journal} {\bibinfo  {journal} {Commun. Math. Phys.}\ }\textbf {\bibinfo
  {volume} {43}},\ \bibinfo {pages} {199} (\bibinfo {year} {1975})},\ \bibinfo
  {note} {[Erratum: Commun.Math.Phys. 46, 206 (1976)]}\BibitemShut {NoStop}%
\bibitem [{\citenamefont {Fredenhagen}\ and\ \citenamefont
  {Haag}(1990)}]{Fredenhagen:1989kr}%
  \BibitemOpen
  \bibfield  {author} {\bibinfo {author} {\bibfnamefont {K.}~\bibnamefont
  {Fredenhagen}}\ and\ \bibinfo {author} {\bibfnamefont {R.}~\bibnamefont
  {Haag}},\ }\href {https://doi.org/10.1007/BF02096757} {\bibfield  {journal}
  {\bibinfo  {journal} {Commun. Math. Phys.}\ }\textbf {\bibinfo {volume}
  {127}},\ \bibinfo {pages} {273} (\bibinfo {year} {1990})}\BibitemShut
  {NoStop}%
\bibitem [{\citenamefont {Witten}(2025)}]{Witten:2024upt}%
  \BibitemOpen
  \bibfield  {author} {\bibinfo {author} {\bibfnamefont {E.}~\bibnamefont
  {Witten}},\ }\href {https://doi.org/10.1140/epjp/s13360-025-06288-y}
  {\bibfield  {journal} {\bibinfo  {journal} {Eur. Phys. J. Plus}\ }\textbf
  {\bibinfo {volume} {140}},\ \bibinfo {pages} {430} (\bibinfo {year}
  {2025})},\ \Eprint {https://arxiv.org/abs/2412.16795} {arXiv:2412.16795
  [hep-th]} \BibitemShut {NoStop}%
\bibitem [{\citenamefont {Strominger}\ and\ \citenamefont
  {Vafa}(1996)}]{Strominger:1996sh}%
  \BibitemOpen
  \bibfield  {author} {\bibinfo {author} {\bibfnamefont {A.}~\bibnamefont
  {Strominger}}\ and\ \bibinfo {author} {\bibfnamefont {C.}~\bibnamefont
  {Vafa}},\ }\href {https://doi.org/10.1016/0370-2693(96)00345-0} {\bibfield
  {journal} {\bibinfo  {journal} {Phys. Lett. B}\ }\textbf {\bibinfo {volume}
  {379}},\ \bibinfo {pages} {99} (\bibinfo {year} {1996})},\ \Eprint
  {https://arxiv.org/abs/hep-th/9601029} {arXiv:hep-th/9601029} \BibitemShut
  {NoStop}%
\bibitem [{\citenamefont {David}\ \emph {et~al.}(2002)\citenamefont {David},
  \citenamefont {Mandal},\ and\ \citenamefont {Wadia}}]{David:2002wn}%
  \BibitemOpen
  \bibfield  {author} {\bibinfo {author} {\bibfnamefont {J.~R.}\ \bibnamefont
  {David}}, \bibinfo {author} {\bibfnamefont {G.}~\bibnamefont {Mandal}},\ and\
  \bibinfo {author} {\bibfnamefont {S.~R.}\ \bibnamefont {Wadia}},\ }\href
  {https://doi.org/10.1016/S0370-1573(02)00271-5} {\bibfield  {journal}
  {\bibinfo  {journal} {Phys. Rept.}\ }\textbf {\bibinfo {volume} {369}},\
  \bibinfo {pages} {549} (\bibinfo {year} {2002})},\ \Eprint
  {https://arxiv.org/abs/hep-th/0203048} {arXiv:hep-th/0203048} \BibitemShut
  {NoStop}%
\bibitem [{\citenamefont {Maldacena}(1998)}]{Maldacena:1997re}%
  \BibitemOpen
  \bibfield  {author} {\bibinfo {author} {\bibfnamefont {J.~M.}\ \bibnamefont
  {Maldacena}},\ }\href {https://doi.org/10.4310/ATMP.1998.v2.n2.a1} {\bibfield
   {journal} {\bibinfo  {journal} {Adv. Theor. Math. Phys.}\ }\textbf {\bibinfo
  {volume} {2}},\ \bibinfo {pages} {231} (\bibinfo {year} {1998})},\ \Eprint
  {https://arxiv.org/abs/hep-th/9711200} {arXiv:hep-th/9711200} \BibitemShut
  {NoStop}%
\bibitem [{\citenamefont {Witten}(1998)}]{Witten:1998qj}%
  \BibitemOpen
  \bibfield  {author} {\bibinfo {author} {\bibfnamefont {E.}~\bibnamefont
  {Witten}},\ }\href {https://doi.org/10.4310/ATMP.1998.v2.n2.a2} {\bibfield
  {journal} {\bibinfo  {journal} {Adv. Theor. Math. Phys.}\ }\textbf {\bibinfo
  {volume} {2}},\ \bibinfo {pages} {253} (\bibinfo {year} {1998})},\ \Eprint
  {https://arxiv.org/abs/hep-th/9802150} {arXiv:hep-th/9802150} \BibitemShut
  {NoStop}%
\bibitem [{\citenamefont {Gubser}\ \emph {et~al.}(1998)\citenamefont {Gubser},
  \citenamefont {Klebanov},\ and\ \citenamefont {Polyakov}}]{Gubser:1998bc}%
  \BibitemOpen
  \bibfield  {author} {\bibinfo {author} {\bibfnamefont {S.~S.}\ \bibnamefont
  {Gubser}}, \bibinfo {author} {\bibfnamefont {I.~R.}\ \bibnamefont
  {Klebanov}},\ and\ \bibinfo {author} {\bibfnamefont {A.~M.}\ \bibnamefont
  {Polyakov}},\ }\href {https://doi.org/10.1016/S0370-2693(98)00377-3}
  {\bibfield  {journal} {\bibinfo  {journal} {Phys. Lett. B}\ }\textbf
  {\bibinfo {volume} {428}},\ \bibinfo {pages} {105} (\bibinfo {year}
  {1998})},\ \Eprint {https://arxiv.org/abs/hep-th/9802109}
  {arXiv:hep-th/9802109} \BibitemShut {NoStop}%
\bibitem [{\citenamefont {Ryu}\ and\ \citenamefont
  {Takayanagi}(2006)}]{Ryu:2006bv}%
  \BibitemOpen
  \bibfield  {author} {\bibinfo {author} {\bibfnamefont {S.}~\bibnamefont
  {Ryu}}\ and\ \bibinfo {author} {\bibfnamefont {T.}~\bibnamefont
  {Takayanagi}},\ }\href {https://doi.org/10.1103/PhysRevLett.96.181602}
  {\bibfield  {journal} {\bibinfo  {journal} {Phys. Rev. Lett.}\ }\textbf
  {\bibinfo {volume} {96}},\ \bibinfo {pages} {181602} (\bibinfo {year}
  {2006})},\ \Eprint {https://arxiv.org/abs/hep-th/0603001}
  {arXiv:hep-th/0603001} \BibitemShut {NoStop}%
\bibitem [{\citenamefont {Hubeny}\ \emph {et~al.}(2007)\citenamefont {Hubeny},
  \citenamefont {Rangamani},\ and\ \citenamefont {Takayanagi}}]{Hubeny:2007xt}%
  \BibitemOpen
  \bibfield  {author} {\bibinfo {author} {\bibfnamefont {V.~E.}\ \bibnamefont
  {Hubeny}}, \bibinfo {author} {\bibfnamefont {M.}~\bibnamefont {Rangamani}},\
  and\ \bibinfo {author} {\bibfnamefont {T.}~\bibnamefont {Takayanagi}},\
  }\href {https://doi.org/10.1088/1126-6708/2007/07/062} {\bibfield  {journal}
  {\bibinfo  {journal} {JHEP}\ }\textbf {\bibinfo {volume} {07}},\ \bibinfo
  {pages} {062}},\ \Eprint {https://arxiv.org/abs/0705.0016} {arXiv:0705.0016
  [hep-th]} \BibitemShut {NoStop}%
\bibitem [{\citenamefont {Lewkowycz}\ and\ \citenamefont
  {Maldacena}(2013)}]{Lewkowycz:2013nqa}%
  \BibitemOpen
  \bibfield  {author} {\bibinfo {author} {\bibfnamefont {A.}~\bibnamefont
  {Lewkowycz}}\ and\ \bibinfo {author} {\bibfnamefont {J.}~\bibnamefont
  {Maldacena}},\ }\href {https://doi.org/10.1007/JHEP08(2013)090} {\bibfield
  {journal} {\bibinfo  {journal} {JHEP}\ }\textbf {\bibinfo {volume} {08}},\
  \bibinfo {pages} {090}},\ \Eprint {https://arxiv.org/abs/1304.4926}
  {arXiv:1304.4926 [hep-th]} \BibitemShut {NoStop}%
\bibitem [{\citenamefont {Faulkner}\ \emph {et~al.}(2013)\citenamefont
  {Faulkner}, \citenamefont {Lewkowycz},\ and\ \citenamefont
  {Maldacena}}]{Faulkner:2013ana}%
  \BibitemOpen
  \bibfield  {author} {\bibinfo {author} {\bibfnamefont {T.}~\bibnamefont
  {Faulkner}}, \bibinfo {author} {\bibfnamefont {A.}~\bibnamefont
  {Lewkowycz}},\ and\ \bibinfo {author} {\bibfnamefont {J.}~\bibnamefont
  {Maldacena}},\ }\href {https://doi.org/10.1007/JHEP11(2013)074} {\bibfield
  {journal} {\bibinfo  {journal} {JHEP}\ }\textbf {\bibinfo {volume} {11}},\
  \bibinfo {pages} {074}},\ \Eprint {https://arxiv.org/abs/1307.2892}
  {arXiv:1307.2892 [hep-th]} \BibitemShut {NoStop}%
\bibitem [{\citenamefont {Engelhardt}\ and\ \citenamefont
  {Wall}(2015)}]{Engelhardt:2014gca}%
  \BibitemOpen
  \bibfield  {author} {\bibinfo {author} {\bibfnamefont {N.}~\bibnamefont
  {Engelhardt}}\ and\ \bibinfo {author} {\bibfnamefont {A.~C.}\ \bibnamefont
  {Wall}},\ }\href {https://doi.org/10.1007/JHEP01(2015)073} {\bibfield
  {journal} {\bibinfo  {journal} {JHEP}\ }\textbf {\bibinfo {volume} {01}},\
  \bibinfo {pages} {073}},\ \Eprint {https://arxiv.org/abs/1408.3203}
  {arXiv:1408.3203 [hep-th]} \BibitemShut {NoStop}%
\bibitem [{\citenamefont {Penington}(2020)}]{Penington:2019npb}%
  \BibitemOpen
  \bibfield  {author} {\bibinfo {author} {\bibfnamefont {G.}~\bibnamefont
  {Penington}},\ }\href {https://doi.org/10.1007/JHEP09(2020)002} {\bibfield
  {journal} {\bibinfo  {journal} {JHEP}\ }\textbf {\bibinfo {volume} {09}},\
  \bibinfo {pages} {002}},\ \Eprint {https://arxiv.org/abs/1905.08255}
  {arXiv:1905.08255 [hep-th]} \BibitemShut {NoStop}%
\bibitem [{\citenamefont {Almheiri}\ \emph
  {et~al.}(2019{\natexlab{a}})\citenamefont {Almheiri}, \citenamefont
  {Engelhardt}, \citenamefont {Marolf},\ and\ \citenamefont
  {Maxfield}}]{Almheiri:2019psf}%
  \BibitemOpen
  \bibfield  {author} {\bibinfo {author} {\bibfnamefont {A.}~\bibnamefont
  {Almheiri}}, \bibinfo {author} {\bibfnamefont {N.}~\bibnamefont
  {Engelhardt}}, \bibinfo {author} {\bibfnamefont {D.}~\bibnamefont {Marolf}},\
  and\ \bibinfo {author} {\bibfnamefont {H.}~\bibnamefont {Maxfield}},\ }\href
  {https://doi.org/10.1007/JHEP12(2019)063} {\bibfield  {journal} {\bibinfo
  {journal} {JHEP}\ }\textbf {\bibinfo {volume} {12}},\ \bibinfo {pages}
  {063}},\ \Eprint {https://arxiv.org/abs/1905.08762} {arXiv:1905.08762
  [hep-th]} \BibitemShut {NoStop}%
\bibitem [{\citenamefont {Almheiri}\ \emph {et~al.}(2020)\citenamefont
  {Almheiri}, \citenamefont {Mahajan}, \citenamefont {Maldacena},\ and\
  \citenamefont {Zhao}}]{Almheiri:2019hni}%
  \BibitemOpen
  \bibfield  {author} {\bibinfo {author} {\bibfnamefont {A.}~\bibnamefont
  {Almheiri}}, \bibinfo {author} {\bibfnamefont {R.}~\bibnamefont {Mahajan}},
  \bibinfo {author} {\bibfnamefont {J.}~\bibnamefont {Maldacena}},\ and\
  \bibinfo {author} {\bibfnamefont {Y.}~\bibnamefont {Zhao}},\ }\href
  {https://doi.org/10.1007/JHEP03(2020)149} {\bibfield  {journal} {\bibinfo
  {journal} {JHEP}\ }\textbf {\bibinfo {volume} {03}},\ \bibinfo {pages}
  {149}},\ \Eprint {https://arxiv.org/abs/1908.10996} {arXiv:1908.10996
  [hep-th]} \BibitemShut {NoStop}%
\bibitem [{\citenamefont {Almheiri}\ \emph
  {et~al.}(2019{\natexlab{b}})\citenamefont {Almheiri}, \citenamefont
  {Mahajan},\ and\ \citenamefont {Maldacena}}]{Almheiri:2019yqk}%
  \BibitemOpen
  \bibfield  {author} {\bibinfo {author} {\bibfnamefont {A.}~\bibnamefont
  {Almheiri}}, \bibinfo {author} {\bibfnamefont {R.}~\bibnamefont {Mahajan}},\
  and\ \bibinfo {author} {\bibfnamefont {J.}~\bibnamefont {Maldacena}},\
  }\href@noop {} {\bibinfo {title} {{Islands outside the horizon}}} (\bibinfo
  {year} {2019}{\natexlab{b}}),\ \Eprint {https://arxiv.org/abs/1910.11077}
  {arXiv:1910.11077 [hep-th]} \BibitemShut {NoStop}%
\bibitem [{\citenamefont {Penington}\ \emph {et~al.}(2022)\citenamefont
  {Penington}, \citenamefont {Shenker}, \citenamefont {Stanford},\ and\
  \citenamefont {Yang}}]{Penington:2019kki}%
  \BibitemOpen
  \bibfield  {author} {\bibinfo {author} {\bibfnamefont {G.}~\bibnamefont
  {Penington}}, \bibinfo {author} {\bibfnamefont {S.~H.}\ \bibnamefont
  {Shenker}}, \bibinfo {author} {\bibfnamefont {D.}~\bibnamefont {Stanford}},\
  and\ \bibinfo {author} {\bibfnamefont {Z.}~\bibnamefont {Yang}},\ }\href
  {https://doi.org/10.1007/JHEP03(2022)205} {\bibfield  {journal} {\bibinfo
  {journal} {JHEP}\ }\textbf {\bibinfo {volume} {03}},\ \bibinfo {pages}
  {205}},\ \Eprint {https://arxiv.org/abs/1911.11977} {arXiv:1911.11977
  [hep-th]} \BibitemShut {NoStop}%
\bibitem [{\citenamefont {Almheiri}\ \emph {et~al.}(2021)\citenamefont
  {Almheiri}, \citenamefont {Hartman}, \citenamefont {Maldacena}, \citenamefont
  {Shaghoulian},\ and\ \citenamefont {Tajdini}}]{Almheiri:2020cfm}%
  \BibitemOpen
  \bibfield  {author} {\bibinfo {author} {\bibfnamefont {A.}~\bibnamefont
  {Almheiri}}, \bibinfo {author} {\bibfnamefont {T.}~\bibnamefont {Hartman}},
  \bibinfo {author} {\bibfnamefont {J.}~\bibnamefont {Maldacena}}, \bibinfo
  {author} {\bibfnamefont {E.}~\bibnamefont {Shaghoulian}},\ and\ \bibinfo
  {author} {\bibfnamefont {A.}~\bibnamefont {Tajdini}},\ }\href
  {https://doi.org/10.1103/RevModPhys.93.035002} {\bibfield  {journal}
  {\bibinfo  {journal} {Rev. Mod. Phys.}\ }\textbf {\bibinfo {volume} {93}},\
  \bibinfo {pages} {035002} (\bibinfo {year} {2021})},\ \Eprint
  {https://arxiv.org/abs/2006.06872} {arXiv:2006.06872 [hep-th]} \BibitemShut
  {NoStop}%
\bibitem [{\citenamefont {Leutheusser}\ and\ \citenamefont
  {Liu}(2023{\natexlab{a}})}]{Leutheusser:2021qhd}%
  \BibitemOpen
  \bibfield  {author} {\bibinfo {author} {\bibfnamefont {S.}~\bibnamefont
  {Leutheusser}}\ and\ \bibinfo {author} {\bibfnamefont {H.}~\bibnamefont
  {Liu}},\ }\href {https://doi.org/10.1103/PhysRevD.108.086019} {\bibfield
  {journal} {\bibinfo  {journal} {Phys. Rev. D}\ }\textbf {\bibinfo {volume}
  {108}},\ \bibinfo {pages} {086019} (\bibinfo {year} {2023}{\natexlab{a}})},\
  \Eprint {https://arxiv.org/abs/2110.05497} {arXiv:2110.05497 [hep-th]}
  \BibitemShut {NoStop}%
\bibitem [{\citenamefont {Leutheusser}\ and\ \citenamefont
  {Liu}(2023{\natexlab{b}})}]{Leutheusser:2021frk}%
  \BibitemOpen
  \bibfield  {author} {\bibinfo {author} {\bibfnamefont {S.~A.~W.}\
  \bibnamefont {Leutheusser}}\ and\ \bibinfo {author} {\bibfnamefont
  {H.}~\bibnamefont {Liu}},\ }\href
  {https://doi.org/10.1103/PhysRevD.108.086020} {\bibfield  {journal} {\bibinfo
   {journal} {Phys. Rev. D}\ }\textbf {\bibinfo {volume} {108}},\ \bibinfo
  {pages} {086020} (\bibinfo {year} {2023}{\natexlab{b}})},\ \Eprint
  {https://arxiv.org/abs/2112.12156} {arXiv:2112.12156 [hep-th]} \BibitemShut
  {NoStop}%
\bibitem [{\citenamefont {Papadodimas}\ and\ \citenamefont
  {Raju}(2013)}]{Papadodimas:2012aq}%
  \BibitemOpen
  \bibfield  {author} {\bibinfo {author} {\bibfnamefont {K.}~\bibnamefont
  {Papadodimas}}\ and\ \bibinfo {author} {\bibfnamefont {S.}~\bibnamefont
  {Raju}},\ }\href {https://doi.org/10.1007/JHEP10(2013)212} {\bibfield
  {journal} {\bibinfo  {journal} {JHEP}\ }\textbf {\bibinfo {volume} {10}},\
  \bibinfo {pages} {212}},\ \Eprint {https://arxiv.org/abs/1211.6767}
  {arXiv:1211.6767 [hep-th]} \BibitemShut {NoStop}%
\bibitem [{\citenamefont {Papadodimas}\ and\ \citenamefont
  {Raju}(2014{\natexlab{a}})}]{Papadodimas:2013wnh}%
  \BibitemOpen
  \bibfield  {author} {\bibinfo {author} {\bibfnamefont {K.}~\bibnamefont
  {Papadodimas}}\ and\ \bibinfo {author} {\bibfnamefont {S.}~\bibnamefont
  {Raju}},\ }\href {https://doi.org/10.1103/PhysRevLett.112.051301} {\bibfield
  {journal} {\bibinfo  {journal} {Phys. Rev. Lett.}\ }\textbf {\bibinfo
  {volume} {112}},\ \bibinfo {pages} {051301} (\bibinfo {year}
  {2014}{\natexlab{a}})},\ \Eprint {https://arxiv.org/abs/1310.6334}
  {arXiv:1310.6334 [hep-th]} \BibitemShut {NoStop}%
\bibitem [{\citenamefont {Papadodimas}\ and\ \citenamefont
  {Raju}(2014{\natexlab{b}})}]{Papadodimas:2013jku}%
  \BibitemOpen
  \bibfield  {author} {\bibinfo {author} {\bibfnamefont {K.}~\bibnamefont
  {Papadodimas}}\ and\ \bibinfo {author} {\bibfnamefont {S.}~\bibnamefont
  {Raju}},\ }\href {https://doi.org/10.1103/PhysRevD.89.086010} {\bibfield
  {journal} {\bibinfo  {journal} {Phys. Rev. D}\ }\textbf {\bibinfo {volume}
  {89}},\ \bibinfo {pages} {086010} (\bibinfo {year} {2014}{\natexlab{b}})},\
  \Eprint {https://arxiv.org/abs/1310.6335} {arXiv:1310.6335 [hep-th]}
  \BibitemShut {NoStop}%
\bibitem [{\citenamefont {Andersson}\ \emph {et~al.}(1992)\citenamefont
  {Andersson}, \citenamefont {Chrusciel},\ and\ \citenamefont
  {Friedrich}}]{Andersson:1992yk}%
  \BibitemOpen
  \bibfield  {author} {\bibinfo {author} {\bibfnamefont {L.}~\bibnamefont
  {Andersson}}, \bibinfo {author} {\bibfnamefont {P.}~\bibnamefont
  {Chrusciel}},\ and\ \bibinfo {author} {\bibfnamefont {H.}~\bibnamefont
  {Friedrich}},\ }\href {https://doi.org/10.1007/BF02096944} {\bibfield
  {journal} {\bibinfo  {journal} {Commun. Math. Phys.}\ }\textbf {\bibinfo
  {volume} {149}},\ \bibinfo {pages} {587} (\bibinfo {year}
  {1992})}\BibitemShut {NoStop}%
\bibitem [{\citenamefont {Andersson}\ and\ \citenamefont
  {Chrusciel}(1996)}]{Andersson:1996xd}%
  \BibitemOpen
  \bibfield  {author} {\bibinfo {author} {\bibfnamefont {L.}~\bibnamefont
  {Andersson}}\ and\ \bibinfo {author} {\bibfnamefont {P.~T.}\ \bibnamefont
  {Chrusciel}},\ }\href@noop {} {\bibfield  {journal} {\bibinfo  {journal}
  {Diss. Math.}\ }\textbf {\bibinfo {volume} {355}},\ \bibinfo {pages} {1}
  (\bibinfo {year} {1996})}\BibitemShut {NoStop}%
\bibitem [{\citenamefont {Sakovich}(2010)}]{Sakovich:2009nb}%
  \BibitemOpen
  \bibfield  {author} {\bibinfo {author} {\bibfnamefont {A.}~\bibnamefont
  {Sakovich}},\ }\href {https://doi.org/10.1088/0264-9381/27/24/245019}
  {\bibfield  {journal} {\bibinfo  {journal} {Class. Quant. Grav.}\ }\textbf
  {\bibinfo {volume} {27}},\ \bibinfo {pages} {245019} (\bibinfo {year}
  {2010})},\ \Eprint {https://arxiv.org/abs/0910.4178} {arXiv:0910.4178
  [gr-qc]} \BibitemShut {NoStop}%
\bibitem [{\citenamefont {Chru\'sciel}\ and\ \citenamefont
  {Galloway}(2022)}]{Chrusciel:2022cjz}%
  \BibitemOpen
  \bibfield  {author} {\bibinfo {author} {\bibfnamefont {P.~T.}\ \bibnamefont
  {Chru\'sciel}}\ and\ \bibinfo {author} {\bibfnamefont {G.~J.}\ \bibnamefont
  {Galloway}},\ }\href@noop {} {\bibinfo {title} {{Maximal hypersurfaces in
  asymptotically Anti-de Sitter spacetime}}} (\bibinfo {year} {2022}),\ \Eprint
  {https://arxiv.org/abs/2208.09893} {arXiv:2208.09893 [gr-qc]} \BibitemShut
  {NoStop}%
\bibitem [{\citenamefont {Witten}(2023)}]{Witten:2022xxp}%
  \BibitemOpen
  \bibfield  {author} {\bibinfo {author} {\bibfnamefont {E.}~\bibnamefont
  {Witten}},\ }\href {https://doi.org/10.4310/ATMP.2023.v27.n1.a6} {\bibfield
  {journal} {\bibinfo  {journal} {Adv. Theor. Math. Phys.}\ }\textbf {\bibinfo
  {volume} {27}},\ \bibinfo {pages} {311} (\bibinfo {year} {2023})},\ \Eprint
  {https://arxiv.org/abs/2212.08270} {arXiv:2212.08270 [hep-th]} \BibitemShut
  {NoStop}%
\bibitem [{\citenamefont {Witten}(2017)}]{Wittentalk}%
  \BibitemOpen
  \bibfield  {author} {\bibinfo {author} {\bibfnamefont {E.}~\bibnamefont
  {Witten}},\ }\href@noop {} {\bibinfo {title} {{\emph{Canonical Quantization
  in Anti de Sitter Space}}}},\ \bibinfo {howpublished}
  {[\href{http://www.kaltura.com/index.php/extwidget/preview/partner_id/1449362/uiconf_id/14292322/entry_id/1_iy6rwyi1/embed/auto?&flashvars[streamerType]=auto}{\textsc{link}}]}
  (\bibinfo {year} {2017}),\ \bibinfo {note} {{Talk at Princeton Center for
  Theoretical Science}}\BibitemShut {NoStop}%
\bibitem [{Note1()}]{Note1}%
  \BibitemOpen
  \bibinfo {note} {The time $\protect \bar {t}$ can be identified with the
  formula for time presented in \cite {Kaushal:2024xob}, where it was shown
  that the notion of time derived from the Einstein-Hamilton-Jacobi equation in
  terms of the configuration space data precisely matches with the time in ADM
  decomposition of the spacetime metric.}\BibitemShut {Stop}%
\bibitem [{\citenamefont {Unruh}(1976)}]{Unruh:1976db}%
  \BibitemOpen
  \bibfield  {author} {\bibinfo {author} {\bibfnamefont {W.~G.}\ \bibnamefont
  {Unruh}},\ }\href {https://doi.org/10.1103/PhysRevD.14.870} {\bibfield
  {journal} {\bibinfo  {journal} {Phys. Rev. D}\ }\textbf {\bibinfo {volume}
  {14}},\ \bibinfo {pages} {870} (\bibinfo {year} {1976})}\BibitemShut
  {NoStop}%
\bibitem [{\citenamefont {Maldacena}(2003)}]{Maldacena:2001kr}%
  \BibitemOpen
  \bibfield  {author} {\bibinfo {author} {\bibfnamefont {J.~M.}\ \bibnamefont
  {Maldacena}},\ }\href {https://doi.org/10.1088/1126-6708/2003/04/021}
  {\bibfield  {journal} {\bibinfo  {journal} {JHEP}\ }\textbf {\bibinfo
  {volume} {04}},\ \bibinfo {pages} {021}},\ \Eprint
  {https://arxiv.org/abs/hep-th/0106112} {arXiv:hep-th/0106112} \BibitemShut
  {NoStop}%
\bibitem [{\citenamefont {Shenker}\ and\ \citenamefont
  {Stanford}(2014)}]{Shenker:2013pqa}%
  \BibitemOpen
  \bibfield  {author} {\bibinfo {author} {\bibfnamefont {S.~H.}\ \bibnamefont
  {Shenker}}\ and\ \bibinfo {author} {\bibfnamefont {D.}~\bibnamefont
  {Stanford}},\ }\href {https://doi.org/10.1007/JHEP03(2014)067} {\bibfield
  {journal} {\bibinfo  {journal} {JHEP}\ }\textbf {\bibinfo {volume} {03}},\
  \bibinfo {pages} {067}},\ \Eprint {https://arxiv.org/abs/1306.0622}
  {arXiv:1306.0622 [hep-th]} \BibitemShut {NoStop}%
\bibitem [{\citenamefont {Maldacena}\ \emph {et~al.}(2016)\citenamefont
  {Maldacena}, \citenamefont {Shenker},\ and\ \citenamefont
  {Stanford}}]{Maldacena:2015waa}%
  \BibitemOpen
  \bibfield  {author} {\bibinfo {author} {\bibfnamefont {J.}~\bibnamefont
  {Maldacena}}, \bibinfo {author} {\bibfnamefont {S.~H.}\ \bibnamefont
  {Shenker}},\ and\ \bibinfo {author} {\bibfnamefont {D.}~\bibnamefont
  {Stanford}},\ }\href {https://doi.org/10.1007/JHEP08(2016)106} {\bibfield
  {journal} {\bibinfo  {journal} {JHEP}\ }\textbf {\bibinfo {volume} {08}},\
  \bibinfo {pages} {106}},\ \Eprint {https://arxiv.org/abs/1503.01409}
  {arXiv:1503.01409 [hep-th]} \BibitemShut {NoStop}%
\bibitem [{\citenamefont {Maldacena}\ and\ \citenamefont
  {Milekhin}(2021)}]{Maldacena:2019ufo}%
  \BibitemOpen
  \bibfield  {author} {\bibinfo {author} {\bibfnamefont {J.}~\bibnamefont
  {Maldacena}}\ and\ \bibinfo {author} {\bibfnamefont {A.}~\bibnamefont
  {Milekhin}},\ }\href {https://doi.org/10.1007/JHEP04(2021)258} {\bibfield
  {journal} {\bibinfo  {journal} {JHEP}\ }\textbf {\bibinfo {volume} {04}},\
  \bibinfo {pages} {258}},\ \Eprint {https://arxiv.org/abs/1912.03276}
  {arXiv:1912.03276 [hep-th]} \BibitemShut {NoStop}%
\bibitem [{\citenamefont {Almheiri}\ \emph {et~al.}(2024)\citenamefont
  {Almheiri}, \citenamefont {Milekhin},\ and\ \citenamefont
  {Swingle}}]{Almheiri:2019jqq}%
  \BibitemOpen
  \bibfield  {author} {\bibinfo {author} {\bibfnamefont {A.}~\bibnamefont
  {Almheiri}}, \bibinfo {author} {\bibfnamefont {A.}~\bibnamefont {Milekhin}},\
  and\ \bibinfo {author} {\bibfnamefont {B.}~\bibnamefont {Swingle}},\ }\href
  {https://doi.org/10.1007/JHEP08(2024)034} {\bibfield  {journal} {\bibinfo
  {journal} {JHEP}\ }\textbf {\bibinfo {volume} {08}},\ \bibinfo {pages}
  {034}},\ \Eprint {https://arxiv.org/abs/1912.04912} {arXiv:1912.04912
  [hep-th]} \BibitemShut {NoStop}%
\bibitem [{\citenamefont {Chen}\ \emph {et~al.}(2020)\citenamefont {Chen},
  \citenamefont {Qi},\ and\ \citenamefont {Zhang}}]{Chen:2020wiq}%
  \BibitemOpen
  \bibfield  {author} {\bibinfo {author} {\bibfnamefont {Y.}~\bibnamefont
  {Chen}}, \bibinfo {author} {\bibfnamefont {X.-L.}\ \bibnamefont {Qi}},\ and\
  \bibinfo {author} {\bibfnamefont {P.}~\bibnamefont {Zhang}},\ }\href
  {https://doi.org/10.1007/JHEP06(2020)121} {\bibfield  {journal} {\bibinfo
  {journal} {JHEP}\ }\textbf {\bibinfo {volume} {06}},\ \bibinfo {pages}
  {121}},\ \Eprint {https://arxiv.org/abs/2003.13147} {arXiv:2003.13147
  [hep-th]} \BibitemShut {NoStop}%
\bibitem [{\citenamefont {Gaikwad}\ \emph {et~al.}(2023)\citenamefont
  {Gaikwad}, \citenamefont {Kaushal}, \citenamefont {Mandal},\ and\
  \citenamefont {Wadia}}]{Gaikwad:2022jar}%
  \BibitemOpen
  \bibfield  {author} {\bibinfo {author} {\bibfnamefont {A.}~\bibnamefont
  {Gaikwad}}, \bibinfo {author} {\bibfnamefont {A.}~\bibnamefont {Kaushal}},
  \bibinfo {author} {\bibfnamefont {G.}~\bibnamefont {Mandal}},\ and\ \bibinfo
  {author} {\bibfnamefont {S.~R.}\ \bibnamefont {Wadia}},\ }\href
  {https://doi.org/10.1007/JHEP08(2023)171} {\bibfield  {journal} {\bibinfo
  {journal} {JHEP}\ }\textbf {\bibinfo {volume} {08}},\ \bibinfo {pages}
  {171}},\ \Eprint {https://arxiv.org/abs/2210.15579} {arXiv:2210.15579
  [hep-th]} \BibitemShut {NoStop}%
\bibitem [{\citenamefont {Estabrook}\ \emph {et~al.}(1973)\citenamefont
  {Estabrook}, \citenamefont {Wahlquist}, \citenamefont {Christensen},
  \citenamefont {DeWitt}, \citenamefont {Smarr},\ and\ \citenamefont
  {Tsiang}}]{Estabrook:1973ue}%
  \BibitemOpen
  \bibfield  {author} {\bibinfo {author} {\bibfnamefont {F.}~\bibnamefont
  {Estabrook}}, \bibinfo {author} {\bibfnamefont {H.}~\bibnamefont
  {Wahlquist}}, \bibinfo {author} {\bibfnamefont {S.}~\bibnamefont
  {Christensen}}, \bibinfo {author} {\bibfnamefont {B.}~\bibnamefont {DeWitt}},
  \bibinfo {author} {\bibfnamefont {L.}~\bibnamefont {Smarr}},\ and\ \bibinfo
  {author} {\bibfnamefont {E.}~\bibnamefont {Tsiang}},\ }\href
  {https://doi.org/10.1103/PhysRevD.7.2814} {\bibfield  {journal} {\bibinfo
  {journal} {Phys. Rev. D}\ }\textbf {\bibinfo {volume} {7}},\ \bibinfo {pages}
  {2814} (\bibinfo {year} {1973})}\BibitemShut {NoStop}%
\bibitem [{\citenamefont {Kaushal}\ \emph
  {et~al.}(2025{\natexlab{a}})\citenamefont {Kaushal}, \citenamefont
  {Prabhakar},\ and\ \citenamefont {Wadia}}]{KPWI}%
  \BibitemOpen
  \bibfield  {author} {\bibinfo {author} {\bibfnamefont {A.}~\bibnamefont
  {Kaushal}}, \bibinfo {author} {\bibfnamefont {N.~S.}\ \bibnamefont
  {Prabhakar}},\ and\ \bibinfo {author} {\bibfnamefont {S.~R.}\ \bibnamefont
  {Wadia}},\ }\href@noop {} {\bibinfo {title} {2+1 dimensional gravity in aads
  spacetimes with spatial wormhole slices: Reduced phase space dynamics and the
  btz black hole}} (\bibinfo {year} {2025}{\natexlab{a}})\BibitemShut {NoStop}%
\bibitem [{\citenamefont {Banados}\ \emph {et~al.}(1992)\citenamefont
  {Banados}, \citenamefont {Teitelboim},\ and\ \citenamefont
  {Zanelli}}]{Banados:1992wn}%
  \BibitemOpen
  \bibfield  {author} {\bibinfo {author} {\bibfnamefont {M.}~\bibnamefont
  {Banados}}, \bibinfo {author} {\bibfnamefont {C.}~\bibnamefont
  {Teitelboim}},\ and\ \bibinfo {author} {\bibfnamefont {J.}~\bibnamefont
  {Zanelli}},\ }\href {https://doi.org/10.1103/PhysRevLett.69.1849} {\bibfield
  {journal} {\bibinfo  {journal} {Phys. Rev. Lett.}\ }\textbf {\bibinfo
  {volume} {69}},\ \bibinfo {pages} {1849} (\bibinfo {year} {1992})},\ \Eprint
  {https://arxiv.org/abs/hep-th/9204099} {arXiv:hep-th/9204099} \BibitemShut
  {NoStop}%
\bibitem [{\citenamefont {Banados}\ \emph {et~al.}(1993)\citenamefont
  {Banados}, \citenamefont {Henneaux}, \citenamefont {Teitelboim},\ and\
  \citenamefont {Zanelli}}]{Banados:1992gq}%
  \BibitemOpen
  \bibfield  {author} {\bibinfo {author} {\bibfnamefont {M.}~\bibnamefont
  {Banados}}, \bibinfo {author} {\bibfnamefont {M.}~\bibnamefont {Henneaux}},
  \bibinfo {author} {\bibfnamefont {C.}~\bibnamefont {Teitelboim}},\ and\
  \bibinfo {author} {\bibfnamefont {J.}~\bibnamefont {Zanelli}},\ }\href
  {https://doi.org/10.1103/PhysRevD.48.1506} {\bibfield  {journal} {\bibinfo
  {journal} {Phys. Rev. D}\ }\textbf {\bibinfo {volume} {48}},\ \bibinfo
  {pages} {1506} (\bibinfo {year} {1993})},\ \bibinfo {note} {[Erratum:
  Phys.Rev.D 88, 069902 (2013)]},\ \Eprint
  {https://arxiv.org/abs/gr-qc/9302012} {arXiv:gr-qc/9302012} \BibitemShut
  {NoStop}%
\bibitem [{\citenamefont {Gourgoulhon}(2007)}]{Gourgoulhon:2007ue}%
  \BibitemOpen
  \bibfield  {author} {\bibinfo {author} {\bibfnamefont {E.}~\bibnamefont
  {Gourgoulhon}},\ }\href@noop {} {\bibinfo {title} {{3+1 formalism and bases
  of numerical relativity}}} (\bibinfo {year} {2007}),\ \Eprint
  {https://arxiv.org/abs/gr-qc/0703035} {arXiv:gr-qc/0703035} \BibitemShut
  {NoStop}%
\bibitem [{\citenamefont {Baumgarte}\ and\ \citenamefont
  {Shapiro}(2010)}]{Baumgarte:2010ndz}%
  \BibitemOpen
  \bibfield  {author} {\bibinfo {author} {\bibfnamefont {T.~W.}\ \bibnamefont
  {Baumgarte}}\ and\ \bibinfo {author} {\bibfnamefont {S.~L.}\ \bibnamefont
  {Shapiro}},\ }\href {https://doi.org/10.1017/CBO9781139193344} {\emph
  {\bibinfo {title} {{Numerical Relativity: Solving Einstein's Equations on the
  Computer}}}}\ (\bibinfo  {publisher} {Cambridge University Press},\ \bibinfo
  {year} {2010})\BibitemShut {NoStop}%
\bibitem [{Note2()}]{Note2}%
  \BibitemOpen
  \bibinfo {note} {This is consistent with the observations in \cite
  {Bahiru:2022oas,Antonini:2025sur} regarding the existence of local
  gauge-invariant observables in perturbation theory about backgrounds which do
  not have isometries.}\BibitemShut {Stop}%
\bibitem [{\citenamefont {Avis}\ \emph {et~al.}(1978)\citenamefont {Avis},
  \citenamefont {Isham},\ and\ \citenamefont {Storey}}]{Avis:1977yn}%
  \BibitemOpen
  \bibfield  {author} {\bibinfo {author} {\bibfnamefont {S.~J.}\ \bibnamefont
  {Avis}}, \bibinfo {author} {\bibfnamefont {C.~J.}\ \bibnamefont {Isham}},\
  and\ \bibinfo {author} {\bibfnamefont {D.}~\bibnamefont {Storey}},\ }\href
  {https://doi.org/10.1103/PhysRevD.18.3565} {\bibfield  {journal} {\bibinfo
  {journal} {Phys. Rev. D}\ }\textbf {\bibinfo {volume} {18}},\ \bibinfo
  {pages} {3565} (\bibinfo {year} {1978})}\BibitemShut {NoStop}%
\bibitem [{\citenamefont {Breitenlohner}\ and\ \citenamefont
  {Freedman}(1982{\natexlab{a}})}]{Breitenlohner:1982bm}%
  \BibitemOpen
  \bibfield  {author} {\bibinfo {author} {\bibfnamefont {P.}~\bibnamefont
  {Breitenlohner}}\ and\ \bibinfo {author} {\bibfnamefont {D.~Z.}\ \bibnamefont
  {Freedman}},\ }\href {https://doi.org/10.1016/0370-2693(82)90643-8}
  {\bibfield  {journal} {\bibinfo  {journal} {Phys. Lett. B}\ }\textbf
  {\bibinfo {volume} {115}},\ \bibinfo {pages} {197} (\bibinfo {year}
  {1982}{\natexlab{a}})}\BibitemShut {NoStop}%
\bibitem [{\citenamefont {Breitenlohner}\ and\ \citenamefont
  {Freedman}(1982{\natexlab{b}})}]{Breitenlohner:1982jf}%
  \BibitemOpen
  \bibfield  {author} {\bibinfo {author} {\bibfnamefont {P.}~\bibnamefont
  {Breitenlohner}}\ and\ \bibinfo {author} {\bibfnamefont {D.~Z.}\ \bibnamefont
  {Freedman}},\ }\href {https://doi.org/10.1016/0003-4916(82)90116-6}
  {\bibfield  {journal} {\bibinfo  {journal} {Annals Phys.}\ }\textbf {\bibinfo
  {volume} {144}},\ \bibinfo {pages} {249} (\bibinfo {year}
  {1982}{\natexlab{b}})}\BibitemShut {NoStop}%
\bibitem [{\citenamefont {Hawking}(1983)}]{Hawking:1983mx}%
  \BibitemOpen
  \bibfield  {author} {\bibinfo {author} {\bibfnamefont {S.~W.}\ \bibnamefont
  {Hawking}},\ }\href {https://doi.org/10.1016/0370-2693(83)90585-3} {\bibfield
   {journal} {\bibinfo  {journal} {Phys. Lett. B}\ }\textbf {\bibinfo {volume}
  {126}},\ \bibinfo {pages} {175} (\bibinfo {year} {1983})}\BibitemShut
  {NoStop}%
\bibitem [{\citenamefont {Henneaux}\ and\ \citenamefont
  {Teitelboim}(1984)}]{Henneaux:1984xu}%
  \BibitemOpen
  \bibfield  {author} {\bibinfo {author} {\bibfnamefont {M.}~\bibnamefont
  {Henneaux}}\ and\ \bibinfo {author} {\bibfnamefont {C.}~\bibnamefont
  {Teitelboim}},\ }\href {https://doi.org/10.1016/0370-2693(84)91339-X}
  {\bibfield  {journal} {\bibinfo  {journal} {Phys. Lett. B}\ }\textbf
  {\bibinfo {volume} {142}},\ \bibinfo {pages} {355} (\bibinfo {year}
  {1984})}\BibitemShut {NoStop}%
\bibitem [{\citenamefont {Mezincescu}\ and\ \citenamefont
  {Townsend}(1985)}]{Mezincescu:1984ev}%
  \BibitemOpen
  \bibfield  {author} {\bibinfo {author} {\bibfnamefont {L.}~\bibnamefont
  {Mezincescu}}\ and\ \bibinfo {author} {\bibfnamefont {P.~K.}\ \bibnamefont
  {Townsend}},\ }\href {https://doi.org/10.1016/0003-4916(85)90150-2}
  {\bibfield  {journal} {\bibinfo  {journal} {Annals Phys.}\ }\textbf {\bibinfo
  {volume} {160}},\ \bibinfo {pages} {406} (\bibinfo {year}
  {1985})}\BibitemShut {NoStop}%
\bibitem [{\citenamefont {Henneaux}\ and\ \citenamefont
  {Teitelboim}(1985)}]{Henneaux:1985tv}%
  \BibitemOpen
  \bibfield  {author} {\bibinfo {author} {\bibfnamefont {M.}~\bibnamefont
  {Henneaux}}\ and\ \bibinfo {author} {\bibfnamefont {C.}~\bibnamefont
  {Teitelboim}},\ }\href {https://doi.org/10.1007/BF01205790} {\bibfield
  {journal} {\bibinfo  {journal} {Commun. Math. Phys.}\ }\textbf {\bibinfo
  {volume} {98}},\ \bibinfo {pages} {391} (\bibinfo {year} {1985})}\BibitemShut
  {NoStop}%
\bibitem [{\citenamefont {Lifschytz}\ and\ \citenamefont
  {Ortiz}(1994)}]{Lifschytz:1993eb}%
  \BibitemOpen
  \bibfield  {author} {\bibinfo {author} {\bibfnamefont {G.}~\bibnamefont
  {Lifschytz}}\ and\ \bibinfo {author} {\bibfnamefont {M.}~\bibnamefont
  {Ortiz}},\ }\href {https://doi.org/10.1103/PhysRevD.49.1929} {\bibfield
  {journal} {\bibinfo  {journal} {Phys. Rev. D}\ }\textbf {\bibinfo {volume}
  {49}},\ \bibinfo {pages} {1929} (\bibinfo {year} {1994})},\ \Eprint
  {https://arxiv.org/abs/gr-qc/9310008} {arXiv:gr-qc/9310008} \BibitemShut
  {NoStop}%
\bibitem [{\citenamefont {Ichinose}\ and\ \citenamefont
  {Satoh}(1995)}]{Ichinose:1994rg}%
  \BibitemOpen
  \bibfield  {author} {\bibinfo {author} {\bibfnamefont {I.}~\bibnamefont
  {Ichinose}}\ and\ \bibinfo {author} {\bibfnamefont {Y.}~\bibnamefont
  {Satoh}},\ }\href {https://doi.org/10.1016/0550-3213(95)00197-Z} {\bibfield
  {journal} {\bibinfo  {journal} {Nucl. Phys. B}\ }\textbf {\bibinfo {volume}
  {447}},\ \bibinfo {pages} {340} (\bibinfo {year} {1995})},\ \Eprint
  {https://arxiv.org/abs/hep-th/9412144} {arXiv:hep-th/9412144} \BibitemShut
  {NoStop}%
\bibitem [{\citenamefont {Balasubramanian}\ \emph {et~al.}(1999)\citenamefont
  {Balasubramanian}, \citenamefont {Kraus},\ and\ \citenamefont
  {Lawrence}}]{Balasubramanian:1998sn}%
  \BibitemOpen
  \bibfield  {author} {\bibinfo {author} {\bibfnamefont {V.}~\bibnamefont
  {Balasubramanian}}, \bibinfo {author} {\bibfnamefont {P.}~\bibnamefont
  {Kraus}},\ and\ \bibinfo {author} {\bibfnamefont {A.~E.}\ \bibnamefont
  {Lawrence}},\ }\href {https://doi.org/10.1103/PhysRevD.59.046003} {\bibfield
  {journal} {\bibinfo  {journal} {Phys. Rev. D}\ }\textbf {\bibinfo {volume}
  {59}},\ \bibinfo {pages} {046003} (\bibinfo {year} {1999})},\ \Eprint
  {https://arxiv.org/abs/hep-th/9805171} {arXiv:hep-th/9805171} \BibitemShut
  {NoStop}%
\bibitem [{\citenamefont {Keski-Vakkuri}(1999)}]{Keski-Vakkuri:1998gmz}%
  \BibitemOpen
  \bibfield  {author} {\bibinfo {author} {\bibfnamefont {E.}~\bibnamefont
  {Keski-Vakkuri}},\ }\href {https://doi.org/10.1103/PhysRevD.59.104001}
  {\bibfield  {journal} {\bibinfo  {journal} {Phys. Rev. D}\ }\textbf {\bibinfo
  {volume} {59}},\ \bibinfo {pages} {104001} (\bibinfo {year} {1999})},\
  \Eprint {https://arxiv.org/abs/hep-th/9808037} {arXiv:hep-th/9808037}
  \BibitemShut {NoStop}%
\bibitem [{\citenamefont {Ishibashi}\ and\ \citenamefont
  {Wald}(2004)}]{Ishibashi:2004wx}%
  \BibitemOpen
  \bibfield  {author} {\bibinfo {author} {\bibfnamefont {A.}~\bibnamefont
  {Ishibashi}}\ and\ \bibinfo {author} {\bibfnamefont {R.~M.}\ \bibnamefont
  {Wald}},\ }\href {https://doi.org/10.1088/0264-9381/21/12/012} {\bibfield
  {journal} {\bibinfo  {journal} {Class. Quant. Grav.}\ }\textbf {\bibinfo
  {volume} {21}},\ \bibinfo {pages} {2981} (\bibinfo {year} {2004})},\ \Eprint
  {https://arxiv.org/abs/hep-th/0402184} {arXiv:hep-th/0402184} \BibitemShut
  {NoStop}%
\bibitem [{Note3()}]{Note3}%
  \BibitemOpen
  \bibinfo {note} {The coordinates $(Q,P)$ are related to by a canonical
  transformation to Kuchar's mass and time shift variables \cite
  {Kuchar:1994zk}, for the BTZ black hole.}\BibitemShut {Stop}%
\bibitem [{Note4()}]{Note4}%
  \BibitemOpen
  \bibinfo {note} {There may be other choices for the smearing functions which
  achieve similar results. For instance, we can take Hermite functions $\psi
  _n(\omega / \alpha )$ for any positive real parameter $\alpha $. The Hermite
  functions $\psi _n(z)$ are defined in terms of the Hermite polynomials
  $H_n(z)$ as \begin {align*} \psi _n(z) &= \protect \frac {1}{(\protect \sqrt
  {\pi } 2^n n!)^{1/2}} \protect \text {e}^{-z^2/2} H_n(z)\protect \, ,\\
  H_n(z) &= (-1)^n \protect \text {e}^{z^2} \protect \frac {\protect \text
  {d}^n}{\protect \text {d}z^n} \protect \text {e}^{-z^2}\protect \, . \end
  {align*} They satisfy the orthogonality and completeness relations \begin
  {align*} &\DOTSI \intop \ilimits@ _{-\infty }^\infty \protect \text
  {d}z\protect \, \psi _n(z)\psi _m(z) = \delta _{nm}\protect \, , \\ &\DOTSB
  \sum@ \slimits@ _{n=0}^\infty \psi _n(z) \psi _n(z') = \delta (z-z')\protect
  \, . \end {align*}}\BibitemShut {NoStop}%
\bibitem [{\citenamefont {Hamilton}\ \emph {et~al.}(2006)\citenamefont
  {Hamilton}, \citenamefont {Kabat}, \citenamefont {Lifschytz},\ and\
  \citenamefont {Lowe}}]{Hamilton:2006az}%
  \BibitemOpen
  \bibfield  {author} {\bibinfo {author} {\bibfnamefont {A.}~\bibnamefont
  {Hamilton}}, \bibinfo {author} {\bibfnamefont {D.~N.}\ \bibnamefont {Kabat}},
  \bibinfo {author} {\bibfnamefont {G.}~\bibnamefont {Lifschytz}},\ and\
  \bibinfo {author} {\bibfnamefont {D.~A.}\ \bibnamefont {Lowe}},\ }\href
  {https://doi.org/10.1103/PhysRevD.74.066009} {\bibfield  {journal} {\bibinfo
  {journal} {Phys. Rev. D}\ }\textbf {\bibinfo {volume} {74}},\ \bibinfo
  {pages} {066009} (\bibinfo {year} {2006})},\ \Eprint
  {https://arxiv.org/abs/hep-th/0606141} {arXiv:hep-th/0606141} \BibitemShut
  {NoStop}%
\bibitem [{Note5()}]{Note5}%
  \BibitemOpen
  \bibinfo {note} {We can split the $x$ integral as a bulk term and two terms
  near the boundary at $x=\pm \infty $, \begin {equation*} \DOTSI \intop
  \ilimits@ _{-\infty }^\infty \protect \text {d}x = \DOTSI \intop \ilimits@
  _{-\infty }^{-X} \protect \text {d}x + \DOTSI \intop \ilimits@ _{-X}^{X}
  \protect \text {d}x + \DOTSI \intop \ilimits@ _{X}^\infty \protect \text {d}x
  \end {equation*} for some large and positive value $X$. The bulk term is
  finite since all the quantities are smooth and finite. Near the boundary, the
  mass term in $A_{nn'qq'}$ goes as, \begin {equation*} m^2\DOTSI \intop
  \ilimits@ _{X}^\infty \protect \text {d}x \protect \sqrt {g}N \protect \hat
  {g}^*_{n'q'} \protect \hat {g}_{nq} \propto \DOTSI \intop \ilimits@
  _{X}^\infty \protect \text {d}x\protect \, x^{1-2\Delta _+} \sim
  x^{2(1-\Delta _+)} \protect \Big |_X^\infty , \end {equation*} which is
  finite for $\Delta _+ >1$. Similarly contribution from all the other terms is
  also finite.}\BibitemShut {Stop}%
\bibitem [{Note6()}]{Note6}%
  \BibitemOpen
  \bibinfo {note} {The time evolution operator satisfies the differential
  equation \begin {equation*} \partial _{\protect \bar t} U(\protect \bar t,0)
  = -i U(\protect \bar t,0) H(\protect \bar t) , \end {equation*} where
  $H(\protect \bar t)$ is the Hamiltonian in the Heisenberg picture.
  Integrating this equation gives \begin {equation*} U(\protect \bar t,0) = 1 -
  \protect \text {i}\DOTSI \intop \ilimits@ _{0}^{\protect \bar t} \protect
  \text {d}t' U(\protect \bar t',0) H(\protect \bar t') . \end {equation*} We
  can iterate this further to get the Dyson series \begin {align*} U(\protect
  \bar t,0) =& 1 - \protect \text {i}\DOTSI \intop \ilimits@ _{0}^{\protect
  \bar t} \protect \text {d}\protect \bar t' H(\protect \bar t') + (-\protect
  \text {i})^2 \DOTSI \intop \ilimits@ _{0}^{\protect \bar t} \protect \text
  {d}\protect \bar t' \DOTSI \intop \ilimits@ _{0}^{\protect \bar t'} \protect
  \text {d}\protect \bar t'' H(\protect \bar t'') H(\protect \bar t') \\ &+
  \protect \cdots + (-\protect \text {i})^n \DOTSI \intop \ilimits@
  _{0}^{\protect \bar t} \protect \text {d}\protect \bar t' \DOTSI \intop
  \ilimits@ _{0}^{\protect \bar t'} \protect \text {d}\protect \bar t''
  \protect \cdots \DOTSI \intop \ilimits@ _{0}^{\protect \bar t^{(n-1)}}
  \protect \text {d}\protect \bar t^{(n)} \times \\ & \times H(\protect \bar
  t^{(n)}) \protect \cdots H(\protect \bar t'') H(\protect \bar t') + \protect
  \cdots , \end {align*} which is written compactly as anti time-ordered
  exponential in \protect \eqref {UOperator}.}\BibitemShut {Stop}%
\bibitem [{Note7()}]{Note7}%
  \BibitemOpen
  \bibinfo {note} {$h_r$ and $h_l$ are the bulk Hamiltonians that evolve in the
  Schwarzschild time $t$ and $\protect \tilde {t}$ in Region I and Region II
  respectively, see equation 2.7 of \cite {Witten:2021unn} for the precise
  definitions.}\BibitemShut {Stop}%
\bibitem [{Note8()}]{Note8}%
  \BibitemOpen
  \bibinfo {note} {It is important that the Cauchy slice is not split, since
  there can be divergent fluctuations near the splitting point.}\BibitemShut
  {Stop}%
\bibitem [{Note9()}]{Note9}%
  \BibitemOpen
  \bibinfo {note} {This is analogous to evolution with the Minkowski
  Hamiltonian as opposed to the Rindler time evolution. Minkowski time slices
  enter the Milne region while the Rindler slices do not.}\BibitemShut {Stop}%
\bibitem [{Note10()}]{Note10}%
  \BibitemOpen
  \bibinfo {note} {Another justification for the bound on the range of $n$ is
  by numerically studying the mode functions $g_{nq}$. For example, using the
  near horizon expressions, one can check that their magnitude decreases with
  increasing $n$.}\BibitemShut {Stop}%
\bibitem [{\citenamefont {Witten}(2022)}]{Witten:2021unn}%
  \BibitemOpen
  \bibfield  {author} {\bibinfo {author} {\bibfnamefont {E.}~\bibnamefont
  {Witten}},\ }\href {https://doi.org/10.1007/JHEP10(2022)008} {\bibfield
  {journal} {\bibinfo  {journal} {JHEP}\ }\textbf {\bibinfo {volume} {10}},\
  \bibinfo {pages} {008}},\ \Eprint {https://arxiv.org/abs/2112.12828}
  {arXiv:2112.12828 [hep-th]} \BibitemShut {NoStop}%
\bibitem [{Note11()}]{Note11}%
  \BibitemOpen
  \bibinfo {note} {An example of a smooth function with compact support is
  given by \begin {equation} f(x) = \begin {cases} \exp \left (-\protect \frac
  {1}{x^2-a^2}\right ) & |x|<a \\ 0 & |x|>a , \end {cases} \end {equation}
  where $a$ is positive number that controls the support of the function
  $f(x)$.}\BibitemShut {Stop}%
\bibitem [{Note12()}]{Note12}%
  \BibitemOpen
  \bibinfo {note} {This is because the scalar field $\phi (\protect \bar
  t,x,\varphi )$ in \protect \eqref {scalargnx} itself satisfies the
  Klein-Gordon equation \protect \eqref {KGwave}.}\BibitemShut {Stop}%
\bibitem [{Note13()}]{Note13}%
  \BibitemOpen
  \bibinfo {note} {See also \cite {Mandal:2014wfa} for computation of the
  left-right boundary correlator using the bulk geodesic approximation in the
  large scalar field mass limit.}\BibitemShut {Stop}%
\bibitem [{Note14()}]{Note14}%
  \BibitemOpen
  \bibinfo {note} {Note that any correlator with insertions in region $R_0$ can
  expressed in terms of boundary operators. We consider this correlator as a
  specific example.}\BibitemShut {Stop}%
\bibitem [{Note15()}]{Note15}%
  \BibitemOpen
  \bibinfo {note} {See also the textbook \cite {Baumgarte:2010ndz} for a
  detailed discussion}\BibitemShut {NoStop}%
\bibitem [{Note16()}]{Note16}%
  \BibitemOpen
  \bibinfo {note} {It is convenient to use a Mathematica package such as
  diffgeo.m by Matthew Headrick.}\BibitemShut {Stop}%
\bibitem [{{\relax DLMF}()}]{NIST:DLMF}%
  \BibitemOpen
  {\relax DLMF},\ \href {https://dlmf.nist.gov/} {\bibinfo {title} {{\it NIST
  Digital Library of Mathematical Functions}}},\ \bibinfo {howpublished}
  {\url{https://dlmf.nist.gov/}, Release 1.2.3 of 2024-12-15},\ \bibinfo {note}
  {f.~W.~J. Olver, A.~B. {Olde Daalhuis}, D.~W. Lozier, B.~I. Schneider, R.~F.
  Boisvert, C.~W. Clark, B.~R. Miller, B.~V. Saunders, H.~S. Cohl, and M.~A.
  McClain, eds.}\BibitemShut {Stop}%
\bibitem [{Note17()}]{Note17}%
  \BibitemOpen
  \bibinfo {note} {To obtain this expression, we have absorbed the phase
  $\protect \text {e}^{-\protect \text {i}\pi 2\alpha }$ into $(1-\rho
  )^{-2\alpha }$ to obtain $(\rho -1)^{-2\alpha }$: \begin {equation*} \protect
  \text {e}^{-\protect \text {i}\pi 2\alpha } (1-\rho )^{-2\alpha } = \protect
  \text {e}^{-2\alpha \log \protect \big ((1-\rho ) \protect \text
  {e}^{-\protect \text {i}\pi }\protect \big )} = (\rho -1)^{-2\alpha }\protect
  \, . \end {equation*}}\BibitemShut {NoStop}%
\bibitem [{\citenamefont {Kaushal}\ \emph
  {et~al.}(2025{\natexlab{b}})\citenamefont {Kaushal}, \citenamefont
  {Prabhakar},\ and\ \citenamefont {Wadia}}]{Kaushal:2024xob}%
  \BibitemOpen
  \bibfield  {author} {\bibinfo {author} {\bibfnamefont {A.}~\bibnamefont
  {Kaushal}}, \bibinfo {author} {\bibfnamefont {N.~S.}\ \bibnamefont
  {Prabhakar}},\ and\ \bibinfo {author} {\bibfnamefont {S.~R.}\ \bibnamefont
  {Wadia}},\ }\href {https://doi.org/10.1103/PhysRevD.111.106006} {\bibfield
  {journal} {\bibinfo  {journal} {Phys. Rev. D}\ }\textbf {\bibinfo {volume}
  {111}},\ \bibinfo {pages} {106006} (\bibinfo {year} {2025}{\natexlab{b}})},\
  \Eprint {https://arxiv.org/abs/2405.18486} {arXiv:2405.18486 [hep-th]}
  \BibitemShut {NoStop}%
\bibitem [{\citenamefont {Bahiru}\ \emph {et~al.}(2023)\citenamefont {Bahiru},
  \citenamefont {Belin}, \citenamefont {Papadodimas}, \citenamefont {Sarosi},\
  and\ \citenamefont {Vardian}}]{Bahiru:2022oas}%
  \BibitemOpen
  \bibfield  {author} {\bibinfo {author} {\bibfnamefont {E.}~\bibnamefont
  {Bahiru}}, \bibinfo {author} {\bibfnamefont {A.}~\bibnamefont {Belin}},
  \bibinfo {author} {\bibfnamefont {K.}~\bibnamefont {Papadodimas}}, \bibinfo
  {author} {\bibfnamefont {G.}~\bibnamefont {Sarosi}},\ and\ \bibinfo {author}
  {\bibfnamefont {N.}~\bibnamefont {Vardian}},\ }\href
  {https://doi.org/10.1103/PhysRevD.108.086035} {\bibfield  {journal} {\bibinfo
   {journal} {Phys. Rev. D}\ }\textbf {\bibinfo {volume} {108}},\ \bibinfo
  {pages} {086035} (\bibinfo {year} {2023})},\ \Eprint
  {https://arxiv.org/abs/2209.06845} {arXiv:2209.06845 [hep-th]} \BibitemShut
  {NoStop}%
\bibitem [{\citenamefont {Antonini}\ \emph {et~al.}(2025)\citenamefont
  {Antonini}, \citenamefont {Chen}, \citenamefont {Maxfield},\ and\
  \citenamefont {Penington}}]{Antonini:2025sur}%
  \BibitemOpen
  \bibfield  {author} {\bibinfo {author} {\bibfnamefont {S.}~\bibnamefont
  {Antonini}}, \bibinfo {author} {\bibfnamefont {C.-H.}\ \bibnamefont {Chen}},
  \bibinfo {author} {\bibfnamefont {H.}~\bibnamefont {Maxfield}},\ and\
  \bibinfo {author} {\bibfnamefont {G.}~\bibnamefont {Penington}},\ }\href@noop
  {} {\bibinfo {title} {{An apologia for islands}}} (\bibinfo {year} {2025}),\
  \Eprint {https://arxiv.org/abs/2506.04311} {arXiv:2506.04311 [hep-th]}
  \BibitemShut {NoStop}%
\bibitem [{\citenamefont {Kuchar}(1994)}]{Kuchar:1994zk}%
  \BibitemOpen
  \bibfield  {author} {\bibinfo {author} {\bibfnamefont {K.~V.}\ \bibnamefont
  {Kuchar}},\ }\href {https://doi.org/10.1103/PhysRevD.50.3961} {\bibfield
  {journal} {\bibinfo  {journal} {Phys. Rev. D}\ }\textbf {\bibinfo {volume}
  {50}},\ \bibinfo {pages} {3961} (\bibinfo {year} {1994})},\ \Eprint
  {https://arxiv.org/abs/gr-qc/9403003} {arXiv:gr-qc/9403003} \BibitemShut
  {NoStop}%
\bibitem [{\citenamefont {Mandal}\ \emph {et~al.}(2015)\citenamefont {Mandal},
  \citenamefont {Sinha},\ and\ \citenamefont {Sorokhaibam}}]{Mandal:2014wfa}%
  \BibitemOpen
  \bibfield  {author} {\bibinfo {author} {\bibfnamefont {G.}~\bibnamefont
  {Mandal}}, \bibinfo {author} {\bibfnamefont {R.}~\bibnamefont {Sinha}},\ and\
  \bibinfo {author} {\bibfnamefont {N.}~\bibnamefont {Sorokhaibam}},\ }\href
  {https://doi.org/10.1007/JHEP01(2015)036} {\bibfield  {journal} {\bibinfo
  {journal} {JHEP}\ }\textbf {\bibinfo {volume} {01}},\ \bibinfo {pages}
  {036}},\ \Eprint {https://arxiv.org/abs/1405.6695} {arXiv:1405.6695 [hep-th]}
  \BibitemShut {NoStop}%
\end{thebibliography}%

\end{document}